\documentclass[12pt,letterpaper,a4paper]{article}
\usepackage[includeheadfoot,
marginratio={1:1,2:3},
width=500pt,
height=720pt,]{geometry}

\usepackage{amsmath}
\usepackage{amsfonts}
\usepackage{amssymb}
\usepackage{graphicx}
\usepackage{cancel}
\usepackage{empheq}
\usepackage{color}
\usepackage{hyperref}

\numberwithin{equation}{section}
\usepackage{float}
\restylefloat{table}
\restylefloat{figure}
\usepackage[utf8]{inputenc}
\usepackage{amsmath, amsthm, amssymb, amsfonts}
\usepackage[font=small, labelfont=bf]{caption}
\usepackage{amsmath, amsthm, amssymb, amsfonts}
\usepackage{multirow}
\usepackage{graphicx}
\usepackage{float}
\usepackage[numbers,sort&compress]{natbib}
\usepackage{enumerate}
\usepackage{amsmath}
\usepackage{amsfonts}
\usepackage{amssymb}
\usepackage{graphicx}
\usepackage{cancel}
\usepackage{empheq}
\usepackage{color}
\usepackage{hyperref}
\usepackage{float}
\restylefloat{table}
\restylefloat{figure}

\newcommand{\nc}{\newcommand}
\nc{\beq}{\begin{equation}}
\nc{\eeq}{\end{equation}}
\nc{\bea}{\begin{eqnarray}}
\nc{\eea}{\end{eqnarray}}

\def\ov{\overline}

\begin{document}
	{\hfill
		%
		arXiv:2407.15822}

	\vspace{1.0cm}
	\begin{center}
		{\Large
			Reading-off the non-geometric scalar potentials with U-dual fluxes}
		\vspace{0.4cm}
	\end{center}

	\vspace{0.35cm}
	\begin{center}
		Sayan  Biswas$^\diamond$, George K. Leontaris$^{\dagger}$, and Pramod Shukla$^\diamond$ \footnote{Email: sayanbis9920@gmail.com, leonta@uoi.gr, pshukla@jcbose.ac.in}
	\end{center}

\vspace{0.1cm}
\begin{center}
{$^\diamond$ Department of Physical Sciences, Bose Institute,\\
Unified Academic Campus, EN 80, Sector V, Bidhannagar, Kolkata 700091, India.}\\
 \vskip0.5cm
{$^\dagger$ Physics Department, University of Ioannina, \\
University Campus, Ioannina 45110, Greece.}
\end{center}
\vspace{1cm}

\abstract{In the context of four-dimensional ${\cal N} = 1$ type IIB superstring compactifications, the U-dual completion of the holomorphic flux superpotential leads to four S-dual pairs of fluxes, namely $(F, H), \, (Q, P), \, (P^\prime, Q^\prime)$ and $(H^\prime, F^\prime)$. It has been observed that the scalar potentials induced by such generalized superpotentials typically have an enormous amount of terms, making it hard to study any phenomenological aspects such as moduli stabilization. In  this regard, we present a set of generic master formulae which not only formulate the scalar potential in a compact way but also enable one to {\it read-off} the various scalar potential pieces by simply knowing a set of topological data of the compactifying Calabi Yau and its mirror threefold. We demonstrate the applicability of our master formulae by {\it reading-off} the scalar potentials for five explicit models, and using a set of {\it axionic flux} combinations we show that 76276 terms arising from the flux superpotential in a ${\mathbb T}^6/({\mathbb Z}_2\times{\mathbb Z}_2)$-based model can be equivalently expressed by using 2816 terms, while 11212 terms arising from the flux superpotential in a Quintic-based model can be equivalently expressed by 668 terms! We argue that the master formulae presented in this work can be useful in an analytic exploration of the rich landscape of the non-geometric flux vacua.
}

\clearpage

\tableofcontents


\section{Introduction}
\label{sec_intro}

In the context of superstring compactifications, the study of effective potentials induced by fluxes has attracted a lot of attention during the last two decades \cite{Angelantonj:2002ct, Sagnotti:1987tw, Derendinger:2004jn,Shelton:2005cf,Grana:2005jc, Douglas:2006es, Blumenhagen:2006ci, Grana:2012rr,Dibitetto:2012rk, Danielsson:2012by, Blaback:2013ht, Damian:2013dq, Damian:2013dwa, Hassler:2014mla, Ihl:2007ah, deCarlos:2009qm, Danielsson:2009ff, Blaback:2015zra, Dibitetto:2011qs, Plauschinn:2018wbo,Damian:2023ote}, and backgrounds with non-geometric fluxes have emerged as interesting playgrounds for initiating some alternate phenomenological model building \cite{Hassler:2014mla, Blumenhagen:2015qda, Blumenhagen:2015kja,Blumenhagen:2015jva, Blumenhagen:2015xpa,  Li:2015taa, Blumenhagen:2015xpa}. In this regard, toroidal orientifolds despite being simple have served as a promising toolkit for model building studies such as moduli stabilization and the search of physical vacua \cite{Kachru:2003aw, Balasubramanian:2005zx, Grana:2005jc, Blumenhagen:2006ci, Douglas:2006es, Denef:2005mm, Blumenhagen:2007sm}. During the initial phase, model building efforts using non-geometric fluxes have been made mostly via considering the 4D effective scalar potentials arising from the K\"ahler potentials and the flux superpotentials \cite{Danielsson:2012by, Blaback:2013ht, Damian:2013dq, Damian:2013dwa, Blumenhagen:2013hva, Villadoro:2005cu, Robbins:2007yv, Ihl:2007ah, Gao:2015nra}, and a proper understanding of the higher-dimensional origin of such 4D non-geometric scalar potentials was lacking until recent developments in \cite{Villadoro:2005cu, Blumenhagen:2013hva, Gao:2015nra,Shukla:2015rua, Shukla:2015bca,Gao:2017gxk,Shukla:2019wfo,Leontaris:2023lfc,Leontaris:2023mmm}. 

In fact, the main obstacle in understanding the higher dimensional origin of the 4D effective potentials in the beyond toroidal cases, e.g. those based on the Calabi Yau (CY) orientifolds, lies in the fact that the explicit form of the metric for a CY threefold is not known. However, given the existence of some close connections between the 4D effective potentials of type II supergravities and the symplectic geometries, the requirement of CY metric can be bypassed by using (a combination of) period vectors and moduli space metrics \cite{Ceresole:1995ca, Taylor:1999ii, D'Auria:2007ay}. This strategy of bypassing the need of CY metric has been subsequently adopted for a series of type IIA/IIB models with various (non-)geometric fluxes, e.g. see \cite{Shukla:2015hpa,Blumenhagen:2015lta,Shukla:2016hyy,Shukla:2019wfo,Shukla:2019dqd,Shukla:2019akv,Leontaris:2023lfc,Leontaris:2023mmm} for type IIB based models, \cite{Gao:2017gxk,Shukla:2019wfo,Marchesano:2020uqz,Prieto:2024shz} for type IIA based models, and \cite{Marchesano:2021gyv} for F-theory based models. For example, the generic 4D scalar potential in type IIB model induced by the standard RR and NS-NS flux pair $(F, H)$ can be equivalently computed via considering the flux superpotential  \cite{Gukov:1999ya,Dasgupta:1999ss} or via the dimensional reduction of the 10D kinetic pieces by using the period matrices \cite{Taylor:1999ii, Blumenhagen:2003vr}. One can generalize this  background by a consistent inclusion of more and more fluxes, say via invoking the S/T-duality arguments leading to  four S-dual pairs of fluxes, namely $(F, H), \, (Q, P), \, (P^\prime, Q^\prime)$ and $(H^\prime, F^\prime)$ in the generalized superpotential \cite{Shelton:2005cf, Aldazabal:2006up,Aldazabal:2008zza,Font:2008vd,Guarino:2008ik, Hull:2004in, Kumar:1996zx, Hull:2003kr,Aldazabal:2010ef, Shukla:2015rua, Lombardo:2016swq, Lombardo:2017yme}. Such fluxes act as some parameters in the 4D effective supergravity dynamics and can induce a diverse set of superpotential couplings in the model building context \cite{Derendinger:2004jn,Grana:2012rr,Dibitetto:2012rk, Danielsson:2012by, Blaback:2013ht, Damian:2013dq, Damian:2013dwa, Hassler:2014mla, Ihl:2007ah, deCarlos:2009qm, Danielsson:2009ff, Blaback:2015zra, Dibitetto:2011qs}. For that purpose, learning more about the effective scalar potentials and their 10D origins have been explored in \cite{Blumenhagen:2013hva, Gao:2015nra,Shukla:2015rua, Shukla:2015bca,Leontaris:2023lfc,Leontaris:2023mmm} as a direct generalization to \cite{Taylor:1999ii, Blumenhagen:2003vr,Villadoro:2005cu}.
 
In the context of model building the major challenges any non-geometric setup faces can be encoded in asking two questions: 
\begin{itemize}

\item 
How many and which types of fluxes can be consistently turned-on in a given explicit orientifold construction ?

\item

How to handle the enormously huge size of the scalar potential induced via the generalized flux superpotential in typical models ?

\end{itemize}
The first challenge is usually encoded in consistently satisfying the tadpole cancellation conditions and Bianchi identities \cite{Robbins:2007yv, Ihl:2007ah,Shukla:2016xdy,Plauschinn:2021hkp}. In this regard it is worth mentioning that there are two inequivalent formulation of Bianchi identities (known as cohomology formulation and standard formulation \cite{Robbins:2007yv, Ihl:2007ah,Shukla:2016xdy}) and addressing this for the beyond toroidal cases with prime-fluxes is still an open question. The second challenge regarding the huge size of scalar potential can be demonstrated by considering concrete setups. For example, implementing the successive chain of T- and S-dualities for a toroidal type IIB $\mathbb T^6/(\mathbb Z_2 \times \mathbb Z_2)$ orientifold model leads to a U-dual completed holomorphic flux superpotential with 128 flux parameters \cite{Aldazabal:2006up, Aldazabal:2008zza, Aldazabal:2010ef, Lombardo:2016swq, Lombardo:2017yme}. It has been recently found that the scalar potential induced by this superpotential results in 76276 terms \cite{Leontaris:2023lfc}. Following the prescription of \cite{Villadoro:2005cu, Blumenhagen:2013hva}, a detailed taxonomy of the various pieces of this effective scalar potential has been presented using the internal toroidal metric in \cite{Leontaris:2023lfc}, and it has been shown that one can reformulate the same scalar potential using some specific combination of fluxes and RR axions (also called as axionic fluxes) which reduces the size of the scalar potential to 10888 terms. Furthermore, as a next step, the scalar potential arising from U-dual superpotentials in beyond toroidal models, such as those based on CY orientifolds, have been presented in a symplectic formulation in \cite{Leontaris:2023mmm}. Although the symplectic formulation  \cite{Leontaris:2023mmm} bypasses the need of knowing the CY metric, one still needs to compute the symplectic ingredients such as period metrics which are to be subsequently used in a very involved scalar potential computation. Therefore this symplectic formulation does not significantly help in reducing the task to the extent of the possibility of analytically exploring the vast landscape of non-geometric flux vacua in generic models.

In order to address this issue, in this current article we aim to present another formulation of the generic scalar potential arising from the generalized flux superpotential with U-dual fluxes. This has basically two main advantages as compared to the previous formulation \cite{Leontaris:2023mmm}:
\begin{itemize}

\item
The scalar potential is expressed compactly in terms of a couple of equivalent `master formulae' using a set of {\it new axionic fluxes}. Unlike the symplectic formulation \cite{Leontaris:2023mmm}, the new axionic fluxes also include the complex structure axions and subsequently all the axionic dependencies are encoded in axionic fluxes. After applying this formulation we find that the 76276 terms arising in toroidal model as mentioned above reduces to 2816 terms which is significantly less than 10888 as resulted in the symplectic proposal.

\item
Another important aspect of this new formulation is the fact that one can `{\it read-off}" the scalar potential pieces via simply knowing a set of topological data (such as triple intersection numbers, second Chern class, Euler characteristics) of the compactifying CY orientifold and its mirror.

\end{itemize}

\noindent
We will demonstrate the utility of our master-formulae for five explicit models, including a couple of those which have been known in the literature \cite{Shukla:2015hpa, Blumenhagen:2015lta, Shukla:2016hyy, Shukla:2019wfo, Leontaris:2023mmm}. This also includes cases like isotropic limit of the toroidal model which leads to 224 terms in the scalar potential when expressed in terms of our new axionic fluxes. We also apply our master formula for Quintic-like models arising from the compactification with CY threefold having $h^{1,1}=1$ and we also establish an equivalence of this model with the isotropic case as both result in 224 terms. These effective potential studies and their compact formulations in terms of presenting the master formulae can be directly relevant for making an analytic exploration of physical vacua in non-geometric models, e.g. on the lines of studies presented in \cite{Shukla:2016xdy, Shukla:2019dqd, Shukla:2019akv, Marchesano:2020uqz,Shukla:2022srx,Prieto:2024shz}.

The article is organized as follows: In Section \ref{sec_setup} we review the necessary ingredients for U-dual completion of the flux superpotential. Section \ref{sec_AxionicFluxes} focuses on finding the {\it new axionic fluxes} in two-steps; first we present a useful combination of axions and fluxes to encode all the RR axions $(c_0, \rho_\alpha)$, and then in the second step we present a set of axionic flux orbits which encode the complex-structure axions ($v^i$) and have a peculiar structure useful for writing down the scalar potential in a compact way. This is presented in the form of a couple of master formulae in Section \ref{sec_bilinear}. Subsequently we demonstrate the {\it reading-off} process for three concrete examples by generalizing the known results in Section \ref{sec_reading-off}. In addition, Section \ref{sec_Demonstration} presents two more explicit models in which we demonstrate the taxonomy of the various scalar potential pieces for toroidal model (with 76276 terms), and for Quintic-based models (with 11212 terms) where we show how one could reformulate such huge scalar potentials into 2816 terms and 668 terms respectively. Finally we present a summary and conclusion with an outlook in Section \ref{sec_conclusions}. In addition, two appendices, \ref{sec_AlternateFormulations} and \ref{sec_Quintic-pieces}, are added to provide the necessary details about master formulae and the 224 terms of the scalar potentials in Quintic-based model.


\section{On effective potentials with $U$-dual fluxes}
\label{sec_setup}


The F-term scalar potential governing the dynamics of the ${\cal N}=1$ low energy effective supergravity can be computed from the K\"ahler potential ($K$) and the flux induced holomorphic superpotential $(W)$ by considering the following well known relation,
\bea
\label{eq:Vtot}
& & V=e^{K}\Big(K^{{A} \ov{B}} \, (D_{A} W) \, (\ov D_{\ov {B}} \ov{W}) -3 |W|^2 \Big)  \,,\,
\eea
where the covariant derivatives are defined with respect to all the chiral variables on which the K\"ahler potential ($K$) and the holomorphic superpotential ($W$) generically depend on. This general expression had resulted in a series of the so-called ``master-formulae" for the scalar potential for a given set of K\"ahler- and the super-potentials; e.g.~see \cite{Cicoli:2007xp,Shukla:2015hpa,Shukla:2016hyy,Cicoli:2017shd,Gao:2017gxk,Shukla:2019wfo,AbdusSalam:2020ywo,Cicoli:2021dhg,Cicoli:2021tzt,Marchesano:2020uqz,Leontaris:2022rzj,Leontaris:2023lfc}. Let us briefly review the necessary ingredients we need for the purpose of the current work.


\subsection{Fluxes, moduli, and the K\"ahler potential}


The massless states in the four dimensional effective theory are in one-to-one correspondence with harmonic forms which are either  even or odd under the action of an isometric, holomorphic involution ($\sigma$) acting on the internal compactifying CY threefolds ($X$), and these do generate the equivariant  cohomology groups $H^{p,q}_\pm (X)$. For that purpose, let us fix our conventions, and denote the bases  of even/odd two-forms as $(\mu_\alpha, \, \nu_a)$ while four-forms as $(\tilde{\mu}_\alpha, \, \tilde{\nu}_a)$ where $\alpha\in h^{1,1}_+(X), \, a\in h^{1,1}_-(X)$. Also, we denote the zero- and six- even forms as ${\bf 1}$ and $\Phi_6$ respectively. In addition, the bases for the even and odd cohomologies  of three-forms $H^3_\pm(X)$ are denoted as the symplectic pairs $(a_K, b^J)$ and $({\cal A}_\Lambda, {\cal B}^\Delta)$ respectively. Here we fix the normalization in the various cohomology bases as,
\bea
\label{eq:intersection}
& & \int_X \, \mu_\alpha \wedge \tilde{\mu}^\beta = {\delta}_\alpha^{\, \, \, \beta} , \quad \int_X \, \mu_\alpha \wedge \mu_\beta \wedge \mu_\gamma = \kappa_{\alpha \beta \gamma},\quad \int_X {\cal A}_\Lambda \wedge {\cal B}^\Delta = \delta_\Lambda{}^\Delta.
\eea
For the orientifold choice with $O3/O7$-planes, $K\in \{1, ..., h^{2,1}_+\}$ and $\Lambda\in \{0, ..., h^{2,1}_-\}$ while for $O5/O9$-planes, one has $K\in \{0, ..., h^{2,1}_+\}$ and $\Lambda\in \{1, ..., h^{2,1}_-\}$. It has been observed that setups with odd-moduli $G^a$ corresponding to $h^{1,1}_-(X) \neq 0$ are usually less studied as compared to the relatively simpler case of $h^{1,1}_-(X) = 0$, and explicit construction of such CY orientifolds with odd two-cycles can be found in \cite{Lust:2006zg,Lust:2006zh,Blumenhagen:2008zz,Cicoli:2012vw,Gao:2013rra,Gao:2013pra}. However, let us mention the outset that in our current work we will consider the CY orientifolds having $h^{1,1}_-(X) = 0 = h^{2,1}_+(X)$, and therefore our current setup does not include the odd moduli ($G^a$) and any D-term fluxes \cite{Robbins:2007yv, Shukla:2015bca, Shukla:2015rua, Shukla:2015hpa, Blumenhagen:2015lta}.

Now, the various field ingredients can be expanded in appropriate bases of the equivariant cohomologies. For example, the K\"ahler form $J$, and the RR four-form $C_4$ can be expanded as \cite{Grimm:2004uq}
\bea
\label{eq:fieldExpansions}
& & J = t^\alpha\, \mu_\alpha, \quad C_4 = {\rho}_{\alpha} \, \tilde\mu^\alpha + \cdots, 
\eea
where $t^\alpha$, and $\rho_\alpha$ denote the Einstein-frame two-cycle volume moduli, and a set of axions descending from the 4-form potential $C_4$, while dots $\cdots$ encode the information of a dual pair of spacetime one-forms, and two-form dual to the scalar field $\rho_\alpha$ which are not relevant for the current analysis. 
In addition, we consider the choice of involution $\sigma$ to be such that $\sigma^*\Omega_3 = - \Omega_3$, where $\Omega_3$ denotes the nowhere vanishing holomorphic three-form depending on the complex structure moduli $U^{i}$ counted in the $h^{2,1}_-(X)$ cohomology. Using these pieces of information one defines a set ($S, T_\alpha, U^i$) of the chiral coordinates as below \cite{Benmachiche:2006df},
\bea
\label{eq:N=1_coords}
& & \hskip-2cm S \equiv C_0 + \, i \, e^{-\phi} = C_0 + i\, s, \quad T_\alpha= {\rho}_\alpha   - i \, \tau_\alpha, \quad U^i = v^i - i\, u^i,
\eea
where $\tau_\alpha = \frac{1}{2} \, \kappa_{\alpha\beta\gamma} t^\beta t^\gamma$ and the triple intersection numbers $\kappa_{\alpha\beta\gamma}$ are defined in Eq.~(\ref{eq:intersection}). 

\subsubsection*{The K\"ahler potential}
Using appropriate chiral variables ($S, T_\alpha,U^i$) as defined in (\ref{eq:N=1_coords}), a generic form of the tree-level K\"{a}hler potential can be written as below,
\bea
\label{eq:K}
& & \hskip-2.0cm K = -\ln\left(-i\int_{X}\Omega_3\wedge{\bar\Omega_3}\right) - \ln\left(-i(S-\ov S)\right) -2\ln{\cal V} \,.
\eea
Here, ${\cal V}$ denotes the overall volume of the Calabi Yau threefold, and $\Omega_3\, \equiv  {\cal X}^\Lambda \, {\cal A}_\Lambda - \, {\cal F}_{\Lambda} \, {\cal B}^\Lambda$ where the period vectors $({\cal X}^\Lambda, {\cal F}_\Lambda)$ are encoded in a pre-potential (${\cal F}$) of the following form,
\bea
\label{eq:prepotential}
& & \hskip-1cm {\cal F} = ({\cal X}^0)^2 \, \, f({U^i}) \,, \qquad  f({U^i}) = -\frac{1}{6}\,{\ell_{ijk} \, U^i\, U^j \, U^k} +  \frac{1}{2} \,{{a}_{ij} \, U^i\, U^j} +  \,{{b}_{i} \, U^i} +  \frac{1}{2} \,{i\, \gamma},
\eea
where $U^i =\frac{\delta^i_\Lambda \, {\cal X}^\Lambda}{{\cal X}^0}$. In fact, the function $f({U^i})$ can generically have an infinite series of non-perturbative contributions which may be ignored in the large complex structure limit. The quantities ${a}_{ij}, {b}_i$ and $\gamma$ are real numbers where $\gamma$ is related to the perturbative $(\alpha^\prime)^3$-corrections on the mirror side \cite{Hosono:1994av,Arends:2014qca, Blumenhagen:2014nba}.

With these pieces of information, the K\"ahler potential (\ref{eq:K}) takes the following explicit form in terms of the respective ``saxions" of the chiral variables defined in (\ref{eq:N=1_coords}),
\bea
\label{eq:K-explicit}
& & \hskip-1.5cm K = -\ln\left(8\,{\cal U} + 2 \gamma\right)\, - \, \ln (2\,s) - 2 \ln{\cal V},
\eea
where T-dual pair $({\cal U}, {\cal V})$ is defined as below,
\bea
\label{eq:vol-calU}
& & {\cal U} = \frac{1}{6} \ell_{ijk} u^i u^j u^k, \quad \sigma_i = \partial_i{\cal U}, \qquad {\cal V} = \frac{1}{6} \,{\kappa_{\alpha \beta \gamma} t^\alpha t^\beta t^{\gamma}}, \quad \tau_\alpha = \partial_\alpha{\cal V},
\eea
where $\kappa_{\alpha\beta\gamma}$ and $\ell_{ijk}$ are triple intersection numbers on the CY threefolds and its mirror.


\subsection{U-dual fluxes and the superpotential}
\label{sec_Udual}


It turns out that making successive applications of T/S-dualities results in the need of introducing more and more fluxes. In fact, one needs a total of four S-dual pairs of fluxes, commonly denotes as:  $(F, H), \, (Q, P), \, (P^\prime, Q^\prime)$ and $(H^\prime, F^\prime)$ \cite{Aldazabal:2006up,Aldazabal:2008zza,Shukla:2015rua,Aldazabal:2006up,Aldazabal:2008zza,Aldazabal:2010ef,Lombardo:2016swq,Lombardo:2017yme,Leontaris:2023lfc}. This leads to the following generalized flux superpotential,
\bea
\label{eq:W-all-4}
& & \hskip-1cm W = \int_{X} \, \biggl[\left(F - S\, H \right) + \left(Q - S\, P \right)\triangleright {\cal J} + \left(P^\prime - S\, Q^\prime \right)\diamond {\cal J}^2 \\
& & \hskip2.5cm + \left(H^\prime - S\, F^\prime \right)\odot {\cal J}^3\biggr] \wedge \Omega_3\,,\nonumber
\eea
where ${\cal J} = T_\alpha\, \mu^\alpha$ the dual complexified four-form dual to the K\"ahler form, and the symplectic form of the various flux actions are defined as below \cite{Leontaris:2023mmm},
\bea
\label{eq:FluxActions-3}
&& \left(Q\triangleright {\cal J} \right) = T_\alpha \, {Q}^{\alpha}, \qquad \qquad \qquad \qquad \, \, \left(P\triangleright {\cal J} \right) = T_\alpha \, {P}^{\alpha} \, ,\\
&& \left(P^\prime \diamond {\cal J}^2 \right)  = \frac{1}{2} \, {P}^{\prime\beta\gamma}\, T_\beta \, T_\gamma \,, \qquad  \qquad \quad \left(Q^\prime \diamond {\cal J}^2 \right)  = \frac{1}{2} \, {Q}^{\prime\beta\gamma}\, T_\beta \, T_\gamma\,, \nonumber\\
&& \left(H^\prime \odot {\cal J}^3 \right) = \frac{1}{3 !} \, {H}^{\prime\alpha\beta\gamma}\,T_\alpha \, T_\beta \, T_\gamma\, , \qquad \, \, \, \left(F^\prime \odot {\cal J}^3 \right) = \frac{1}{3 !} \, {F}^{\prime\alpha\beta\gamma}\,T_\alpha \, T_\beta \, T_\gamma. \nonumber
\eea
Here ${Q}^{\alpha}, {P}^{\alpha}, {P}^{\prime\beta\gamma}, {Q}^{\prime\beta\gamma}, {H}^{\prime\alpha\beta\gamma}$ and ${F}^{\prime\alpha\beta\gamma}$ denote the 3-forms similar to the standard $F_3/H_3$ fluxes and can be expanded in the symplectic basis $\{{\cal A}_\Lambda, {\cal B}^\Delta\}$ in the following manner,
\bea
\label{eq:FluxActions-2}
&& F = F^\Lambda \, {\cal A}_\Lambda - F_{\Lambda} \, {\cal B}^\Lambda\,,\qquad \qquad \qquad \qquad H = H_\Lambda \, {\cal A}_\Lambda - H_{\Lambda} \, {\cal B}^\Lambda\,,\\
&& {Q}^{\alpha} = {Q}^{\alpha{}\Lambda} \, {\cal A}_\Lambda - {Q}^{\alpha}{}_{\Lambda} \, {\cal B}^\Lambda\,,\qquad \qquad \qquad {P}^{\alpha} = {P}^{\alpha{}\Lambda} \, {\cal A}_\Lambda - {P}^{\alpha}{}_{\Lambda} \, {\cal B}^\Lambda\,,\nonumber\\
&& {P}^{\prime\beta\gamma}  = {P}^{\prime\beta\gamma{}\Lambda} \, {\cal A}_\Lambda - {P}^{\prime\beta\gamma}{}_{\Lambda} \, {\cal B}^\Lambda\,, \qquad \qquad {Q}^{\prime\beta\gamma}\,= {Q}^{\prime\beta\gamma{}\Lambda} \, {\cal A}_\Lambda - {Q}^{\prime\beta\gamma}{}_{\Lambda} \, {\cal B}^\Lambda\,, \nonumber\\
&& H^{\prime\alpha\beta\gamma} =  {H}^{\prime\alpha\beta\gamma{}\Lambda} \, {\cal A}_\Lambda - {H}^{\prime\alpha\beta\gamma}{}_{\Lambda} \, {\cal B}^\Lambda\, , \qquad F^{\prime\alpha\beta\gamma}\, = {F}^{\prime\alpha\beta\gamma{}\Lambda} \, {\cal A}_\Lambda - {F}^{\prime\alpha\beta\gamma}{}_{\Lambda} \, {\cal B}^\Lambda\,. \nonumber
\eea
Note that the flux components $Q^\alpha$ and $P^\alpha$ having a single index $\alpha \in h^{1,1}_+$ (and $\Lambda \in h^{2,1}_-$ indices) are counted accordingly. However, counting of the prime flux components is a bit tricky because their appearance as ${P}^{\prime\beta\gamma}, {Q}^{\prime\beta\gamma}, {H}^{\prime\alpha\beta\gamma}$ and ${F}^{\prime\alpha\beta\gamma}$ in (\ref{eq:FluxActions-3}) does not seem to have any constraint except for being symmetric in indices $\{\alpha, \beta, \gamma\} \in h^{1,1}_+$. For that purpose, motivated by the previous toroidal studies \cite{Aldazabal:2010ef, Lombardo:2016swq, Lombardo:2017yme, Leontaris:2023lfc}, one can equivalently define \cite{Leontaris:2023mmm},
\bea
\label{eq:primefluxIIBa}
& & P^\prime_{\alpha\Lambda}, \, {P^\prime}_\alpha{}^\Lambda, \qquad Q^\prime_{\alpha\Lambda}, \, {Q^\prime}_\alpha{}^\Lambda, \qquad H^\prime_{\Lambda}, \, {H^\prime}^\Lambda, \qquad F^\prime_{\Lambda}, \, {F^\prime}^\Lambda\,,
\eea
which are related to the earlier set of prime fluxes as below
\bea
\label{eq:primefluxIIBc}
& & {P}^{\prime\beta\gamma{}\Lambda}= {P^\prime}_{\alpha}{}^{\Lambda}\, \kappa^{\alpha\beta\gamma}, \qquad \, {P}^{\prime\beta\gamma}{}_{\Lambda} = {P^\prime}_{\alpha \Lambda} \, \kappa^{\alpha\beta\gamma}\,,\\
& & {Q}^{\prime\beta\gamma{}\Lambda}= {Q^\prime}_{\alpha}{}^{\Lambda}\, \kappa^{\alpha\beta\gamma}, \qquad \, {Q}^{\prime\beta\gamma}{}_{\Lambda} = {Q^\prime}_{\alpha \Lambda} \, \kappa^{\alpha\beta\gamma}\,,\nonumber\\
& & {H}^{\prime\alpha\beta\gamma{}\Lambda}= {H^\prime}^{\Lambda}\, \kappa^{\alpha\beta\gamma}, \qquad {H}^{\prime\alpha\beta\gamma}{}_{\Lambda} = {H^\prime}_{\Lambda} \, \kappa^{\alpha\beta\gamma}\,, \nonumber\\
& & {F}^{\prime\alpha\beta\gamma{}\Lambda}= {F^\prime}^{\Lambda}\, \kappa^{\alpha\beta\gamma}, \qquad \, \, {F}^{\prime\alpha\beta\gamma}{}_{\Lambda} = {F^\prime}_{\Lambda} \, \kappa^{\alpha\beta\gamma}.\nonumber
\eea
Here $\kappa^{\alpha\beta\gamma}$ is defined as
\bea
\label{eq:Inv-lijk}
& & \kappa^{\alpha\beta\gamma} = {\cal V}^3\,\kappa_{\alpha^\prime\beta^\prime\gamma^\prime}\,{\cal G}^{\alpha\alpha^\prime}\, {\cal G}^{\beta\beta^\prime} \, {\cal G}^{\gamma\gamma^\prime}\,,
\eea
where the tree-level metric ${\cal G}^{\alpha\beta}$ and its inverse ${\cal G}_{\alpha\beta}$ are defined as,
\begin{eqnarray}
\label{eq:genMetrices}
& & \hskip-1cm {\cal G}^{\alpha \beta} = \frac{1}{4{\cal V}^2} \left(2 \, t^\alpha \, t^\beta - \,4 \, {\cal V} \, \kappa^{\alpha \beta} \right), \qquad {\cal G}_{\alpha \beta} =  \tau_\alpha \,\tau_\beta - \, {\cal V} \, \kappa_{\alpha\beta}.
\end{eqnarray}
We also note that the metrics defined in (\ref{eq:genMetrices}) result in the following useful identities,
\bea
\label{eq:metric-identities}
& & {\cal G}_{\alpha \beta} t^\alpha = {\cal V}\, \tau_\beta, \quad {\cal G}_{\alpha \beta} t^\alpha t^\beta = 3 {\cal V}^2, \quad {\cal G}^{\alpha \beta} \tau_\alpha = \frac{t^\alpha}{\cal V}, \quad {\cal G}^{\alpha \beta} \tau_\alpha \tau_\beta = 3.
\eea
Moreover we find that the definition (\ref{eq:Inv-lijk}) leads to the following interesting identities \cite{Leontaris:2023mmm},
\bea
\label{eq:Jidentity-cohom}
& & \hskip-1.0cm t^\alpha = \frac{1}{2} \, \kappa^{\alpha\beta\gamma}\, \tau_\beta \, \tau_\gamma, \qquad \quad {\cal V} = \frac{1}{3!}\, \kappa^{\alpha\beta\gamma} \, \tau_\alpha \, \tau_\beta\, \tau_\gamma \,,
\eea
where $\tau_\alpha$ corresponds to the volume of the 4-cycle and can be written in terms of the 2-cycle volumes as: $\tau_\alpha = \frac{1}{2}\, \kappa_{\alpha\beta\gamma}\, t^\beta\, t^\gamma = \frac{1}{2}\, \kappa_\alpha$. Here we have used the shorthand notations $\kappa_{\alpha} = \kappa_{\alpha\beta}\,t^\beta = \kappa_{\alpha\beta\gamma}\,t^\beta\, t^\gamma$ etc.

From this approach it is clear that counting of prime fluxes is also similar to those of the non-prime fluxes, i.e. $\{H', F'\}$ are counted by the $h^{2,1}_-({\rm CY})$ indices similar to the standard $H_3/F_3$ fluxes while $\{P', Q'\}$ are counted via one $\alpha$ index and one $\Lambda$ index. However, the crucial thing to take into account is  the holomorphicity of the flux superpotential which shows that fluxes used in actions (\ref{eq:FluxActions-3}) and (\ref{eq:FluxActions-2}) are the ones to be considered in the superpotential along with the chiral variables $S, T_\alpha$ and $U^i$. Subsequently one has the following generalized flux superpotential,
\bea
\label{eq:W-all-1}
& & \hskip-1.0cm W  = \int_{X} \Biggl[\left({F}  + {Q}^{\alpha}\, T_\alpha + \frac{1}{2} {P}^{\prime\alpha\beta}\, T_\alpha\, T_\beta \, +  \, \frac{1}{6} H^{\prime\alpha\beta\gamma}\, T_\alpha \, T_\beta \, T_\gamma \right) \\
& & \hskip1cm - \, S \, \left({H}  + {P}^{\alpha}\, T_\alpha +  \frac{1}{2} {Q}^{\alpha\beta}\, T_\alpha\, T_\beta \, +\, \frac{1}{6} \,F^{\prime\alpha\beta\gamma}\, T_\alpha \, T_\beta \, T_\gamma \right) \Biggr]_3 \wedge \Omega_3 \,,\nonumber
\eea
where all the terms appearing inside the bracket $[...]_3$ denote a collection of three-forms as expanded in (\ref{eq:FluxActions-2}). Let us also make a side remark that for the holomorphicity of the flux superpotential all these 3-form flux ingredients in (\ref{eq:W-all-1}), or equivalently in (\ref{eq:FluxActions-2}), are independent of the volume moduli and therefore such fluxes are used for expressing the superpotential for the purpose of computing the scalar potential using the ${\cal N} = 1$ formula (\ref{eq:Vtot}). However, after the scalar potential is computed one can always reshuffle and rewrite the scalar potential pieces in order to achieve a more compact way of expressing the full potential, which a priori is a complicated task.

Finally, using the flux actions in Eq.~(\ref{eq:FluxActions-2}), the generalized flux superpotential $W$ in (\ref{eq:W-all-1}) can be equivalently given as below,
\bea
\label{eq:W-gen}
& & \hskip-1.0cm W \equiv \int_{X} [{\rm Flux}]_3 \wedge \Omega_3 = e_\Lambda \, {\cal X}^\Lambda - m^\Lambda \, {\cal F}_\Lambda\,,
\eea
where the symplectic electromagnetic vectors $(e_\Lambda, m^\Lambda)$ are given as,
\begin{eqnarray}
\label{eq:eANDm}
& &  e_\Lambda = \left({F}_\Lambda - S \, {H}_\Lambda\right)  + T_\alpha \left({Q}^{\alpha}{}_\Lambda\,  - S \, {P}^{\alpha}{}_\Lambda \right) + \frac{1}{2} T_\alpha\, T_\beta \left({P}^{\prime\alpha\beta}{}_\Lambda\, - S\, {Q}^{\prime\alpha\beta}{}_\Lambda \right) \, \\
& & \hskip2cm + \, \frac{1}{6} \, T_\alpha \, T_\beta \, T_\gamma \left(H^{\prime\alpha\beta\gamma}{}_\Lambda\, - S \, F^{\prime\alpha\beta\gamma}{}_\Lambda\right), \, \nonumber\\
& &  m^\Lambda = \left({F}^\Lambda - S \, {H}^\Lambda\right)  + T_\alpha \left({Q}^{\alpha}{}^\Lambda\,  - S \, {P}^{\alpha}{}^\Lambda \right) + \frac{1}{2} T_\alpha\, T_\beta \left({P}^{\prime\alpha\beta}{}^\Lambda\, - S\, {Q}^{\prime\alpha\beta}{}^\Lambda \right) \, \nonumber\\
& & \hskip2cm + \, \frac{1}{6} \, T_\alpha \, T_\beta \, T_\gamma \left(H^{\prime\alpha\beta\gamma}{}^\Lambda\, - S \, F^{\prime\alpha\beta\gamma}{}^\Lambda\right).\nonumber
\end{eqnarray}
Thus the generic superpotential (\ref{eq:W-gen}) is linear in the axio-dilaton modulus $S$ and has cubic dependence in moduli $U^i$ and $T_\alpha$ both. Using the symplectic vectors $(e_\Lambda, m^\Lambda)$ of (\ref{eq:eANDm}) in the flux superpotential (\ref{eq:W-gen}), one can compute the derivatives with respect to chiral variables, $S$ and $T_\alpha$ which are given as below,
\begin{eqnarray}
\label{eq:derW_gen}
& & W_S = {(e_1)}_\Lambda \, {\cal X}^\Lambda - {(m_1)}^\Lambda \, {\cal F}_\Lambda, \qquad W_{T_\alpha} = {(e_2)}^\alpha_\Lambda \, {\cal X}^\Lambda - {(m_2)}^\alpha{}^\Lambda \, {\cal F}_\Lambda,
\end{eqnarray}
where the two new pairs of symplectic vectors $(e_1, m_1)$ and $(e_2, m_2)$ are given as:
\begin{eqnarray}
\label{eq:e1ANDm1}
& &  {(e_1)}_\Lambda = - \biggl[{H}_\Lambda  + T_\alpha \, {P}^{\alpha}{}_\Lambda + \frac{1}{2} T_\alpha\, T_\beta \, {Q}^{\prime\alpha\beta}{}_\Lambda + \, \frac{1}{6} \, T_\alpha \, T_\beta \, T_\gamma \, F^{\prime\alpha\beta\gamma}{}_\Lambda \biggr], \,\\
& &  {(m_1)}^\Lambda = -\biggl[{H}^\Lambda + T_\alpha \, {P}^{\alpha}{}^\Lambda + \frac{1}{2} T_\alpha\, T_\beta \, {Q}^{\prime\alpha\beta}{}^\Lambda  + \, \frac{1}{6} \, T_\alpha \, T_\beta \, T_\gamma  \, F^{\prime\alpha\beta\gamma}{}^\Lambda\, \biggr]\,,\nonumber
\end{eqnarray}
and
\begin{eqnarray}
\label{eq:e2ANDm2}
& & \hskip-1.5cm {(e_2)}^\alpha_\Lambda = \left({Q}^{\alpha}{}_\Lambda - S \, {P}^{\alpha}{}_\Lambda \right) + T_\beta \left({P}^{\prime\alpha\beta}{}_\Lambda\, - S\, {Q}^{\prime\alpha\beta}{}_\Lambda \right) + \, \frac{1}{2} T_\beta \, T_\gamma \left(H^{\prime\alpha\beta\gamma}{}_\Lambda\, - S \, F^{\prime\alpha\beta\gamma}{}_\Lambda\right),\\
& & \hskip-1.5cm {(m_2)}^\alpha{}^\Lambda = \left({Q}^{\alpha}{}^\Lambda\,  - S \, {P}^{\alpha}{}^\Lambda \right) + \, T_\beta \left({P}^{\prime\alpha\beta}{}^\Lambda\, - S\, {Q}^{\prime\alpha\beta}{}^\Lambda \right) + \, \frac{1}{2} \, T_\beta \, T_\gamma \left(H^{\prime\alpha\beta\gamma}{}^\Lambda\, - S \, F^{\prime\alpha\beta\gamma}{}^\Lambda\right).\nonumber
\end{eqnarray}


\section{Invoking the axionic flux polynomials}
\label{sec_AxionicFluxes}


\subsection{Step 1: Absorbing the RR axions}


Having the concrete form of the holomorphic superpotential now we define a set of the so-called {\it axionic-flux orbits} which will turn out to be extremely useful for rearranging the scalar potential pieces into a compact form. We consider the following combination of axions and fluxes to begin with
\bea
\label{eq:AxionicFlux}
& & \hskip-0.5cm {\mathbb F}_\Lambda = \left({F}_\Lambda  + \rho_\alpha \, {Q}^{\alpha}{}_\Lambda + \frac{1}{2} \rho_\alpha\, \rho_\beta \, {P}^{\prime\alpha\beta}{}_\Lambda + \, \frac{1}{6} \, \rho_\alpha \, \rho_\beta \, \rho_\gamma \, H^{\prime\alpha\beta\gamma}{}_\Lambda\right) \\
& & \hskip0.5cm - \, C_0 \,\left({H}_\Lambda  + \rho_\alpha \, {P}^{\alpha}{}_\Lambda + \frac{1}{2} \rho_\alpha\, \rho_\beta \, {Q}^{\prime\alpha\beta}{}_\Lambda + \, \frac{1}{6} \, \rho_\alpha \, \rho_\beta \, \rho_\gamma \, F^{\prime\alpha\beta\gamma}{}_\Lambda\right),\nonumber\\
& & \hskip-0.5cm {\mathbb F}^\Lambda = \left({F}^\Lambda + \rho_\alpha \, {Q}^{\alpha}{}^\Lambda + \frac{1}{2} \rho_\alpha\, \rho_\beta \, {P}^{\prime\alpha\beta}{}^\Lambda  + \, \frac{1}{6} \, \rho_\alpha \, \rho_\beta \, \rho_\gamma  \, H^{\prime\alpha\beta\gamma}{}^\Lambda\right) \nonumber\\
& & \hskip0.5cm - \, C_0 \, \left({H}^\Lambda + \rho_\alpha \, {P}^{\alpha}{}^\Lambda + \frac{1}{2} \rho_\alpha\, \rho_\beta \, {Q}^{\prime\alpha\beta}{}^\Lambda  + \, \frac{1}{6} \, \rho_\alpha \, \rho_\beta \, \rho_\gamma  \, F^{\prime\alpha\beta\gamma}{}^\Lambda\right), \nonumber
\eea
and subsequently we define
\bea
\label{eq:AxionicFlux1}
& &  {\mathbb H}_\Lambda = - \partial_{C_0} {\mathbb F}_\Lambda = {H}_\Lambda  + \rho_\alpha \, {P}^{\alpha}{}_\Lambda + \frac{1}{2} \rho_\alpha\, \rho_\beta \, {Q}^{\prime\alpha\beta}{}_\Lambda + \, \frac{1}{6} \, \rho_\alpha \, \rho_\beta \, \rho_\gamma \, F^{\prime\alpha\beta\gamma}{}_\Lambda\,,\\
& & {\mathbb H}^\Lambda = - \partial_{C_0} {\mathbb F}^\Lambda = {H}^\Lambda + \rho_\alpha \, {P}^{\alpha}{}^\Lambda + \frac{1}{2} \rho_\alpha\, \rho_\beta \, {Q}^{\prime\alpha\beta}{}^\Lambda  + \, \frac{1}{6} \, \rho_\alpha \, \rho_\beta \, \rho_\gamma  \, F^{\prime\alpha\beta\gamma}{}^\Lambda\,, \nonumber\\
& & \nonumber\\
& & { {\mathbb P}}^{\alpha}{}_\Lambda = - \partial_{\rho^\alpha}\partial_{C_0} {\mathbb F}_\Lambda = \, {P}^{\alpha}{}_\Lambda + \, \rho_\beta \, {Q}^{\prime\alpha\beta}{}_\Lambda + \, \frac{1}{2} \, \rho_\beta \, \rho_\gamma \, F^{\prime\alpha\beta\gamma}{}_\Lambda\,, \nonumber\\
& & {{\mathbb P}}^{\alpha}{}^\Lambda = - \partial_{\rho^\alpha}\partial_{C_0} {\mathbb F}^\Lambda = {P}^{\alpha}{}^\Lambda +  \rho_\beta \, {Q}^{\prime\alpha\beta}{}^\Lambda  + \, \frac{1}{2} \, \rho_\alpha \, \rho_\beta \, \rho_\gamma  \, F^{\prime\alpha\beta\gamma}{}^\Lambda\,, \nonumber\\
& & { {\mathbb Q}}^{\alpha}{}_\Lambda = \partial_{\rho^\alpha} {\mathbb F}_\Lambda = \, {Q}^{\alpha}{}_\Lambda + \, \rho_\beta \, {P}^{\prime\alpha\beta}{}_\Lambda + \, \frac{1}{2} \, \rho_\beta \, \rho_\gamma \, H^{\prime\alpha\beta\gamma}{}_\Lambda -\, C_0\, { {\mathbb P}}^{\alpha}{}_\Lambda\,, \nonumber\\
& & {{\mathbb Q}}^{\alpha}{}^\Lambda = \partial_{\rho^\alpha} {\mathbb F}^\Lambda = {Q}^{\alpha}{}^\Lambda +  \rho_\beta \, {P}^{\prime\alpha\beta}{}^\Lambda  + \, \frac{1}{2} \, \rho_\alpha \, \rho_\beta \, \rho_\gamma  \, H^{\prime\alpha\beta\gamma}{}^\Lambda\, - C_0 \, {{\mathbb P}}^{\alpha}{}^\Lambda\,, \nonumber\\
& & \nonumber\\
& & { {\mathbb Q}}^{\prime\alpha\beta}{}_\Lambda = - \partial_{\rho^\alpha} \partial_{\rho^\beta} \partial_{C_0} {\mathbb F}_\Lambda =  {Q}^{\prime\alpha\beta}{}_\Lambda + \, \rho_\gamma \, F^{\prime\alpha\beta\gamma}{}_\Lambda\,, \nonumber\\
& & { {\mathbb Q}}^{\prime\alpha\beta}{}^\Lambda = - \partial_{\rho^\alpha} \partial_{\rho^\beta} \partial_{C_0} {\mathbb F}^\Lambda =  {Q}^{\prime\alpha\beta}{}^\Lambda  + \, \rho_\gamma  \, F^{\prime\alpha\beta\gamma}{}^\Lambda\,, \nonumber\\
& & { {\mathbb P}}^{\prime\alpha\beta}{}_\Lambda = \partial_{\rho^\alpha} \partial_{\rho^\beta} {\mathbb F}_\Lambda =  {P}^{\prime\alpha\beta}{}_\Lambda + \, \rho_\gamma \, H^{\prime\alpha\beta\gamma}{}_\Lambda\, - C_0\, { {\mathbb Q}}^{\prime\alpha\beta}{}_\Lambda\,, \nonumber\\
& & { {\mathbb P}}^{\prime\alpha\beta}{}^\Lambda = \partial_{\rho^\alpha} \partial_{\rho^\beta} {\mathbb F}^\Lambda =  {P}^{\prime\alpha\beta}{}^\Lambda  + \, \rho_\gamma  \, H^{\prime\alpha\beta\gamma}{}^\Lambda\, - \, C_0\, { {\mathbb Q}}^{\prime\alpha\beta}{}^\Lambda\,, \nonumber\\
& & \nonumber\\
& & {\mathbb F}^{\prime\alpha\beta\gamma}{}_\Lambda = - \partial_{\rho^\alpha} \partial_{\rho^\beta} \partial_{\rho^\gamma} \partial_{C_0} {\mathbb F}_\Lambda = \, F^{\prime\alpha\beta\gamma}{}_\Lambda\,, \nonumber\\
& & {\mathbb F}^{\prime\alpha\beta\gamma}{}^\Lambda = - \partial_{\rho^\alpha} \partial_{\rho^\beta} \partial_{\rho^\gamma} \partial_{C_0} {\mathbb F}^\Lambda = \, F^{\prime\alpha\beta\gamma}{}^\Lambda\,, \nonumber\\
& & {\mathbb H}^{\prime\alpha\beta\gamma}{}_\Lambda = \partial_{\rho^\alpha} \partial_{\rho^\beta} \partial_{\rho^\gamma} {\mathbb F}_\Lambda = \, H^{\prime\alpha\beta\gamma}{}_\Lambda\, - \, C_0\,  {\mathbb F}^{\prime\alpha\beta\gamma}{}_\Lambda, \nonumber\\
& & {\mathbb H}^{\prime\alpha\beta\gamma}{}^\Lambda = \partial_{\rho^\alpha} \partial_{\rho^\beta} \partial_{\rho^\gamma} {\mathbb F}^\Lambda = \, H^{\prime\alpha\beta\gamma}{}^\Lambda\,- C_0\, {\mathbb F}^{\prime\alpha\beta\gamma}{}^\Lambda. \nonumber
\eea
Thus we see that all the axionic fluxes in (\ref{eq:AxionicFlux1}) can be derived by taking axionic derivatives of ${\mathbb F}_\Lambda$ and ${\mathbb F}^\Lambda$  defined in (\ref{eq:AxionicFlux}). Using these axionic-flux combinations (\ref{eq:AxionicFlux})-(\ref{eq:AxionicFlux1}) along with the definitions of chiral variables in Eq.~(\ref{eq:N=1_coords}), the three pairs of symplectic vectors, namely $(e, m), \, (e_1, m_1)$ and $(e_2, m_2)$ which are respectively given in Eqs.~(\ref{eq:eANDm}), (\ref{eq:e1ANDm1}) and (\ref{eq:e2ANDm2}), can be expressed in the following compact form,
\bea
\label{eq:em}
& & e_\Lambda = \left({\mathbb F}_\Lambda - s \, {\mathbb P}^{}{}_{\Lambda} - {\mathbb P}^{\prime}{}_{\Lambda} + s\, {\mathbb F}^{\prime}{}_{\Lambda} \right)  + i \, \left( - s \, {\mathbb H}_\Lambda - {{\mathbb Q}}^{}{}_{\Lambda} + s \, {{\mathbb Q}}^{\prime}{}_{\Lambda} + {{\mathbb H}}^{\prime}{}_{\Lambda}\right),\\
& & m^\Lambda = \left({\mathbb F}^\Lambda - s \, {\mathbb P}^{}{}^{\Lambda} - {\mathbb P}^{\prime}{}^{\Lambda} + s\, {\mathbb F}^{\prime}{}^{\Lambda} \right)  + i \, \left( -\, s \, {\mathbb H}^\Lambda - {{\mathbb Q}}^{}{}^{\Lambda} + s \, {{\mathbb Q}}^{\prime}{}^{\Lambda} + {{\mathbb H}}^{\prime}{}^{\Lambda}\right), \nonumber
\eea
\bea
\label{eq:e1m1}
& & \hskip-1.5cm {(e_1)}_\Lambda = \left(- \, {\mathbb H}_\Lambda + \, {{\mathbb Q}}^{\prime}{}_{\Lambda} \right) + i \left({\mathbb P}^{}{}_{\Lambda} - \, {\mathbb F}^{\prime}{}_{\Lambda} \right)\,, \quad {(m_1)}^\Lambda = \left(-\, {\mathbb H}^\Lambda + \, {{\mathbb Q}}^{\prime}{}^{\Lambda}\right) + i \left({\mathbb P}^{}{}^{\Lambda}  - \, {\mathbb F}^{\prime}{}^{\Lambda} \right)\,,
\eea
\bea
\label{eq:e2m2}
& & {(e_2)}^\alpha{}_\Lambda = \left({{\mathbb Q}}^{\alpha}{}_{\Lambda} - \, s \, {{\mathbb Q}}^{\prime\alpha}{}_{\Lambda} - \, {{\mathbb H}}^{\prime\alpha}{}_{\Lambda}\right) + i \left(- s \, {\mathbb P}^{\alpha}{}_{\Lambda} - {\mathbb P}^{\prime\alpha}{}_{\Lambda} + s\, {\mathbb F}^{\prime\alpha}{}_{\Lambda} \right)\,,\\
& & {(m_2)}^\alpha{}^\Lambda = \left({{\mathbb Q}}^{\alpha}{}^{\Lambda} - s \, {{\mathbb Q}}^{\prime\alpha}{}^{\Lambda} - {{\mathbb H}}^{\prime\alpha}{}^{\Lambda}\right) + i \left( - \, s \, {\mathbb P}^{\alpha}{}^{\Lambda} - {\mathbb Q}^{\prime\alpha}{}^{\Lambda} + s\, {\mathbb F}^{\prime\alpha}{}^{\Lambda} \right)\,, \nonumber
\eea
where we have used the shorthand notations like ${\mathbb Q}^{}_\Lambda = \tau_\alpha\,{\mathbb Q}^{\alpha}{}_\Lambda$, ${\mathbb Q}^{\prime\alpha}_\Lambda = \,\tau_\beta{\mathbb Q}^{\prime\alpha\beta}{}_\Lambda$, ${\mathbb Q}^{\prime}_\Lambda = \frac12 \tau_\alpha\,\tau_\beta{\mathbb Q}^{\prime\alpha\beta}{}_\Lambda$,  and ${\mathbb H}^{\prime}_\Lambda = \frac16\,\tau_\alpha\tau_\beta \tau_\gamma{\mathbb H}^{\prime\alpha\beta\gamma}{}_\Lambda$ etc. In addition, we mention that such shorthand notations are applicable only with $\tau_\alpha$ contractions, and not to be (conf)used with axionic ($\rho_\alpha$) contractions. This convention will be used wherever the $(Q, P)$, $(P^\prime, Q^\prime)$ and $(H^\prime, F^\prime)$ fluxes are seen with/without a free index $\alpha \in h^{1,1}_+(X)$. Here we recall that $\tau_\alpha = \frac{1}{2}\, {\kappa}_{\alpha \beta \gamma} t^\beta t^\gamma$.


\subsection{Step 2: Absorbing the complex-structure axions}


Using the pre-potential derivatives one can extract the dependence on the complex-structure moduli $U^i$ which leads to the following form of the superpotential given in (\ref{eq:W-gen}),
\bea
\label{eq:W-gen3}
& & \hskip-1.5cm W \equiv e_\Lambda {\cal X}^\Lambda - m^\Lambda {\cal F}_\Lambda = \widetilde{e}_0 + \widetilde{e}_i \, U^i + m^i \left(\frac{1}{2} \, \ell_{ijk} U^j U^k\right) - m^0 \, \left(\frac{1}{6} \, \ell_{ijk} U^i U^j U^k +\,i \, \gamma\right),
\eea
where the fluxes $\widetilde{e}_0$ and $\widetilde{e}_i$ corresponds to having some rational shift \cite{Shukla:2019wfo, Leontaris:2023mmm},
\bea
\label{eq:rational-shift}
& & \widetilde{e}_0 = e_0 - m^i \, {b}_i, \qquad \widetilde{e}_i = e_i - m^j \, {a}_{ij} - m^0\, {b}_{i}.
\eea
Such rational shifts have been also observed for type IIA models, e.g. see \cite{Escobar:2018rna}. Now, using the three pairs of symplectic vectors, namely $(e, m), \, (e_1, m_1)$ and $(e_2, m_2)$, as defined in Eqs.~(\ref{eq:eANDm}), (\ref{eq:e1ANDm1}) and (\ref{eq:e2ANDm2}), the respective derivatives of the superpotential (\ref{eq:e1m1})-(\ref{eq:e2m2}) are given as below,
\bea
\label{eq:derW_gen2}
& & W_{U^i} = \widetilde{e}_i\, + {m}^j \,\left(\ell_{ijk} U^k\right) - {m}^0 \, \left(\frac{1}{2} \, \ell_{ijk} \,U^j U^k \right),\\
& & W_S = {(\widetilde{e_1})}_0 +  U^i \, {(\widetilde{e_1})}_i\, + {(m_1)}^i \,\left(\frac{1}{2} \, \ell_{ijk} U^j U^k\right) -\, {(m_1)}^0 \, \left(\frac{1}{6} \, \ell_{ijk} U^i U^j U^k +\,i \, \gamma\right),\nonumber\\
& & W_{T_\alpha} = {(\widetilde{e_2^\alpha})}_0 +  U^i \, {(\widetilde{e_2^\alpha})}_i\, + {(\widetilde{m_2^\alpha})}^i \,\left(\frac{1}{2} \, \ell_{ijk} U^j U^k\right) -\, {(\widetilde{m_2^\alpha})}^0 \, \left(\frac{1}{6} \, \ell_{ijk} U^i U^j U^k +\,i \, \gamma\right). \nonumber
\eea
Subsequently using the chiral coordinates (\ref{eq:N=1_coords}) one invokes a {\it new} set of axionic fluxes for all the eight classes of fluxes, namely ${\rm f}, {\rm h}, {\rm q}, {\rm p}, {\rm p}^\prime, {\rm q}^\prime, {\rm h}^\prime$ and ${\rm f}^\prime$, which can absorb all the axionic dependence present in the generalized flux superpotential (\ref{eq:W-gen3}) as well as its various derivatives (\ref{eq:derW_gen2}). These most generic axionic fluxes take the following explicit form,
\bea
\label{eq:newOrbits1}
& & \widetilde{\rm f}_0 = \widetilde{\mathbb F}_0 + \widetilde{\mathbb F}_i\, v^i + {\mathbb F}^i \, \left(\frac{1}{2} \, \ell_{ijk} v^j v^k\right) - {\mathbb F}^0 \, \left(\frac{1}{6} \, \ell_{ijk} v^i v^j v^k\right),\\
& & \widetilde{\rm f}_i  = \widetilde{\mathbb F}_i\, + \ell_{ijk}\, {\mathbb F}^j \,v^k - {\mathbb F}^0 \, \left(\frac{1}{2} \, \ell_{ijk} v^j v^k\right),\nonumber\\
& & {\rm f}^i =  {\mathbb F}^i - {\mathbb F}^0\, v^i,\nonumber\\
& & {\rm f}^0 = - {\mathbb F}^0,\nonumber\\
& & \nonumber\\
& & \widetilde{\rm h}_0 = \widetilde{\mathbb H}_0 + \widetilde{\mathbb H}_i\, v^i + {\mathbb H}^i \, \left(\frac{1}{2} \, \ell_{ijk} v^j v^k\right) - {\mathbb H}^0 \, \left(\frac{1}{6} \, \ell_{ijk} v^i v^j v^k\right), \nonumber\\
& & \widetilde{\rm h}_i = \widetilde{\mathbb H}_i\, + \ell_{ijk}\, {\mathbb H}^j \,v^k - {\mathbb H}^0 \, \left(\frac{1}{2} \, \ell_{ijk} v^j v^k\right), \nonumber\\
& & {\rm h}^i = {\mathbb H}^i - {\mathbb H}^0\, v^i,\nonumber\\
& & {\rm h}^0 = - {\mathbb H}^0,\nonumber\\
& & \nonumber\\
& & \widetilde{\rm q}^\alpha_0 = \widetilde{\mathbb Q}^\alpha_0 + \widetilde{\mathbb Q}^\alpha_i\, v^i + {\mathbb Q}^{\alpha i} \, \left(\frac{1}{2} \, \ell_{ijk} v^j v^k\right) - {\mathbb Q}^{\alpha 0} \, \left(\frac{1}{6} \, \ell_{ijk} v^i v^j v^k\right), \nonumber\\
& & \widetilde{\rm q}_i = \widetilde{\mathbb Q}^\alpha_i\, + \ell_{ijk}\, {\mathbb Q}^{\alpha j} \,v^k - {\mathbb Q}^{\alpha 0} \, \left(\frac{1}{2} \, \ell_{ijk} v^j v^k\right),\nonumber\\
& & {\rm q}^{\alpha i} = {\mathbb Q}^{\alpha i} - {\mathbb Q}^{\alpha 0}\, v^i,\nonumber\\
& & {\rm q}^{\alpha 0} = - {\mathbb Q}^{\alpha 0},\nonumber
\eea
\bea
\label{eq:newOrbits2}
& & \widetilde{\rm p}^\alpha_0 = \widetilde{\mathbb P}^\alpha_0 + \widetilde{\mathbb P}^\alpha_i\, v^i + {\mathbb P}^{\alpha i} \, \left(\frac{1}{2} \, \ell_{ijk} v^j v^k\right) - {\mathbb P}^{\alpha 0} \, \left(\frac{1}{6} \, \ell_{ijk} v^i v^j v^k\right), \\
& & \widetilde{\rm p}_i = \widetilde{\mathbb P}^\alpha_i\, + \ell_{ijk}\, {\mathbb P}^{\alpha j} \,v^k - {\mathbb P}^{\alpha 0} \, \left(\frac{1}{2} \, \ell_{ijk} v^j v^k\right),\nonumber\\
& & {\rm p}^{\alpha i} = {\mathbb P}^{\alpha i} - {\mathbb P}^{\alpha 0}\, v^i,\nonumber\\
& & {\rm p}^{\alpha 0} = - {\mathbb P}^{\alpha 0},\nonumber\\
& & \nonumber\\
& & \widetilde{\rm p^\prime}^{\alpha\beta}_0 = \widetilde{\mathbb P^\prime}^{\alpha\beta}_0 + \widetilde{\mathbb P^\prime}^{\alpha\beta}_i\, v^i + {\mathbb P^\prime}^{\alpha\beta i} \, \left(\frac{1}{2} \, \ell_{ijk} v^j v^k\right) - {\mathbb P^\prime}^{\alpha\beta 0} \, \left(\frac{1}{6} \, \ell_{ijk} v^i v^j v^k\right),\nonumber\\
& & \widetilde{\rm p^\prime}^{\alpha\beta}_i = \widetilde{\mathbb P^\prime}^{\alpha\beta}_i\, + \ell_{ijk}\, {\mathbb P^\prime}^{\alpha\beta j} \,v^k - {\mathbb P^\prime}^{\alpha\beta 0} \, \left(\frac{1}{2} \, \ell_{ijk} v^j v^k\right),\nonumber\\
& & {\rm p^\prime}^{\alpha\beta i} = {\mathbb P^\prime}^{\alpha\beta i} - {\mathbb P^\prime}^{\alpha\beta 0}\, v^i,\nonumber\\
& & {\rm p^\prime}^{\alpha\beta 0} = - {\mathbb P^\prime}^{\alpha\beta 0},\nonumber\\
& & \nonumber\\
& & \widetilde{\rm q^\prime}^{\alpha\beta}_0 = \widetilde{\mathbb Q^\prime}^{\alpha\beta}_0 + \widetilde{\mathbb Q^\prime}^{\alpha\beta}_i\, v^i + {\mathbb Q^\prime}^{\alpha\beta i} \, \left(\frac{1}{2} \, \ell_{ijk} v^j v^k\right) - {\mathbb Q^\prime}^{\alpha\beta 0} \, \left(\frac{1}{6} \, \ell_{ijk} v^i v^j v^k\right), \nonumber\\
& & \widetilde{\rm q^\prime}^{\alpha\beta}_i = \widetilde{\mathbb Q^\prime}^{\alpha\beta}_i\, + \ell_{ijk}\, {\mathbb Q^\prime}^{\alpha\beta j} \,v^k - {\mathbb Q^\prime}^{\alpha\beta 0} \, \left(\frac{1}{2} \, \ell_{ijk} v^j v^k\right),\nonumber\\
& & {\rm q^\prime}^{\alpha\beta i} = {\mathbb Q^\prime}^{\alpha\beta i} - {\mathbb Q^\prime}^{\alpha\beta 0}\, v^i,\nonumber\\
& & {\rm q^\prime}^{\alpha\beta 0} = - {\mathbb Q^\prime}^{\alpha\beta 0},\nonumber\\
& & \nonumber\\
& & \widetilde{\rm h^\prime}^{\alpha\beta\gamma}_0 = \widetilde{\mathbb H^\prime}^{\alpha\beta\gamma}_0 + \widetilde{\mathbb H^\prime}^{\alpha\beta\gamma}_i\, v^i + {\mathbb H^\prime}^{\alpha\beta\gamma i} \left(\frac{1}{2} \, \ell_{ijk} v^j v^k\right) - {\mathbb H^\prime}^{\alpha\beta\gamma 0} \, \left(\frac{1}{6} \, \ell_{ijk} v^i v^j v^k\right), \nonumber\\
& & \widetilde{\rm h^\prime}^{\alpha\beta\gamma}_i = \widetilde{\mathbb H^\prime}^{\alpha\beta\gamma}_i\, + \ell_{ijk}\, {\mathbb H^\prime}^{\alpha\beta\gamma j} \,v^k - {\mathbb H^\prime}^{\alpha\beta\gamma 0} \, \left(\frac{1}{2} \, \ell_{ijk} v^j v^k\right),\nonumber\\
& & {\rm h^\prime}^{\alpha\beta\gamma i} = {\mathbb H^\prime}^{\alpha\beta\gamma i} - {\mathbb H^\prime}^{\alpha\beta\gamma 0}\, v^i,\nonumber\\
& & {\rm h^\prime}^{\alpha\beta\gamma 0} = - {\mathbb H^\prime}^{\alpha\beta\gamma 0},\nonumber\\
& & \nonumber\\
& & \widetilde{\rm f^\prime}^{\alpha\beta\gamma}_0 = \widetilde{\mathbb F^\prime}^{\alpha\beta\gamma}_0 + \widetilde{\mathbb F^\prime}^{\alpha\beta\gamma}_i\, v^i + {\mathbb F^\prime}^{\alpha\beta\gamma i} \left(\frac{1}{2} \, \ell_{ijk} v^j v^k\right) - {\mathbb F^\prime}^{\alpha\beta\gamma 0} \, \left(\frac{1}{6} \, \ell_{ijk} v^i v^j v^k\right), \nonumber\\
& & \widetilde{\rm f^\prime}^{\alpha\beta\gamma}_i = \widetilde{\mathbb F^\prime}^{\alpha\beta\gamma}_i\, + \ell_{ijk}\, {\mathbb F^\prime}^{\alpha\beta\gamma j} \,v^k - {\mathbb F^\prime}^{\alpha\beta\gamma 0} \, \left(\frac{1}{2} \, \ell_{ijk} v^j v^k\right),\nonumber\\
& & {\rm f^\prime}^{\alpha\beta\gamma i} = {\mathbb F^\prime}^{\alpha\beta\gamma i} - {\mathbb F^\prime}^{\alpha\beta\gamma 0}\, v^i,\nonumber\\
& & {\rm f^\prime}^{\alpha\beta\gamma 0} = - {\mathbb F^\prime}^{\alpha\beta\gamma 0},\nonumber
\eea
where tilde components of each of the fluxes include the corresponding rational shifts as defined as (\ref{eq:rational-shift}). Now the explicit dependence on all the non-axionic moduli takes the form:
\bea
\label{eq:W-all-3}
& & \hskip-0.5cm W = \biggl[\widetilde{\rm f}_0 -i\, \widetilde{\rm f}_i u^i - {\rm f}^i \sigma_i + i\, {\rm f}^0 \left({\cal U} + \gamma\right) \biggr] \\
& & - \,i\, s \, \biggl[\widetilde{\rm h}_0 -i \, \widetilde{\rm h}_i u^i - {\rm h}^i \sigma_i + i\, {\rm h}^0 \, \left({\cal U} + \, \gamma\right) \biggr] \nonumber\\
& & -\, i\, \tau_\alpha\, \biggl[\widetilde{\rm q}^\alpha_0 -i \, \widetilde{\rm q}^\alpha_i\,\,u^i - {\rm q}^{\alpha i} \,\sigma_i + i\, {\rm q}^{\alpha 0} \, \left({\cal U} + \, \gamma\right) \biggr]\,\nonumber
\eea
\bea
& & - \, s \, \tau_\alpha\, \biggl[\widetilde{\rm p}^\alpha_0 -i \, \widetilde{\rm p}^\alpha_i\,\,u^i - {\rm p}^{\alpha i} \,\sigma_i + i\, {\rm p}^{\alpha 0} \, \left({\cal U} + \, \gamma\right) \biggr]\, \\
& & - \frac{1}{2}\,\tau_\alpha \tau_\beta \biggl[\widetilde{\rm p}^{\prime\alpha\beta}_0 -i \, \widetilde{\rm p}^{\prime\alpha\beta}_i\,\,u^i - {\rm p}^{\prime\alpha\beta i} \,\sigma_i + i\, {\rm p}^{\prime\alpha\beta 0} \, \left({\cal U} + \, \gamma\right) \biggr]\, \nonumber\\
& & + \frac{i\, s}{2}\, \tau_\alpha \tau_\beta \biggl[\widetilde{\rm q}^{\prime\alpha\beta}_0 -i \, \widetilde{\rm q}^{\prime\alpha\beta}_i\,\,u^i - {\rm q}^{\prime\alpha\beta i} \,\sigma_i + i\, {\rm q}^{\prime\alpha\beta 0} \, \left({\cal U} + \, \gamma \right) \biggr]\, \nonumber\\
& & + \frac{i}{6} \,\tau_\alpha\, \tau_\beta \, \tau_\gamma \biggl[\widetilde{\rm h}^{\prime\alpha\beta\gamma}_0 -i \, \widetilde{\rm h}^{\prime\alpha\beta\gamma}_i\,\,u^i - {\rm h}^{\prime\alpha\beta\gamma i} \,\sigma_i + i\, {\rm h}^{\prime\alpha\beta\gamma 0} \, \left({\cal U} + \, \gamma \right) \biggr] \nonumber\\
& & + \frac{s}{6}\, \tau_\alpha\, \tau_\beta \, \tau_\gamma \biggl[\widetilde{\rm f}^{\prime\alpha\beta\gamma}_0 -i \, \widetilde{\rm f}^{\prime\alpha\beta\gamma}_i\,\,u^i - {\rm f}^{\prime\alpha\beta\gamma i} \,\sigma_i + i\, {\rm f}^{\prime\alpha\beta\gamma 0} \, \left({\cal U} + \, \gamma\right) \biggr] \,, \nonumber
\eea
where the new axionic-flux polynomials which are some combinations of the previous axionic fluxes by incorporating the complex-structure axions $v^i$.


\section{Generic scalar potential with $U$-dual fluxes}
\label{sec_bilinear}


In this section we will invoke a new version of the axionic flux polynomials which also include the complex structure axions $v^i$. Subsequently, using the new axionic flux polynomials we will show that the  generic scalar potential can be not only written in a more compact form but one can literally {\it read-off} the scalar potential pieces by merely knowing some topological quantities such as triple intersection numbers and Hodge number for a given model based on the CY orientifold compactification.

For deriving a master formula relevant for our current non-geometric setup with the generalized flux superpotential of the form (\ref{eq:W-gen}) and the K\"ahler potential (\ref{eq:K-explicit}), using (\ref{eq:Vtot}) we find the need of computing three blocks as mentioned below
\bea
\label{eq:V_gen}
& & V = e^K\biggl[W_A \, K^{{A} \ov {B}}\, \ov{W}_{\ov{B}} + \left(W_A \, K^{{A} \ov{B}} K_{\ov{B}} \, \ov{W} + W\, K_A K^{{A} \ov{B}} {\ov{W}}_{\ov{B}} \right) \\
& & \hskip1cm + \left(K_A\, K^{{A} \ov {B}} K_{\ov{B}} -3\right) |W|^2 \biggr]. \nonumber
\eea
Now we collect the relevant ingredients through a set of intermediate steps.


\subsection{Some identities using the moduli space metric}
For the purpose of deriving a master formula to ``{\it read-off}" the most generic scalar potential arising from the generalized flux superpotential we consider the following expressions for the various K\"ahler derivatives,
\begin{eqnarray}
\label{eq:derK}
& & \hskip-1cm K_{U^i} = -\frac{2\,i\, \ell_i}{{\cal Z}} = - K_{\ov{U}^i}, \quad K_S = \frac{i}{2 \,s } = - K_{\ov S}, \quad K_{T_\alpha} = -\frac{i \, t^\alpha}{2\cal V} = - K_{\ov{T}_\alpha},
\end{eqnarray}
where ${\cal Z} = 8 \, {\cal U} + 2 \gamma$ and the inverse K\"ahler metric is given as,
\begin{eqnarray}
\label{eq:Kmetric}
& & \hskip-1cm K_{U^i \ov{U}^j} = \frac{4\,\ell_i \ell_j}{{\cal Z}^2} - \frac{2}{{\cal Z}}\ell_{ij}, \quad K_{S \ov{S}} =  \frac{1}{4 \, s^2}, \quad  K_{T_\alpha \, \ov{T}_\beta} = \frac{2 \, t^\alpha \, t^\beta - \,4 \, {\cal V} \, \kappa^{\alpha \beta}}{16{\cal V}^2} \equiv \frac{1}{4}{\cal G}^{\alpha\beta},\\
& & \hskip-1cm K_{U^i \, \ov{S}} = 0 = K_{S \ov{U}_i}, \quad K_{T_\alpha \, \ov{S}} = 0 = K_{S \ov{T}_\alpha}, \quad K_{T_\alpha \, \ov{U}^i} = 0 = K_{U^i \ov{T}_\alpha},\nonumber
\end{eqnarray}
which subsequently results in the following inverse K\"ahler metric,
\begin{eqnarray}
\label{eq:InvK}
& & \hskip-1cm K^{U^i \ov{U}^j} = \frac{2\,{\cal Z}\,u^i u^j}{{\cal Z}-6\gamma} - \frac{{\cal Z}}{2} \ell^{ij}, \quad K^{S \ov{S}} =  4 \, s^2, \quad  K^{T_\alpha \, \ov{T}_\beta} = \kappa_\alpha \,\kappa_\beta - 4\, {\cal V} \, \kappa_{\alpha\beta} \equiv 4\, {\cal G}_{\alpha\beta},\\
& & \hskip-1cm K^{U^i \, \ov{S}} = 0 = K^{S \ov{U}_i}, \quad K^{T_\alpha \, \ov{S}} = 0 = K^{S \ov{T}_\alpha}, \quad K^{T_\alpha \, \ov{U}^i} = 0 = K^{U^i \ov{T}_\alpha}.\nonumber
\end{eqnarray}
Here we recall that ${\cal G}_{\alpha\beta}$ and its inverse metric ${\cal G}^{\alpha\beta}$ are defined in (\ref{eq:genMetrices}). In addition, we have introduced $\kappa_0 = 6 {\cal V} = \kappa_\alpha\, t^\alpha$, $\kappa_\alpha = \kappa_{\alpha \beta} \, t^\beta$, and $\kappa_{\alpha\beta} = \kappa_{\alpha\beta\gamma} \, t^\gamma$. Using the pieces of information in Eqs.~(\ref{eq:derK})-(\ref{eq:InvK}) one gets the following important identities\footnote{These identities hold true for more general cases \cite{AbdusSalam:2020ywo,Cicoli:2017shd}, e.g. in the presence of the perturbative ${\alpha^\prime}^3$-corrections of \cite{Becker:2002nn}, and even when the odd-moduli are included \cite{Cicoli:2021tzt}. However, these relations generically do not hold in the presence of string-loop corrections \cite{Cicoli:2021tzt,Leontaris:2022rzj}.},
\begin{eqnarray}
\label{eq:kaehler-identities}
& & {K}_A\, {K}^{{A} \ov {S}} \,  = (S -\ov S) = - {K}^{{S} \ov {B}} \, {K}_{\ov B}\,, \\
& & {K}_A\, {K}^{{A} \ov {T_\alpha}} \, = (T_\alpha -\ov T_\alpha) = - {K}^{{T_\alpha} \ov {B}} \,  {K}_{\ov B}\,, \nonumber\\
& & {K}_A\, {K}^{{A} \ov {U^i}} \,  = \frac{{\cal Z}\,(U^i -\ov U^i)}{{\cal Z}-6\gamma} = - {K}^{{U^i} \ov {B}} \, {K}_{\ov B}\,, \nonumber\\
& & K^{{A} \ov {B}} \, K_A \, K_{\ov {B}} = \,7+\frac{12\gamma}{{\cal Z}-6\gamma}. \nonumber
\end{eqnarray}
Note that, in the absence of odd-axions in type IIB construction, $\tau_\alpha = - \, {\rm Im}(T_\alpha)$, otherwise ${\rm Im}(T_\alpha)$ can generically include odd axions as well. In addition to (\ref{eq:kaehler-identities}), we have the following identities which are useful at the intermediate stage of simplifying the scalar potential,
\bea
\label{eq:kaehler-identities1}
& & {K}^{{U^i} \ov {U^j}} \, \sigma_j = \frac{{\cal Z}^2 \, u^i}{2({\cal Z} - 6 \gamma)}, \quad {K}^{{U^i} \ov {U^j}} \, \sigma_i\, \sigma_j = \frac{3{\cal Z}^2 \, ({\cal Z}-2\gamma)}{16({\cal Z} - 6 \gamma)},\\
& & {K}^{{U^i} \ov {U^j}} \, \ell_{ik} = \frac{{\cal Z}}{({\cal Z} - 6 \gamma)}\left( 4 \,u^j\, \sigma_k - \frac{{\cal Z} - 6 \gamma}{2}\, \delta_k^j\right), \nonumber\\
& & {K}^{{U^i} \ov {U^j}} \, \ell_{ik} \, \ell_{jm} = \frac{{\cal Z}^2}{4}\, {K}_{{U^k} \ov{U^m}} + \frac{{\cal Z}+6\gamma}{{\cal Z} - 6 \gamma} \, (4\, \sigma_k \, \sigma_m). \nonumber
\eea


\subsection{Expressing the superpotential using axionic fluxes}
Using these {\it new} axionic fluxes (\ref{eq:newOrbits1})-(\ref{eq:newOrbits2}), now we introduce the following combinations, namely $\psi$ and $\chi$, which also involve the volume moduli but no CS-saxions,

\bea
\label{eq:chi-def}
& & \hskip-1cm \chi_0 = \left(\widetilde{\rm f}_0 - \frac{1}{2} \tau_\alpha \tau_\beta \, \widetilde{\rm p}^{\prime\alpha\beta}_0 \right)  + i \left(- \tau_\alpha{\widetilde{\rm q}}^{\alpha}_0{} + \frac{1}{6} \tau_\alpha\tau_\beta\tau_\gamma \, {\widetilde{\rm h}_0}^{\prime\alpha\beta\gamma}\right) + \, i\, s\, \psi_0,\\
& & \hskip-1cm \chi_i = \left(\widetilde{\rm f}_i - \frac{1}{2} \tau_\alpha \tau_\beta \, \widetilde{\rm p}^{\prime\alpha\beta}_i \right)  + i \left(- \tau_\alpha{\widetilde{\rm q}}^{\alpha}_i{} + \frac{1}{6} \tau_\alpha\tau_\beta\tau_\gamma \, {\widetilde{\rm h}_i}^{\prime\alpha\beta\gamma}\right) + \, i\, s\, \psi_i, \nonumber\\
& & \hskip-1cm \chi^i = \left({\rm f}^i - \frac{1}{2} \tau_\alpha \tau_\beta \, {\rm p}^{\prime\alpha\beta i} \right)  + i \left(- \tau_\alpha{{\rm q}}^{\alpha i} + \frac{1}{6} \tau_\alpha\tau_\beta\tau_\gamma \, {{\rm h}}^{\prime\alpha\beta\gamma i}\right) + \, i\, s\, \psi^i, \nonumber\\
& & \hskip-1cm \chi^0 = \left({\rm f}^0 - \frac{1}{2} \tau_\alpha \tau_\beta \, {\rm p}^{\prime\alpha\beta 0} \right)  + i \left(-\tau_\alpha{{\rm q}}^{\alpha 0} + \frac{1}{6} \tau_\alpha\tau_\beta\tau_\gamma \, {{\rm h}}^{\prime\alpha\beta\gamma 0}\right) + \, i\, s\, \psi_0, \nonumber
\eea
where
\bea
\label{eq:psi-def}
& & \hskip-1cm \psi_0 = \left(-\widetilde{\rm h}_0 + \frac{1}{2} \tau_\alpha\tau_\beta \, {\widetilde{\rm q}}^{\prime\alpha\beta}_0 \right) + i \left(\tau_\alpha\,\widetilde{\rm p}^\alpha_0 - \frac{1}{6} \tau_\alpha\tau_\beta\tau_\gamma\, \widetilde{\rm f}^{\prime\alpha\beta\gamma}_0{}\right), \\
& & \hskip-1cm \psi_i = \left(-\widetilde{\rm h}_i + \frac{1}{2} \tau_\alpha\tau_\beta \, {\widetilde{\rm q}}^{\prime\alpha\beta}_i \right) + i \left(\tau_\alpha\,\widetilde{\rm p}^\alpha_i - \frac{1}{6} \tau_\alpha\tau_\beta\tau_\gamma\, \widetilde{\rm f}^{\prime\alpha\beta\gamma}_i{}\right), \nonumber\\
& & \hskip-1cm \psi^i = \left(-{\rm h}_i + \frac{1}{2} \tau_\alpha\tau_\beta \, {{\rm q}}^{\prime\alpha\beta i} \right) + i \left(\tau_\alpha\,{\rm p}^{\alpha i} - \frac{1}{6} \tau_\alpha\tau_\beta\tau_\gamma\, {\rm f}^{\prime\alpha\beta\gamma i}{}\right), \nonumber\\
& & \hskip-1cm \psi^0 = \left(-{\rm h}^0 + \frac{1}{2} \tau_\alpha\tau_\beta \, {{\rm q}}^{\prime\alpha\beta 0} \right) + i \left(\tau_\alpha\,{\rm p}^{\alpha 0} - \frac{1}{6} \tau_\alpha\tau_\beta\tau_\gamma\, {\rm f}^{\prime\alpha\beta\gamma 0}{}\right), \nonumber
\eea
From (\ref{eq:chi-def}) and (\ref{eq:psi-def}), let us note that the first two components with lower indices include the rational shifts (denoted with tildes) while those with upper indices do not have such shifts.

These new combinations $\chi$ and $\psi$ as defined in (\ref{eq:chi-def}) and (\ref{eq:psi-def}) plays a very crucial role in rewriting the scalar potential in a compact manner as we will see later. For example, the generalized flux superpotential (\ref{eq:W-all-3}) can be expressed in the following compact form,
\bea
\label{eq:W-gen4}
& & W = {\chi}_0 - i\,  u^i \, {\chi}_i\, - {\chi}^i \,\sigma_i + i\, {\chi}^0 \, \left({\cal U} + \, \gamma\right),
\eea
where let us recall that the new combinations $\chi$'s do not depend on the CS-saxion $u^i$.


\subsection{Master formulae for the scalar potential}
Using the important identities given in Eq.~(\ref{eq:kaehler-identities}), the generic scalar potential (\ref{eq:V_gen}) simplifies to the following form,
\bea
\label{eq:V_gen1}
& & \hskip-0.5cm V = e^K\biggl[W_{U^i} \, K^{U^i \ov{U^j}}\, \ov{W}_{\ov{U}^j} + W_{T_\alpha} \, K^{{T_\alpha} \ov{T}_\beta}\, \ov{W}_{\ov{T}_\beta} + 4\,s^2\, W_S \, \ov{W}_{\ov{S}} \\
& & +\frac{2\, i\, u^i\, {\cal Z}}{{\cal Z} - 6 \gamma} (W_{U^i} \, \ov{W} - \ov{W}_{\ov{U}^i} \,{W}) + 2\, i\, \tau_\alpha (W_{T_\alpha} \, \ov{W} - \ov{W}_{\ov{T}_\alpha} \,{W}) - 2\, i\, s (W_{S} \, \ov{W} - \ov{W}_{\ov{S}} \,{W}) \nonumber\\
& & + \left(4+\frac{12\gamma}{{\cal Z}-6\gamma}\right) |W|^2 \biggr]. \nonumber
\eea
Now using the {\it new} axionic fluxes (\ref{eq:newOrbits1})-(\ref{eq:newOrbits2}), the generalized flux superpotential (\ref{eq:W-all-3}) and its derivatives (\ref{eq:derW_gen2}) can be concisely expressed in the following manner,
\bea
\label{eq:W-derW-compact}
& & W = {\chi}_0 - i\,  u^i \, {\chi}_i\, - {\chi}^i \,\sigma_i + i\, {\chi}^0 \, \left({\cal U} + \, \gamma\right),\\
& & W_{U^i} = {\chi}_i\, -\, i \,\ell_{ij} \, {\chi}^j - {\chi}^0 \, \sigma_i,\nonumber\\
& & W_S = {\psi}_0 -\,i\, u^i \, {\psi}_i\, - {\psi}^i \,\sigma_i + i\, {\psi}^0 \, \left({\cal U} + \, \gamma\right),\nonumber\\
& & W_{T_\alpha} = {\Psi}^\alpha_0 -\,i\, u^i \, {\Psi}^\alpha_i\, - {\Psi}^{\alpha i} \,\sigma_i +\,i\, {\Psi}^{\alpha 0} \, \left({\cal U} + \, \gamma\right),\nonumber
\eea
where similar to the combinations $\chi$ and $\psi$ as defined in (\ref{eq:chi-def})-(\ref{eq:psi-def}), we have introduced another set of combinations $\Psi^\alpha$ defined as below,
\bea
\label{eq:Psi-def}
& & \Psi^\alpha_0 = \left({\widetilde{\rm q}}^{\alpha}_0 - \, s \,\tau_\beta\, {\widetilde{\rm q}}^{\prime\alpha\beta}_0 - \,\frac{1}{2}\, \tau_\beta \tau_\gamma\, {\widetilde{\rm h}}^{\prime\alpha\beta\gamma}_0\right) + i \left(- s \, \widetilde{\rm p}^{\alpha}_0 - \tau_\beta\, \widetilde{\rm p}^{\prime\alpha\beta}_0 + \frac{s}{2}\,\tau_\beta\tau_\gamma \widetilde{\rm f}^{\prime\alpha\beta\gamma}_0\right),\\
& & \Psi^\alpha_i = \left({\widetilde{\rm q}}^{\alpha}_i - \, s \,\tau_\beta\, {\widetilde{\rm q}}^{\prime\alpha\beta}_i - \,\frac{1}{2}\, \tau_\beta \tau_\gamma\, {\widetilde{\rm h}}^{\prime\alpha\beta\gamma}_i\right) + i \left(- s \, \widetilde{\rm p}^{\alpha}_i - \tau_\beta\, \widetilde{\rm p}^{\prime\alpha\beta}_i + \frac{s}{2}\,\tau_\beta\tau_\gamma \widetilde{\rm f}^{\prime\alpha\beta\gamma}_i\right),\nonumber\\
& & \Psi^{\alpha i} = \left({{\rm q}}^{\alpha i} - \, s \,\tau_\beta\, {{\rm q}}^{\prime\alpha\beta i} - \,\frac{1}{2}\, \tau_\beta \tau_\gamma\, {{\rm h}}^{\prime\alpha\beta\gamma i}\right) + i \left(- s \, {\rm p}^{\alpha i} - \tau_\beta\, {\rm p}^{\prime\alpha\beta i} + \frac{s}{2}\,\tau_\beta\tau_\gamma {\rm f}^{\prime\alpha\beta\gamma i}\right),\nonumber\\
& & \Psi^{\alpha 0} = \left({{\rm q}}^{\alpha 0} - \, s \,\tau_\beta\, {{\rm q}}^{\prime\alpha\beta 0} - \,\frac{1}{2}\, \tau_\beta \tau_\gamma\, {{\rm h}}^{\prime\alpha\beta\gamma 0}\right) + i \left(- s \, {\rm p}^{\alpha 0} - \tau_\beta\, {\rm p}^{\prime\alpha\beta 0} + \frac{s}{2}\,\tau_\beta\tau_\gamma {\rm f}^{\prime\alpha\beta\gamma 0}\right),\nonumber
\eea
From (\ref{eq:chi-def}), (\ref{eq:psi-def}) and (\ref{eq:Psi-def}) let us note that the first two components with lower indices include the rational shifts (denoted with tildes) while those with upper indices do not have such shifts. However, we omit writing tildes explicitly for the lower components of $\chi, \psi$ and $\Psi^\alpha$ and it is understood that such a convention applies. In fact one can compactly express the definitions (\ref{eq:chi-def}), (\ref{eq:psi-def}) and (\ref{eq:Psi-def}) in the following shorthand notations,
\bea
\label{eq:chi-psi-Psi}
& & \chi = \left({\rm f} - s \, {\rm p} - {\rm p}^{\prime} + s\, {\rm f}^{\prime}{}\right)  + i \, \left(- s \, {\rm h} - {{\rm q}}^{}{} + s \, {{\rm q}}^{\prime} + {{\rm h}}^{\prime}\right),\\
& & \psi = \, \left(- \, {\rm h} + \, {{\rm q}}^{\prime}\right) + i \, \left({\rm p} - \, {\rm f}^{\prime} \right)\,, \nonumber\\
& & \Psi^\alpha = \left({{\rm q}}^{\alpha} - \, s \, {{\rm q}}^{\prime\alpha} - \, {{\rm h}}^{\prime\alpha}\right) + i \left(- s \, {\rm p}^{\alpha} - {\rm p}^{\prime\alpha} + s\, {\rm f}^{\prime\alpha}\right), \nonumber
\eea
where in the above shorthand notation (\ref{eq:chi-psi-Psi}) the following conventions are taken to be understood:
\begin{itemize}

\item
Each of the three axionic fluxes, namely $\chi, \psi$ and $\Psi$, has four components with appropriate indices which is correlated with the axionic fluxes with same index structure; for example the flux ${\rm f}$ appearing in the expression of $\chi$ has components $\{\widetilde{\rm f}_0, \widetilde{\rm f}_i, {\rm f}^i, {\rm f}^0\}$ which will contribute in the respective definitions of $\{\chi_0, \chi_i, \chi^i, \chi^0\}$  as seen from (\ref{eq:chi-def}), (\ref{eq:psi-def}) and (\ref{eq:Psi-def}).

\item
All the additional $\alpha \in h^{1,1}_+$ indices carried by a flux of particular type are summed over by contracting with $\tau_\alpha$'s and having the appropriate normalization factors; for example ${\rm p} = \tau_\alpha {\rm p}^\alpha$, ${\rm q}^{\prime\alpha} = \tau_\beta {\rm q}^{\prime\alpha\beta}$, ${\rm q}' = \frac{1}{2} \tau_\alpha \tau_\beta {\rm q}^{\prime\alpha\beta}$, ${\rm h}' = \frac{1}{6} \tau_\alpha\tau_\beta\tau_\gamma {\rm h}^{\prime\alpha\beta\gamma}$, ${\rm h}^{\prime \alpha} = \frac{1}{2} \tau_\alpha\tau_\beta\tau_\gamma {\rm h}^{\prime\alpha\beta\gamma}$ etc.

\item
Using the shorthand notations with summed $\alpha$-indices as mentioned above implies that,
\bea
\label{eq:Psi-summed-alpha}
&& \Psi \equiv \tau_\alpha \Psi^\alpha = \left({\rm q} - 2\, s \, {\rm q}^{\prime} - 3\, {\rm h}^{\prime}\right) + i \left(- s \, {\rm p} - 2\,{\rm p}^{\prime} + 3\, s\, {\rm f}^{\prime}\right).
\eea

\end{itemize}

\subsubsection*{First formulation:}

Motivated by the pieces of information in (\ref{eq:W-derW-compact}), we define the following two linear functionals $\Phi[y]$ and $\Theta_i[y]$ as the key ingredients for reading-off the generic scalar potential,
\bea
\label{eq:Phi-Theta}
& & \Theta_i[y] = {y}_i\, -\, i \,\ell_{ij} \, {y}^j - {y}^0 \, \sigma_i,\\
& & \Phi[y] = y_0 - i\,  u^i \, y_i\, - y^i \,\sigma_i + i\, y^0 \, \left({\cal U} + \, \gamma\right), \nonumber
\eea
which means that $\Phi[\chi] = W, \Theta_i[\chi] = W_{U^i},  \Phi[\psi] = W_S$ and $\Phi[\Psi^\alpha] = W_{T_\alpha}$.  Moreover, these functionals are linear in terms of axionic fluxes, e.g. $\Phi[y_1+y_2] = \Phi[y_1] + \Phi[y_2]$, and one has to also remember that $y$'s are generically complex and so $\ov\Phi[y]$ and $\ov\Phi[\ov{y}]$ are generically different.

Using the functionals $\Phi[y]$ and its derivative $\Theta_i[y]$ defined in (\ref{eq:Phi-Theta}) the generic scalar potential (\ref{eq:V_gen1}) in the ``First formulation" can be expressed as,
\bea
\label{eq:V_gen2}
& & \hskip-1cm V_{\rm I} = e^K \biggl[K^{i\ov{j}}\,\Theta_i[\chi] \cdot  \ov\Theta_j[\ov\chi] + \frac{4\,{\cal U}+\gamma}{2\,{\cal U}-\gamma} (i\, u^i)\, \left(\Theta_i[\chi] \cdot \ov\Phi[\ov\chi] - \Phi[\chi]\cdot \ov\Theta_i[\ov\chi] \right) \\
& & \hskip-0.5cm + \frac{8\,{\cal U}-\gamma}{2\,{\cal U}-\gamma} \Phi[\chi]\cdot \ov\Phi[\ov\chi] + 4\, s^2\, \Phi[\psi] \cdot \ov\Phi[\ov\psi] + K_{\alpha\ov\beta}\, \Phi[\Psi^\alpha] \cdot \ov\Phi[\ov\Psi^\beta]   \nonumber\\
& & \hskip-0.5cm + \, 2\,i\, \left(\Phi[\Psi -s \, \psi] \cdot \ov\Phi[\ov\chi] - \Phi[\chi]\cdot \ov\Phi[\ov\Psi -s \, \ov\psi] \right) \biggr]. \nonumber
\eea
Thus the main task for reading-off the scalar potential boils down to write down the two functional $\Theta_i[\chi]$ and $\Phi[y]$ for the relevant combinations $\chi, \psi$ and $\Psi^\alpha$. Notice that the first three pieces of the master-formula (\ref{eq:V_gen2}) depend on the $\chi$ combinations only, and therefore one can even get rid of the $\Theta_i[\chi]$ functional via explicitly writing things in terms of $\chi$ and $\Phi[y]$. This leads to an equivalent version of the master formula (\ref{eq:V_gen2}) written as below,
\bea
\label{eq:master1}
& & \hskip5.5cm V_{\rm I} = V_{\rm I}^{(1)} + V_{\rm I}^{(2)}, \\
& & \hskip-0.75cm {\rm where} \nonumber\\
& & \hskip-0.5cm V_{\rm I}^{(1)} = e^K \biggl[\frac{8\,{\cal U} - \gamma}{2\,{\cal U} - \gamma} \chi_0 \ov{\chi}_0 + \left(K^{i\ov{j}} -\frac{3\gamma u^i \, u^j}{2{\cal U} - \gamma}\right)  \chi_i \, \ov{\chi}_j + \left((4 {\cal U} + \gamma)^2\,K_{i\ov{j}}+\frac{3\gamma \sigma_i\, \sigma_j}{2{\cal U} - \gamma} \right) \chi^i\, \ov{\chi}^j  \nonumber\\
& & \hskip-0cm + \frac{1}{2}\left(8\,{\cal U}^2 + \gamma\, {\cal U} + 2\, \gamma^2\right) \chi^0\, \ov{\chi}^0 +\frac{3 \gamma \sigma_i}{(2\, {\cal U} - \gamma)} (\chi^i \ov\chi_0 + \ov\chi^i \chi_0)-\frac{3 \gamma \, u^i}{2} (\chi_i \ov\chi^0 + \ov\chi_i \chi^0) \nonumber\\
& & \hskip-0cm - i\, (4\,{\cal U} + \gamma) \left(\chi_i \ov\chi^i - \ov\chi_i \chi^i \right) +\, i\, (2\,{\cal U} - \gamma) \left(\chi_0 \ov\chi^0 - \ov\chi_0 \chi^0 \right)   - 2\, i\, u^i\,(\chi_i \ov\chi_0 - \ov\chi_i \chi_0) \nonumber\\
& & \hskip-0cm + 2\, i\, ({\cal U} + \gamma) \, \sigma_i\,(\chi^i \ov\chi^0 - \ov\chi^i \chi^0) + 2\,i \, u^i\, \sigma_j \, (\chi_i \ov\chi^j - \ov\chi_i \chi^j)\biggr]    \nonumber\\
& & \hskip-0.5cm V_{\rm I}^{(2)} = e^K \biggl[4\, s^2\, \Phi[\psi] \cdot \ov\Phi[\ov\psi] + K_{\alpha\ov\beta}\, \Phi[\Psi^\alpha] \cdot \ov\Phi[\ov\Psi^\beta] + \, 2\,i\, \left(\Phi[\Psi -s \, \psi] \cdot \ov\Phi[\ov\chi] - \Phi[\chi]\cdot \ov\Phi[\ov\Psi -s \, \ov\psi] \right) \biggr]. \nonumber
\eea
We argue that once a set of topological data such as the triple intersection numbers $(\kappa_{\alpha\beta\gamma}, \ell_{ijk})$ of the compactifying CY threefold and its mirror, the Euler characteristics and the second Chern numbers are known for a given concrete model, one can {\it read-off} the generic scalar potential from the master formula (\ref{eq:V_gen2}) or (\ref{eq:master1}) via simply writing down the two functionals $\Phi[y]$ and $\Theta_i[y]$ by using the combinations $\chi, \psi$ and $\Psi^\alpha$ (\ref{eq:chi-psi-Psi}) written out in terms of the {\it new} axionic fluxes (\ref{eq:newOrbits1})-(\ref{eq:newOrbits2}).

As a side remark let us mention that the coefficient metrics corresponding to $(\chi_i \, \ov{\chi}_j)$ and $(\chi^i \, \ov{\chi}^j)$ in $V_{\rm I}^{(1)}$ of (\ref{eq:master1}) do not form inverse metric pair. In fact, we have the following relations,
\bea
\label{eq:identities-2}
& & \frac{1}{4\,{\cal U}+\gamma}\left(K^{i\ov{j}} -\frac{3\gamma u^i \, u^j}{2{\cal U} - \gamma}\right) = \frac{2\, u^i\, u^j}{4\,{\cal U}+\gamma} - \, \ell^{ij}, \\
& & \frac{1}{4\,{\cal U}+\gamma}\left((4 {\cal U} + \gamma)^2\,K_{k\ov{j}}+\frac{3\gamma \sigma_k\, \sigma_j}{2{\cal U} - \gamma}\right) = \frac{(8\,{\cal U} - \gamma)\, \sigma_i\, \sigma_j}{{(4\,{\cal U} + \gamma})\,(2{\cal U} - \gamma)} - \,\ell_{ij},\nonumber\\
& & \left(\frac{2\, u^i\, u^j}{4\,{\cal U}+\gamma} - \, \ell^{ij}\right) \left(\frac{(8\,{\cal U} - \gamma)\, \sigma_j\, \sigma_k}{(4{\cal U} + \gamma)\,(2{\cal U} - \gamma)} - \,\ell_{jk}\right) = \delta^i_k + \frac{9\, \gamma^2 u^i\, \sigma_k}{2\,(4{\cal U} + \gamma)^2\,(2{\cal U} - \gamma)},\nonumber
\eea
where we have used $e^{-K} = 4\,s\,{\cal V}^2\, (4{\cal U} + \gamma)$ to pull out an overall factor $(4{\cal U}+\gamma)^{-1}$ in the first two lines of (\ref{eq:identities-2}). Thus, they form inverse metric pair only for $\gamma = 0$. However another thing to observe is the presence of two cross-terms in the second line of $V_{\rm I}^{(1)}$ and it happens that if one redefines $\chi_0$ and $\chi_i$ with the following shifts,
\bea
\label{eq:shift-chi0-chii}
& & \chi_0^{\rm new} = \chi_0 + \frac{3\gamma}{8{\cal U} - \gamma} \sigma_i\, \chi^i, \qquad \chi_i^{\rm new} = \chi_i -\frac{3\gamma}{8{\cal U} - \gamma} \sigma_i\, \chi^0,
\eea
then the two cross-terms in the second line of $V_{\rm I}^{(1)}$ vanish, and subsequently one finds that the $V_{\rm I}^{(1)} $ piece of (\ref{eq:master1}) can be also expressed as below,
\bea
\label{eq:V_gen3b}
& & V_{\rm I}^{(1)}  = \frac{1}{4\,s\,{\cal V}^2} \biggl[\frac{\chi_0^{\rm new} \, \, \ov\chi_0^{\rm new}}{x_0} + x^{ij}\, \chi_i^{\rm new} \, \, \ov\chi_i^{\rm new} + x_{ij} \, \chi^{i} \, \, \ov\chi^{i} + x_0\, \chi^{0} \, \, \ov\chi^{0} \\
& & \hskip0.5cm + \frac{i\, (4\,{\cal U} + \gamma)}{(8\,{\cal U}-\gamma)} \left(\chi_0^{\rm new} \ov\chi^0 - \ov\chi_0^{\rm new} \chi^0 \right)   - \frac{2\, i\,u^i}{(4\,{\cal U}+\gamma)}\,(\chi_i^{\rm new} \ov\chi_0^{\rm new} - \ov\chi_i^{\rm new} \chi_0^{\rm new}) \nonumber\\
& & \hskip0.5cm + \frac{4\, i\, x_0\, \sigma_i}{(8\,{\cal U} - \gamma)} \, (\chi^i \ov\chi^0 - \ov\chi^i \chi^0) + \frac{4\,i \, u^i\, \sigma_j}{(8\,{\cal U} - \gamma)} \, (\chi_i^{\rm new} \ov\chi^j - \ov\chi_i^{\rm new} \chi^j)- i\,\left(\chi_i^{\rm new} \ov\chi^i - \ov\chi_i^{\rm new} \chi^i \right)\biggr],\nonumber
\eea
where $x_0$ and the inverse pair of metrics $(x_{ij}, x^{ij})$ are given as below,
\bea
\label{eq:x0-xij-invxij-gen}
& & \hskip-1cm x_0 = \frac{(2\, {\cal U} - \gamma) (4\,{\cal U} + \gamma)}{8\,{\cal U} - \gamma}, \quad x^{ij} = \frac{2\, u^i\, u^j}{4\, {\cal U} + \gamma} - \ell^{ij}, \quad x_{ij} = \frac{8\sigma_i\sigma_j}{8\,{\cal U} - \gamma} - \ell_{ij}.
\eea
Thus, after implementing this shift, the pair $(x_{ij}, x^{ij})$ consists of metrics which are indeed inverse of each other, though they are different from the moduli space metric pairs $((4 {\cal U} + \gamma)^2\,K_{i\ov{j}}, K^{i\ov{j}})$. These observations are necessary to make connections with the results of \cite{Shukla:2016hyy} where ingredients defined in Eq.~(\ref{eq:x0-xij-invxij-gen}) are used.

It is also worth noticing that the two approaches are identified for $\gamma = 0$ which leads to $\chi_0^{\rm new} =\chi_0$ and $\chi_i^{\rm new} = \chi_i$ as seen from (\ref{eq:shift-chi0-chii}), and the $V_{\rm I}^{(1)}$ pieces simplifies as
\bea
\label{eq:V_gen3c}
& & \hskip-1cm V_{\rm I}^{(1)}  = \frac{1}{4\,s\,{\cal V}^2} \biggl[\frac{\chi_0^{\rm new} \, \, \ov\chi_0^{\rm new}}{\hat{x}_0} + \hat{x}^{ij}\, \chi_i^{\rm new} \, \, \ov\chi_i^{\rm new} + \hat{x}_{ij} \, \chi^{i} \, \, \ov\chi^{i} + \hat{x}_0\, \chi^{0} \, \, \ov\chi^{0} \\
& & \hskip0.0cm + \frac{i}{2} \left(\chi_0^{\rm new} \ov\chi^0 - \ov\chi_0^{\rm new} \chi^0 \right)  - i\,\left(\chi_i^{\rm new} \ov\chi^i - \ov\chi_i^{\rm new} \chi^i \right) - \frac{i\,u^i}{2\,{\cal U}}\,(\chi_i^{\rm new} \ov\chi_0^{\rm new} - \ov\chi_i^{\rm new} \chi_0^{\rm new}) \nonumber\\
& & \hskip0.0cm + \frac{i\, \sigma_i}{2} \, (\chi^i \ov\chi^0 - \ov\chi^i \chi^0) + \frac{i \, u^i\, \sigma_j}{2\,{\cal U}} \, (\chi_i^{\rm new} \ov\chi^j - \ov\chi_i^{\rm new} \chi^j)\biggr],\nonumber
\eea
where, using the metrics (\ref{eq:Kmetric})-(\ref{eq:InvK}), the definitions in (\ref{eq:x0-xij-invxij-gen}) simplify for $\gamma = 0$ as below,
\bea
\label{eq:x0-xij-invxij}
& & \hskip-1cm \hat{x}_0 = {\cal U}, \quad \hat{x}^{ij} = \frac{2u^i u^j- 4\, \ell^{ij}}{4{\cal U}} = \frac{1}{{4\,{\cal U}}}\, K^{i\ov{j}}_0, \quad \hat{x}_{ij} = \frac{\sigma_i \, \sigma_j - {\cal U} \, \ell{ij}}{{\cal U}} = 4{\cal U}\, K^0_{i\ov{j}}.
\eea
We note that $K^0_{i\ov{j}}$ and $K^{i\ov{j}}_0$ denote the respective tree-level metrics corresponding to $\gamma = 0$. This simplification matches with the conventions and results presented in \cite{Shukla:2019wfo} as we will detail later on.

However, we find that the shifted combinations defined in (\ref{eq:shift-chi0-chii}) are no more better in compactly writing down the scalar potential in the presence of prime-fluxes as the $V_{\rm I}^{(2)}$ piece (\ref{eq:master1}) subsequently induces several additional cross-terms. Therefore we will take (\ref{eq:master1}) as the main master-formula in our current work where the triplet quantities $\{\chi, \psi, \Psi^\alpha\}$, which play the central role, are defined in (\ref{eq:chi-psi-Psi}) using axionic fluxes in (\ref{eq:newOrbits1})-(\ref{eq:newOrbits2}), without any further mixing or the need of taking combinations thereof !

\subsubsection*{Second formulation:}
In fact with some reshuffling of terms one can rewrite the master formula (\ref{eq:V_gen2}) or (\ref{eq:master1}) in some alternative ways as well, depending on whether one wants to use moduli space metrics for the K\"ahler and the complex-structure moduli spaces or simply the quantities like $\ell^{ij}$ and $\kappa_{\alpha\beta}$ in order to read-off the scalar potential. Note that any two of the three ingredients from the set $\{(\tau_\alpha\tau_\beta), \kappa_{\alpha\beta}, K_{\alpha\ov\beta}\}$ can be used as they are related by the metric relation (\ref{eq:InvK}). The first formulation of the scalar potential given in (\ref{eq:V_gen2}) and \ref{eq:master1} uses $\{K_{\alpha\ov\beta}, (\tau_\alpha \tau_\beta)\}$ and one can have an alternate one using $\{\kappa_{\alpha\beta}, \, (\tau_\alpha \tau_\beta)\}$ which we present in the ``Second formulation". Here we note that we want to keep $(\tau_\alpha \tau_\beta)$ in both the  formulations as it helps in writing shorthand quantities like $\Psi = (\tau_\alpha \Psi^\alpha)$ which subsequently is useful in writing the scalar potential in compact way.

Using the expressions of moduli space metrics given in (\ref{eq:InvK}), the generic scalar potential (\ref{eq:V_gen1}) apparently takes a simpler form given as,
\bea
\label{eq:V_gen4}
& & \hskip-1cm V = e^K\biggl[\frac{4\,{\cal U}+ \gamma}{2\,{\cal U} - \gamma} \, \, \bigl|W + \,i\, u^i\, W_{U^i}\bigr|^2 + \bigl|W + 2\, i\, \tau_\alpha W_{T_\alpha}\bigr|^2 + \bigl|W - 2\, i\, s W_{S}\bigr|^2\\
& & -\left(4\, {\cal U} + \gamma\right) \, \ell^{ij} \, W_{U^i}\, \ov{W}_{\ov{U}^j} - 4\,{\cal V}\,\kappa_{\alpha\beta}\, W_{T_\alpha}  \ov{W}_{\ov{T}_\beta} \biggr], \nonumber
\eea
where as we see, three pieces in (\ref{eq:V_gen4}) are positive semi-definite. Using the pieces of information in (\ref{eq:W-derW-compact}), the generic scalar potential (\ref{eq:V_gen4}) can be expressed by the following five pieces,
\bea
\label{eq:master2}
& & V_{\rm II} = e^K \biggl[V_{\rm II}^{(1)} + V_{\rm II}^{(2)} + V_{\rm II}^{(3)} + V_{\rm II}^{(4)} + V_{\rm II}^{(5)}\biggr],
\eea
where
\bea
\label{eq:V_gen5a}
& & V_{\rm II}^{(1)} = \frac{4\,{\cal U}+ \gamma}{2\,{\cal U} - \gamma} \, \, \biggl|{\chi}_0 + {\chi}^i \,\sigma_i -i\, \chi^0 \,(2\, {\cal U} - \gamma)\biggr|^2,\\
& & V_{\rm II}^{(2)} = -\left(4\, {\cal U} + \gamma\right) \, \ell^{ij} \, \left({\chi}_i\, - i\, \ell_{ik}\,{\chi}^k - {\chi}^0 \,\sigma_i\right)\, \left(\ov{\chi}_j\, + i\, \ell_{jm}\,\ov{\chi}^m - \ov{\chi}^0 \, \sigma_j\right), \nonumber\\
& & V_{\rm II}^{(3)} = \biggl|({\chi}_0 - 2 \, i\, s\, \psi_0) - i\,  u^i \, ({\chi}_i- 2 \, i\, s\, \psi_i)\, - ({\chi}^i- 2 \, i\, s\, \psi^i) \,\sigma_i + i\,({\cal U}+\gamma)\,(\chi^0- 2 \, i\, s\, \psi^0)\biggr|^2, \nonumber\\
& & V_{\rm II}^{(4)} = \biggl|({\chi}_0 + 2 \, i\, \Psi_0) - i\,  u^i \, ({\chi}_i+ 2 \, i\, \Psi_i)\, - ({\chi}^i+ 2 \, i\, \Psi^i) \,\sigma_i + i\,({\cal U}+\gamma)({\chi}^0- 2 \, i\, \Psi^0)\biggr|^2, \nonumber\\
& & V_{\rm II}^{(5)} = - 4\,{\cal V}\,\kappa_{\alpha\beta}\, \left({\Psi}^\alpha_0 -\,i\, u^i \, {\Psi}^\alpha_i\, - {\Psi}^{\alpha i} \,\sigma_i +\,i\,({\cal U}+\gamma)\, {\Psi}^{\alpha 0}\right) \nonumber\\
&& \hskip2cm \times \left({\ov\Psi}^\beta_0 +\,i\, u^j \, \ov{\Psi}^\beta_j\, - \ov{\Psi}^{\beta j} \,\sigma_j -\,i\,({\cal U}+\gamma)\, \ov{\Psi}^{\beta 0}\right),\nonumber
\eea
where we used $K_{\alpha\ov\beta} =\,4\, \tau_\alpha\tau_\beta -4\,{\cal V} \kappa_{\alpha\beta}$. Also, note that this formulation can also be expressed using a set of functionals like $\Phi[y]$ and $\Theta_i[y]$ as we used earlier, however for the current formulation we prefer to express all the pieces using $\chi, \psi$ and $\Psi$ only as defined in (\ref{eq:chi-psi-Psi}) via the {\it new} axionic-flux polynomials (\ref{eq:newOrbits1})-(\ref{eq:newOrbits2}).

A couple of additional formulations of the scalar potential with the explicit manifestation of the CS-moduli $u^i$ dependence are also presented in the appendix \ref{sec_AlternateFormulations}.


\section{Reading-off the scalar potentials}
\label{sec_reading-off}
In this section, we will illustrate the process of {\it reading-off} the scalar potentials using our master formula in some concrete models.

\subsection{Model A : $(F, H)$ fluxes only}
As a warm-up let us begin with a setup having $\{F, H\}$ fluxes only which induces the standard flux superpotential \cite{Gukov:1999ya,Dasgupta:1999ss}. Then the simplified triplets $\{\chi, \psi, \Psi^\alpha\}$ are given as (\ref{eq:chi-psi-Psi}),
\bea
\label{eq:}
& & \chi = {\rm f} + i \, \left(- s \, {\rm h}\right), \quad \psi = - \, {\rm h}, \quad \Psi^\alpha =0, \quad \Psi \equiv \tau_\alpha \Psi^\alpha = 0,
\eea
which means $(\Psi - s \psi) = s\, {\rm h} = \frac{i}{2}(\chi - \ov\chi)$ is a real quantity. Also note that the internal structure of ${\rm f}$ and ${\rm h}$ combinations changes accordingly in the sense that $(Q, P)$ as well as prime-fluxes are absent. Subsequently the master-formula (\ref{eq:master1}) simplifies to the following generic scalar potential,
\bea
\label{eq:V_gen-FH}
& & \hskip-1.5cm V_{\rm FH}  = e^K \biggl[\frac{8\,{\cal U} - \gamma}{2\,{\cal U} - \gamma} \chi_0 \ov{\chi}_0 + \left(K^{i\ov{j}} -\frac{3\gamma u^i \, u^j}{2{\cal U} - \gamma}\right)  \chi_i \, \ov{\chi}_j + \left((4 {\cal U} + \gamma)^2\,K_{i\ov{j}}+\frac{3\gamma \sigma_i\, \sigma_j}{2{\cal U} - \gamma} \right) \chi^i\, \ov{\chi}^j  \\
& & \hskip-0cm + \frac{1}{2}\left(8\,{\cal U}^2 + \gamma\, {\cal U} + 2\, \gamma^2\right) \chi^0\, \ov{\chi}^0 +\frac{3 \gamma \sigma_i}{(2\, {\cal U} - \gamma)} (\chi^i \ov\chi_0 + \ov\chi^i \chi_0) -\frac{3 \gamma \, u^i}{2} (\chi_i \ov\chi^0 + \ov\chi_i \chi^0) \nonumber\\
& & \hskip-0cm + i\, (4\,{\cal U} + \gamma) \left(\chi_0 \ov\chi^0- \chi_i \ov\chi^i + \ov\chi_i \chi^i - \ov\chi_0 \chi^0 \right)\biggr].\nonumber
\eea
This is a fairly simple form for the generic $(F, H)$ induced scalar potential expressed using a single combination $\chi = {\rm f} + i \, \left(- s \, {\rm h}\right)$. In terms of the real axionic-flux polynomials $({\rm f}, {\rm h})$, the generic $(F,H)$ scalar potential (\ref{eq:V_gen-FH}) takes the following form,
\bea
\label{eq:GVW-generic-master-formula}
& & \hskip-1.5cm V_{\rm FH} = V_{\rm FH}^{(1)} + V_{\rm FH}^{(2)},
\eea
with the two pieces given as below
\bea
& & \hskip-0.5cm V_{\rm FH}^{(1)} = e^K \biggl[\frac{8\,{\cal U} - \gamma}{2\,{\cal U} - \gamma} \left((\widetilde{\rm f}_0)^2 + s^2 \, (\widetilde{\rm h}_0)^2\right) + \left(K^{i\ov{j}} -\frac{3\gamma u^i \, u^j}{2{\cal U} - \gamma}\right)  \left(\widetilde{\rm f}_i \, \widetilde{\rm f}_j + s^2 \, \widetilde{\rm h}_i \, \widetilde{\rm h}_j\right) \nonumber\\
& & + \left((4 {\cal U} + \gamma)^2\,K_{i\ov{j}}+\frac{3\gamma \sigma_i\, \sigma_j}{2{\cal U} - \gamma} \right) \left({\rm f}^i {\rm f}^j + s^2\, {\rm h}^i {\rm h}^j \right) + \frac{8\,{\cal U}^2 + \gamma\, {\cal U} + 2\, \gamma^2}{2} \left(({\rm f}^0{\rm )}^2 + s^2\, ({\rm h}^0)^2 \right) \nonumber\\
& & + \frac{6\, \gamma\, \sigma_i}{2\,{\cal U} - \gamma} \left(\widetilde{\rm f}_0\, {\rm f}^i + s^2\, \widetilde{\rm h}_0\, {\rm h}^i \right) - 3 \, \gamma \, u^i \, \left(\widetilde{\rm f}_i\, {\rm f}^0 + s^2 \, \widetilde{\rm h}_i {\rm h}^0\right)\biggr], \nonumber\\
& & \hskip-0.5cm V_{\rm FH}^{(2)} = e^K \cdot (2\,s) \, (4\,{\cal U}+ \gamma) \left({\rm f}^0\,\widetilde{\rm h}_0 - {\rm f}^i\,\widetilde{\rm h}_i +{\rm h}^i\,\widetilde{\rm f}_i - {\rm h}^0\,\widetilde{\rm f}_0 \right) = \frac{1}{2\,{\cal V}^2} \left({\rm f}^0\,{\rm h}_0 - {\rm f}^i\,{\rm h}_i +{\rm h}^i\,{\rm f}_i - {\rm h}^0\,{\rm f}_0 \right),\nonumber
\eea
where recall that $\widetilde{\rm f}_0, \widetilde{\rm f}_i, \widetilde{\rm h}_0$ and $\widetilde{\rm h}_i$ receive the rational shifts as mentioned in Eq.~(\ref{eq:rational-shift}). Also, the last term $V_{\rm FH}^{(2)}$ corresponds to the tadpole piece arising from the 10-dimensional Chern-Simons action, expressed as $(F \wedge H)$ piece as seen from the contraction of indices. Here it may be worth mentioning that for the ($F, H$) model, there is a crucial cancellation in the second piece $V_{\rm I}^{(2)}$ of the master-formula (\ref{eq:master1}) as,
\bea
& & \hskip-0.5cm V_{\rm I}^{(2)} = e^K \biggl[4\, s^2\, \Phi[{\rm h}] \cdot \ov\Phi[{\rm h}] + \, 2\,i\,s\, \left(\Phi[{\rm h}] \cdot \ov\Phi[{\rm f} + i \, s \, {\rm h}] - \Phi[{\rm f} - i \, s \, {\rm h}]\cdot \ov\Phi[{\rm h}] \right) \biggr]\nonumber\\
& & = e^K \biggl[2\,i\,s\, \left(\Phi[{\rm h}] \cdot \ov\Phi[{\rm f}] - \Phi[{\rm f}]\cdot \ov\Phi[{\rm h}] \right) \biggr],
\eea
where the pieces with $s^2$ dependence result in a precise cancellation.

\subsection{Model B : $(F, H, Q)$ fluxes only}
If we consider a setup with $\{F, H, Q\}$ fluxes only, e.g. those studied in \cite{Blumenhagen:2013hva,Shukla:2015hpa,Blumenhagen:2015lta,Shukla:2019wfo}, then the simplified triplets $\{\chi, \psi, \Psi^\alpha\}$ are given as (\ref{eq:chi-psi-Psi}),
\bea
\label{eq:chi-psi-Psi-FH}
& & \chi = {\rm f} + i \, \left(- s \, {\rm h} - {\rm q}\right), \quad \psi = - \, {\rm h}, \quad \Psi^\alpha ={\rm q}^\alpha, \quad \Psi \equiv \tau_\alpha \Psi^\alpha = {\rm q},
\eea
which means $(\Psi - s \psi) = s\, {\rm h} + {\rm q} = \frac{i}{2}(\chi - \ov\chi)$ is again a real quantity. Also note that the internal structure of ${\rm f}, {\rm h}$ and ${\rm q}^\alpha$ combinations simplifies accordingly, as compared to (\ref{eq:chi-psi-Psi}), in the sense that $P$-flux as well as all prime-fluxes are absent. Now the full scalar potential takes the following form,
\bea
\label{eq:FHQ}
& & \hskip-1.5cm V_{\rm FHQ} = V_{\rm FHQ}^{(1)} + V_{\rm FHQ}^{(2)},
\eea
where
\bea
\label{eq:FHQa}
& & \hskip-1cm V_{\rm FHQ}^{(1)} = e^K \biggl[\frac{8\,{\cal U} - \gamma}{2\,{\cal U} - \gamma} \left((\widetilde{\rm f}_0)^2 + \, (s\,\widetilde{\rm h}_0 + \widetilde{\rm q}_0)^2\right) \\
& & + \left(K^{i\ov{j}} -\frac{3\gamma u^i \, u^j}{2{\cal U} - \gamma}\right)  \left(\widetilde{\rm f}_i \, \widetilde{\rm f}_j + \, (s\,\widetilde{\rm h}_i + \widetilde{\rm q}_i) \, (s\, \widetilde{\rm h}_j + \widetilde{\rm q}_j) \right) \nonumber\\
& & + \left((4 {\cal U} + \gamma)^2\,K_{i\ov{j}}+\frac{3\gamma \sigma_i\, \sigma_j}{2{\cal U} - \gamma} \right) \left({\rm f}^i {\rm f}^j + (s\, {\rm h}^i +  {\rm q}^i)\, (s\, {\rm h}^j +  {\rm q}^j)\right) \nonumber\\
& & + \frac{8\,{\cal U}^2 + \gamma\, {\cal U} + 2\, \gamma^2}{2} \left(({\rm f}^0{\rm )}^2 + (s\, {\rm h}^0 + {\rm q}^0)^2 \right) \nonumber\\
& & + \frac{6\, \gamma\, \sigma_i}{2\,{\cal U} - \gamma} \left(\widetilde{\rm f}_0\, {\rm f}^i + \, (s\, \widetilde{\rm h}_0 +  \widetilde{\rm q}_0) \, (s\, {\rm h}^i +  {\rm q}^i)\right) - 3 \, \gamma \, u^i \, \left(\widetilde{\rm f}_i\, {\rm f}^0 +\, (s\, \widetilde{\rm h}_i +\, \widetilde{\rm q}_i ) (s\, {\rm h}^0 + {\rm q}^0)\right) \nonumber\\
& & -8\, s\, \left((\widetilde{\rm h}_0 - {\rm h}^{i} \sigma_i)(\widetilde{\rm q}_0 - {\rm q}^{j} \sigma_j) + (\widetilde{\rm h}_i u^i - ({\cal U} + \gamma){\rm h}^{0})(\widetilde{\rm q}_j u^j - ({\cal U} + \gamma){\rm q}^{0}) \right)\nonumber\\
& & 
-4\,{\cal V} \kappa_{\alpha\beta}\left((\widetilde{\rm q}^\alpha_0 - {\rm q}^{\alpha i} \sigma_i)(\widetilde{\rm q}^\beta_0 - {\rm q}^{\beta j} \sigma_j) + (\widetilde{\rm q}^\alpha_i u^i - ({\cal U} + \gamma){\rm q}^{\alpha 0})(\widetilde{\rm q}^\beta_j u^j - ({\cal U} + \gamma){\rm q}^{\beta 0}) \right)\biggr], \nonumber\\
& & \hskip-1cm V_{\rm FHQ}^{(2)} = e^K \cdot (2) \, (4\,{\cal U}+ \gamma) \biggl[{\rm f}^0\,(s\, \widetilde{\rm h}_0 + \widetilde{\rm q}_0) - {\rm f}^i\,(s\, \widetilde{\rm h}_i + \widetilde{\rm q}_i) +(s\, {\rm h}^i + {\rm q}^i)\,\widetilde{\rm f}_i - (s\, {\rm h}^0 + {\rm q}^0)\,\widetilde{\rm f}_0 \biggr] \nonumber\\
& &  = \frac{1}{2\,{\cal V}^2} \left({\rm f}^0\,{\rm h}_0 - {\rm f}^i\,{\rm h}_i +{\rm h}^i\,{\rm f}_i - {\rm h}^0\,{\rm f}_0 \right) + \frac{1}{2\,s\,{\cal V}^2} \left({\rm f}^0\,{\rm q}_0 - {\rm f}^i\,{\rm q}_i +{\rm q}^i\,{\rm f}_i - {\rm q}^0\,{\rm f}_0 \right), \nonumber
\eea
while we recall that $K_{\alpha\ov\beta} =\,4\, \tau_\alpha\tau_\beta -4\,{\cal V} \kappa_{\alpha\beta}$, and the fact that fluxes with lower indices, namely $\widetilde{\rm f}_0, \widetilde{\rm f}_i, \widetilde{\rm h}_0, \widetilde{\rm h}_i, \widetilde{\rm q}_0$ and $\widetilde{\rm q}_i$, receive the rational shifts as mentioned in Eq.~(\ref{eq:rational-shift}). Also, the piece $V_{\rm FHQ}^{(2)}$ corresponds to the tadpole contributions arising from the D3/D7-brane tadpoles appearing as $(F \wedge H)$ and $(F \wedge Q)$ Chern-Simons action from a higher dimensional theory \cite{Blumenhagen:2013hva}. As we discussed previously, there is a crucial cancellation in the second piece $V_{\rm I}^{(2)}$ of the master-formula (\ref{eq:master1}) as,
\bea
& & \hskip-1cm V_{\rm I}^{(2)} 
= e^K \biggl[\left(K_{\alpha\ov\beta} -\,4\, \tau_\alpha\tau_\beta\right)\, \Phi[{\rm q}^\alpha] \cdot \ov\Phi[{\rm q}^\beta] + 2\,i \left(\Phi[s\, {\rm h} +  {\rm q}] \cdot \ov\Phi[{\rm f}] - \Phi[{\rm f}]\cdot \ov\Phi[s\,{\rm h}+\,{\rm q}] \right)\biggr],
\eea
where the pieces with $s^2$ dependence result in a precise cancellation. The master formula in (\ref{eq:FHQ}) generalizes the results of \cite{Shukla:2019wfo} which were reported for $\gamma= 0$ case with the presence of $\{F, H, Q\}$ fluxes. This corresponds to a form given as below which matches with that of \cite{Shukla:2019wfo},
\bea
\label{eq:FHQ-simp}
& & \hskip-1cm V_{\rm FHQ}^{\gamma = 0} = \frac{1}{4\,s\, {\cal U}\,{\cal V}^2} \biggl[\left((\widetilde{\rm f}_0)^2 + \, (s\,\widetilde{\rm h}_0 + \widetilde{\rm q}_0)^2\right) + \frac{1}{4}\,K^{i\ov{j}} \left(\widetilde{\rm f}_i \, \widetilde{\rm f}_j + \, (s\,\widetilde{\rm h}_i + \widetilde{\rm q}_i) \, (s\, \widetilde{\rm h}_j + \widetilde{\rm q}_j) \right) \\
& & + \, 4 \,{\cal U}^2\,K_{i\ov{j}}+\left({\rm f}^i {\rm f}^j + (s\, {\rm h}^i +  {\rm q}^i)\, (s\, {\rm h}^j +  {\rm q}^j)\right) + {\cal U}^2\, \left(({\rm f}^0{\rm )}^2 + (s\, {\rm h}^0 + {\rm q}^0)^2 \right) \nonumber\\
& & -\,2\, s\, \left((\widetilde{\rm h}_0 - {\rm h}^{i} \sigma_i)(\widetilde{\rm q}_0 - {\rm q}^{j} \sigma_j) + (\widetilde{\rm h}_i u^i - {\cal U}{\rm h}^{0})(\widetilde{\rm q}_j u^j - {\cal U}{\rm q}^{0}) \right)\nonumber\\
& & 
-\,{\cal V} \kappa_{\alpha\beta}\left((\widetilde{\rm q}^\alpha_0 - {\rm q}^{\alpha i} \sigma_i)(\widetilde{\rm q}^\beta_0 - {\rm q}^{\beta j} \sigma_j) + (\widetilde{\rm q}^\alpha_i u^i - {\cal U}{\rm q}^{\alpha 0})(\widetilde{\rm q}^\beta_j u^j - {\cal U}{\rm q}^{\beta 0}) \right)\biggr] \nonumber\\
& & + \, \frac{1}{2\,s\,{\cal V}^2} \biggl[s\,\left({\rm f}^0\,{\rm h}_0 - {\rm f}^i\,{\rm h}_i -{\rm h}^i\,{\rm f}_i - {\rm h}^0\,{\rm f}_0 \right) + \left({\rm f}^0\,{\rm q}_0 - {\rm f}^i\,{\rm q}_i -{\rm q}^i\,{\rm f}_i - {\rm q}^0\,{\rm f}_0 \right)\biggr], \nonumber
\eea
where $K_0^{i\ov{j}}$ and $K^0_{i\ov{j}}$ denote the respective metrics for $\gamma = 0$, i.e.
\bea
\label{eq:Kmetric-gamma-zero}
& & K_0^{i\ov{j}} = 2u^i u^j- 4\, {\cal U} \, \ell^{ij}, \qquad  K^0_{i\ov{j}} = \frac{4\, \sigma_i \, \sigma_j - 4\, {\cal U} \, \ell_{ij}}{16\, {\cal U}^2}.
\eea
The above form of the scalar potential induced via the presence of $F, H$ and $Q$ fluxes given in (\ref{eq:FHQ-simp}) turns out to be even more compact and simpler than the one presented in \cite{Shukla:2019wfo}.


\subsection{Model C : $(F, H, Q, P)$ fluxes only}
As a third scenario, now we will show that the master-formula (\ref{eq:master1}) recovers the results of \cite{Gao:2015nra,Shukla:2015rua,Shukla:2016hyy} as a particular case of having no prime-fluxes, i.e. including only two of the four S-dual pairs of fluxes, namely $(F, H)$ and $(Q, P)$. For that purpose, we consider the triplets $\{\chi, \psi, \Psi^\alpha\}$ as defined in (\ref{eq:chi-psi-Psi}) which are simplified as
\bea
\label{eq:chi-psi-Psi-FHQP}
& & \hskip-1cm \chi = \left({\rm f} -s \, {\rm p}\right) + i \, \left(- s \, {\rm h} - {\rm q}\right), \quad \psi = - \, {\rm h} + i\, {\rm p}, \quad \Psi^\alpha ={\rm q}^\alpha + i\, (-s\, {\rm p}^\alpha),
\eea
which means $\Psi \equiv \tau_\alpha \Psi^\alpha = ({\rm q} - i\, s\, {\rm p})$ and $(\Psi - s \psi) = (s\, {\rm h} + {\rm q} - 2\, i\, s\, {\rm p})$ is no more a real quantity like previous cases. Also, recall that the internal structure of ${\rm f}, {\rm h}, {\rm q}^\alpha$ and ${\rm p}^\alpha$ combinations simplifies accordingly, as compared to (\ref{eq:chi-psi-Psi}), in the sense that all the prime-fluxes are still absent. Now the full scalar potential takes the following form,
\bea
\label{eq:FHQP}
& & \hskip-1.5cm V_{\rm FHQP} = V_{\rm FHQP}^{(1)} + V_{\rm FHQP}^{(2)},
\eea
where
\bea
\label{eq:FHQPa}
& & \hskip-0.75cm V_{\rm FHQP}^{(1)} = e^K \biggl[\frac{8\,{\cal U} - \gamma}{2\,{\cal U} - \gamma} \left((\widetilde{\rm f}_0 - s\,\widetilde{\rm p}_0)^2 + \, (s\,\widetilde{\rm h}_0 + \widetilde{\rm q}_0)^2\right) \\
& & + \left(K^{i\ov{j}} -\frac{3\gamma u^i \, u^j}{2{\cal U} - \gamma}\right)  \left((\widetilde{\rm f}_i - s\,\widetilde{\rm p}_i)\, (\widetilde{\rm f}_j -s\,\widetilde{\rm p}_j) + \, (s\,\widetilde{\rm h}_i + \widetilde{\rm q}_i) \, (s\, \widetilde{\rm h}_j + \widetilde{\rm q}_j) \right) \nonumber\\
& & + \left((4 {\cal U} + \gamma)^2\,K_{i\ov{j}}+\frac{3\gamma \sigma_i\, \sigma_j}{2{\cal U} - \gamma} \right) \left(({\rm f}^i-s\, {\rm p}^i) ({\rm f}^j-s\, {\rm h}^j) + (s\, {\rm h}^i +  {\rm q}^i)\, (s\, {\rm h}^j +  {\rm q}^j)\right) \nonumber\\
& & + \frac{8\,{\cal U}^2 + \gamma\, {\cal U} + 2\, \gamma^2}{2} \left(({\rm f}^0 -s\, {\rm p}^0)^2 + (s\, {\rm h}^0 + {\rm q}^0)^2 \right) \nonumber\\
& & + \frac{6\, \gamma\, \sigma_i}{2\,{\cal U} - \gamma} \left((\widetilde{\rm f}_0 - s\,\widetilde{\rm p}_0)\, ({\rm f}^i-s\, {\rm p}^i) + \, (s\, \widetilde{\rm h}_0 +  \widetilde{\rm q}_0) \, (s\, {\rm h}^i +  {\rm q}^i)\right) \nonumber\\
& & - 3 \, \gamma \, u^i \, \left((\widetilde{\rm f}_i - s\,\widetilde{\rm p}_i)\, ({\rm f}^0 -s\, {\rm p}^0) +\, (s\, \widetilde{\rm h}_i +\, \widetilde{\rm q}_i ) (s\, {\rm h}^0 + {\rm q}^0)\right) \nonumber\\
& & -8\, s\, \left((\widetilde{\rm h}_0 - {\rm h}^{i} \sigma_i)(\widetilde{\rm q}_0 - {\rm q}^{j} \sigma_j) + (\widetilde{\rm h}_i u^i - ({\cal U} + \gamma){\rm h}^{0})(\widetilde{\rm q}_j u^j - ({\cal U} + \gamma){\rm q}^{0}) \right) \nonumber\\
& & +8\, s\, \left((\widetilde{\rm f}_0 - {\rm f}^{i} \sigma_i)(\widetilde{\rm p}_0 - {\rm p}^{j} \sigma_j) + (\widetilde{\rm f}_i u^i - ({\cal U} + \gamma){\rm f}^{0})(\widetilde{\rm p}_j u^j - ({\cal U} + \gamma){\rm p}^{0}) \right) \nonumber\\
& & 
-4\,{\cal V} \kappa_{\alpha\beta}\left((\widetilde{\rm q}^\alpha_0 - {\rm q}^{\alpha i} \sigma_i)(\widetilde{\rm q}^\beta_0 - {\rm q}^{\beta j} \sigma_j) + (\widetilde{\rm q}^\alpha_i u^i - ({\cal U} + \gamma){\rm q}^{\alpha 0})(\widetilde{\rm q}^\beta_j u^j - ({\cal U} + \gamma){\rm q}^{\beta 0}) \right) \nonumber\\
& & -4\,s^2\, {\cal V} \kappa_{\alpha\beta}\left((\widetilde{\rm p}^\alpha_0 - {\rm p}^{\alpha i} \sigma_i)(\widetilde{\rm p}^\beta_0 - {\rm p}^{\beta j} \sigma_j) + (\widetilde{\rm p}^\alpha_i u^i - ({\cal U} + \gamma){\rm p}^{\alpha 0})(\widetilde{\rm p}^\beta_j u^j - ({\cal U} + \gamma){\rm p}^{\beta 0}) \right) \nonumber\\
& & -4\,{\cal V} \kappa_{\alpha\beta}\,(2\,s) \left((\widetilde{\rm p}^\alpha_0 - {\rm p}^{\alpha i} \sigma_i)(\widetilde{\rm q}^\beta_j u^j - ({\cal U} + \gamma){\rm q}^{\beta 0}) -(\widetilde{\rm q}^\alpha_0 - {\rm q}^{\alpha i} \sigma_i)(\widetilde{\rm p}^\beta_j u^j - ({\cal U} + \gamma){\rm p}^{\beta 0}) \right)\biggr], \nonumber
\eea
\bea
& & \hskip-0.5cm V_{\rm FHQP}^{(2)} = e^K \cdot (2) \, (4\,{\cal U}+ \gamma) \biggl[({\rm f}^0 -s\, {\rm p}^0)\,(s\, \widetilde{\rm h}_0 + \widetilde{\rm q}_0) - ({\rm f}^i-s\, {\rm p}^i)\,(s\, \widetilde{\rm h}_i + \widetilde{\rm q}_i) \nonumber\\
& & \hskip0.75cm +(s\, {\rm h}^i + {\rm q}^i)\,(\widetilde{\rm f}_i - s\,\widetilde{\rm p}_i) - (s\, {\rm h}^0 + {\rm q}^0)\,(\widetilde{\rm f}_0 - s\,\widetilde{\rm p}_0) \biggr] \nonumber\\
& &  \hskip0.75cm = \frac{1}{2\,{\cal V}^2} \left({\rm f}^0\,{\rm h}_0 - {\rm f}^i\,{\rm h}_i +{\rm h}^i\,{\rm f}_i - {\rm h}^0\,{\rm f}_0 \right) + \frac{1}{2\,s\,{\cal V}^2} \left({\rm f}^0\,{\rm q}_0 - {\rm f}^i\,{\rm q}_i +{\rm q}^i\,{\rm f}_i - {\rm q}^0\,{\rm f}_0 \right)\nonumber\\
& & \hskip0.75cm + \frac{1}{2\,{\cal V}^2} \left({\rm q}^0\,{\rm p}_0 - {\rm q}^i\,{\rm p}_i +{\rm p}^i\,{\rm q}_i - {\rm p}^0\,{\rm q}_0 \right) + \frac{s}{2\,{\cal V}^2} \left({\rm h}^0\,{\rm p}_0 - {\rm h}^i\,{\rm p}_i +{\rm p}^i\,{\rm h}_i - {\rm p}^0\,{\rm h}_0 \right), \nonumber
\eea
while we recall that $K_{\alpha\ov\beta} =\,4\, \tau_\alpha\tau_\beta -4\,{\cal V} \kappa_{\alpha\beta}$, and the fact that fluxes with lower indices, namely $\widetilde{\rm f}_0, \widetilde{\rm f}_i, \widetilde{\rm h}_0, \widetilde{\rm h}_i, \widetilde{\rm q}_0$ and $\widetilde{\rm q}_i$, receive the rational shifts as mentioned in Eq.~(\ref{eq:rational-shift}). Also, the piece $V_{\rm FHQ}^{(2)}$ corresponds to the tadpole contributions arising from the D3/D7-brane tadpoles appearing as $(F \wedge H), (F \wedge Q), (P \wedge Q)$ and $(H \wedge P)$ Chern-Simons action from a higher dimensional theory \cite{Gao:2015nra, Shukla:2016hyy}. Finally, let us mention that in order to manifestly see the equivalence of (\ref{eq:FHQP}) with the formulation presented in \cite{Shukla:2016hyy} one needs to use the shifted $\chi_0$ and $\chi_i$ as defined in (\ref{eq:shift-chi0-chii}).


\section{Scalar potentials with all the fluxes}
\label{sec_Demonstration}


In order to demonstrate the use of axionic fluxes for rewriting the scalar potential in a compact form, we consider a couple of explicit examples. First we discuss a toroidal type IIB model based on $\mathbb T^6/(\mathbb Z_2 \times \mathbb Z_2)$ orientifold compactification, and as a second example we consider the orientifolds of CY threefolds with $h^{1,1} = 1$.


\subsection{Type IIB model using a $\mathbb T^6/(\mathbb Z_2 \times \mathbb Z_2)$ orientifold}
\label{sec_modelD}


Considering the untwisted sector with the standard involution for the non-geometric type IIB setup (e.g. see \cite{Blumenhagen:2013hva, Gao:2015nra, Shukla:2015hpa, Shukla:2016hyy} for details), we can start with the following topological data,
\bea
\label{eq:data-torus}
& & \hskip-1.5cm h^{1,1}_+ = 3, \qquad h^{1,1}_- = 0, \qquad h^{2,1}_+ = 0\,, \qquad  h^{2,1}_- = 3, \\
& & \hskip-1.5cm \kappa_{123} = 1, \qquad \ell_{123} = 1, \qquad a_{ij} = 0, \qquad b_i = 0, \qquad \gamma = 0. \nonumber
\eea
The Hodge numbers show that there would be three $T_\alpha$ moduli and three $U^i$ moduli along with the universal axio-dilaton $S$ in this setup while there are no odd-moduli $G^a$ present in this setup as the odd $(1,1)$-cohomology is trivial. In addition, there are no $D$-terms generated in the scalar potential as the even $(2,1)$-cohomology is trivial which projects out all the $D$-term fluxes. It also turns out that the conventional geometric flux $\omega$ and the non-geometric $R$ flux do not survive the orientifold projection in this setup.

For the toroidal setup under consideration we have only one non-zero component of the triple intersection tensor $\kappa_{123} = 1$, and it turns out that the quantity $\kappa^{\alpha\beta\gamma}$ defined in (\ref{eq:Inv-lijk}) takes a simple form $\kappa^{\alpha\beta\gamma} = {\kappa_{\alpha\beta\gamma}}/{{\cal V}}$, which means that the volume dependence appears only through the overall volume modulus ${\cal V}$, and not in terms of the 4-cycle or 2-cycle volumes. Subsequently, the non-zero components of the various prime fluxes given in Eq.~(\ref{eq:primefluxIIBc}) simplify to take the following form,
\bea
\label{eq:primefluxIIBd}
& & \hskip-1cm {P^\prime}_{\alpha}{}^{\Lambda} = {\cal V}\,{P}^{\prime\beta\gamma}_\Lambda, \qquad {P^\prime}_{\alpha\, \Lambda} = {\cal V}\, {P}^{\prime\beta\gamma}{}_{\Lambda},\qquad {Q^\prime}_{\alpha}{}^{\Lambda} = {\cal V}\,{Q}^{\prime\beta\gamma}_\Lambda, \qquad {Q^\prime}_{\alpha\, \Lambda} = {\cal V}\, {Q}^{\prime\beta\gamma}{}_{\Lambda},\\
& & \hskip-1cm {H^\prime}^{\Lambda} = {\cal V}{H}^{\prime\alpha\beta\gamma{}\Lambda},\qquad {H^\prime}_{\Lambda} = {\cal V} {H}^{\prime\alpha\beta\gamma}{}_\Lambda, \qquad {F^\prime}^{\Lambda} = {\cal V}{F}^{\prime\alpha\beta\gamma{}\Lambda},\qquad {F^\prime}_{\Lambda} = {\cal V} {F}^{\prime\alpha\beta\gamma}{}_\Lambda\,,\nonumber
\eea
where $\alpha \neq \beta\neq\gamma$ for $\{\alpha, \, \beta, \, \gamma\} \in \{1,2,3\}$ and $\Lambda \in \{0, 1, 2, 3 \}$.
This means that we finally have 8 components for each of the fluxes $F, H, H^\prime$ and $F^\prime$ while there are 24 components for each of the $Q, P, P^\prime$ and $Q^\prime$ fluxes, making a total of 128 fluxes being allowed by the current toroidal orientifold. 

The U-dual completion of the holomorphic flux superpotential results in having 128 flux components in total along with 7 complexified moduli, namely $\bigl\{S, T_1, T_2, T_3, U^1, U^2, U^3\bigr\}$, and for this toroidal model, the generalized flux superpotential boils down to the following form,
\bea
\label{eq:W-toroidal}
& & \hskip-0.5cm W = \biggl[F_0 + \sum_{i=1}^3 F_i\, U^i + \frac{1}{2} \, \sum_{i\neq j \neq k} \,F^i \,U^j U^k - \, U^1 U^2 U^3 \, F^0 \biggr]\\
& & \hskip0.5cm - \, S \, \biggl[H_0 + \sum_{i=1}^3 H_i\, U^i + \frac{1}{2} \, \sum_{i \neq j \neq k} \, H^i \,U^j U^k - U^1 U^2 U^3 \, H^0 \biggr] \nonumber\\
& & \hskip0.15cm + \sum_{\alpha =1}^3\, T_\alpha\, \biggl[{Q}^\alpha{}_0 + \sum_{i=1}^3 {Q}^\alpha{}_i\, U^i + \frac{1}{2} \, \sum_{i \neq j \neq k} \, {Q}^{\alpha \, i} \,U^j U^k - U^1 U^2 U^3 \, {Q}^{\alpha 0} \biggr]\,\nonumber\\
& & - \, S \sum_{\alpha =1}^3 \, T_\alpha\, \biggl[{P}^\alpha{}_0 + \sum_{i=1}^3 {P}^\alpha{}_i\, U^i + \frac{1}{2} \, \sum_{i \neq j \neq k} \, {P}^{\alpha \, i} \,U^j U^k - U^1 U^2 U^3 \, {P}^{\alpha 0} \biggr]\, \nonumber\\
& & + \frac{1}{2}\, \sum_{\substack{\alpha, \beta =1 \\ \alpha \neq \beta}}^{3} \biggl[T_\alpha T_\beta \biggl\{{P}^{\prime \alpha\beta}{}_{0} + \sum_{i=1}^3 {P}^{\prime \alpha\beta}{}_{i}\, U^i + \frac{1}{2} \, \sum_{i \neq j \neq k} \, {P}^{\prime \alpha\beta i} \, U^j U^k - U^1 U^2 U^3 \, {P}^{\prime \alpha\beta 0} \biggr\}\biggr]\, \nonumber\\
& & - \frac{S}{2}\, \sum_{\substack{\alpha, \beta =1 \\ \alpha \neq \beta}}^{3} \biggl[T_\alpha T_\beta \biggl\{{Q}^{\prime \alpha\beta}{}_{0} + \sum_{i=1}^3 {Q}^{\prime \alpha\beta}{}_{i}\, U^i + \frac{1}{2} \, \sum_{i \neq j \neq k} \, {Q}^{\prime \alpha\beta i}\, U^j U^k - U^1 U^2 U^3\, {Q}^{\prime \alpha\beta 0} \biggr\} \biggr]\, \nonumber\\
& & + \, T_1\, T_2 \, T_3 \biggl[{H}^{\prime123}{}_0 + \sum_{i=1}^3 {H}^{\prime123}{}_i\, U^i + \frac{1}{2} \, \sum_{i \neq j \neq k} \,  {H}^{{\prime123} \, i} \, U^j U^k - U^1 U^2 U^3 \, {H}^{{\prime123} 0} \biggr] \nonumber\\
& & - S\, T_1\, T_2 \, T_3 \biggl[{F}^{\prime123}{}_0 + \sum_{i=1}^3 {F}^{\prime123}{}_i\, U^i + \frac{1}{2} \, \sum_{i \neq j \neq k} \, {F}^{{\prime123} \, i} \,U^j U^k - U^1 U^2 U^3 \,  {F}^{{\prime123} 0}\biggr] \,. \nonumber
\eea
The K\"ahler potential takes the following simple form,
\bea
\label{eq:K-toroidal}
& & \hskip-1cm K = -\ln(-i(S-\ov{S})) - \sum_{i=1}^3 \ln(i(U^i-\ov{U}^i)) - \sum_{\alpha=1}^3 \ln\left(\frac{i(T_\alpha-\ov{T}_\alpha)}{2}\right).
\eea
This flux superpotential (\ref{eq:W-toroidal}) generically results in a total of 76276 terms at the level of the four-dimensional ${\cal N} = 1$ effective scalar potential \cite{Leontaris:2023lfc,Leontaris:2023mmm}. However, it has been shown in \cite{Leontaris:2023lfc,Leontaris:2023mmm} that using certain combinations of RR axion and fluxes one can absorb all the RR axions to encode their explicit presence only through the axionic fluxes. This subsequently reduces the number of terms in the scalar potential to 10888 \cite{Leontaris:2023lfc,Leontaris:2023mmm}. This is an outcome of what we call the step-1 of the axionic fluxes as given in (\ref{eq:AxionicFlux})-(\ref{eq:AxionicFlux1}). 

In the current work we have also absorbed the CS-axions via considering a more general axionic flux polynomials such that all the axionic presence is encoded in these polynomials, leading to a flux bilinear form of the scalar potential, similar to those of the type IIA and F-theory setups presented in \cite{Bielleman:2015ina,Escobar:2018tiu,Herraez:2018vae,Marchesano:2020uqz, Marchesano:2021gyv}. We find that after using the {\it new axionic fluxes} in Eqs.~(\ref{eq:newOrbits1})-(\ref{eq:newOrbits2}), the scalar potential reduces to 2816 terms which can be used to study the saxionic dynamics pretending that axions are absent, as being absorbed in their respective axionic polynomials! The axion minimization can be separately performed by using the fact that derivatives of one axionic polynomial leads to another, e.g. axionic polynomials in (\ref{eq:AxionicFlux1}) can be derived as derivatives of those given in (\ref{eq:AxionicFlux}). This property can subsequently simplify the on-shell conditions. 

Thus the systematic reformulation and reduction of the size of scalar potential from 76276 terms to 2816 terms creates the hope for any attempts for generic moduli stabilization which a priori does not look a feasible and pragmatic task. In order to appreciate the counting of terms in various (particular) scenarios regarding the systematic reduction of the scalar potential at various stages we present the relevant taxonomy details in Table \ref{tab_term-counting1}.

\begin{table}[h!]
\begin{center}
\begin{tabular}{|c||c|c||c|c||c|c|}
\hline
& &&&&&\\
& Fluxes & $\#$(V) & Axionic-fluxes & $\#$(V) & Axionic-fluxes & $\#$(V) \\
& & & (of step-1) & \cite{Leontaris:2023lfc,Leontaris:2023mmm} & (of step-2) & \\
\hline
1 & $F$  & 76 & ${\mathbb F}$ & 76  & $f$ & 8  \\
\hline
2 & $F, H$ & 361 & ${\mathbb F}, {\mathbb H}$ & 160  & $f, \, h$ & 24  \\
\hline
3 & $F, H, Q$ & 2422 \cite{Blumenhagen:2013hva} & ${\mathbb F}, {\mathbb H}, {\mathbb Q}$ & 772  & $f, \, h, \, q$ & 216  \\
\hline
4 & $F, H, Q, P$ & 9661 \cite{Gao:2015nra,Shukla:2016hyy} & ${\mathbb F}, {\mathbb H}, {\mathbb Q}, {\mathbb P}$ & 2356  & $f, \, h, \, q, \, p$ & 624  \\
\hline
5 & $F, H, Q, P, P'$  & 23314 & ${\mathbb F}, {\mathbb H}, {\mathbb Q}, {\mathbb P}, {\mathbb P}'$ & 4855 & $f, \, h, \, q, \, p, \, p'$ &  1272 \\
\hline
6 & $F, H, Q, P,$  & 50185 & ${\mathbb F}, {\mathbb H}, {\mathbb Q}, {\mathbb P},$ & 8326  & $f, \, h, \, q, \, p,$ &  2136 \\
& $P', Q'$ & &  ${\mathbb P}', {\mathbb Q}'$ & &  $p',\, q'$ & \\
\hline
7 & $F, H, Q, P,$  & 60750 & ${\mathbb F}, {\mathbb H}, {\mathbb Q}, {\mathbb P},$ & 9603  & $f, \, h, \, q, \, p,$ &  2472 \\
& $P', Q', H'$ & &  ${\mathbb P}', {\mathbb Q}', {\mathbb H}'$ & &  $p', \, q', \, h'$ &\\
\hline
8 & $F, H, Q, P, $  & 76276 & ${\mathbb F}, {\mathbb H}, {\mathbb Q}, {\mathbb P},$ & 10888  & $f, \, h, \, q, \, p,$ & 2816  \\
& $P', Q', H', F'$ & \cite{Leontaris:2023lfc,Leontaris:2023mmm} & ${\mathbb P}', {\mathbb Q}', {\mathbb H}', {\mathbb F}'$ & & $p', \, q', \, h', \, f'$ &\\
\hline
\end{tabular}
\end{center}
\caption{Counting of scalar potential terms for a set of fluxes being turned-on at a time. The first and second columns representing the counting have been fully observed in \cite{Leontaris:2023lfc,Leontaris:2023mmm} while the last column is the new result obtained by using {\it new axionic fluxes} defined in (\ref{eq:newOrbits1})-(\ref{eq:newOrbits2}).}
\label{tab_term-counting1}
\end{table}


\noindent
In the large complex structure regime $\gamma \ll {\cal U}$, and one may simply ignore the perturbative effects on the mirror CY threefold, and the generic scalar potential (\ref{eq:master1}) takes the form,
\bea
\label{eq:V_gen5}
& & \hskip-0.5cm V = e^K \biggl[4\,\chi_0 \ov{\chi}_0 + \chi_i \, K^{i\ov{j}} \, \ov{\chi}_j + 16\, {\cal U}^2\, \chi^i \, K_{i\ov{j}} \, \ov{\chi}^j + 4\,{\cal U}^2\,\chi^0\, \ov{\chi}^0 + K_{\alpha\ov\beta}\, \Phi[\Psi^\alpha] \cdot \ov\Phi[\ov\Psi^\beta] \\
& & \hskip-0.5cm + 4\, s^2\, \Phi[\psi] \cdot \ov\Phi[\ov\psi] + \, 2\,i\, \left(\Phi[\Psi -s \, \psi] \cdot \ov\Phi[\ov\chi] - \Phi[\chi]\cdot \ov\Phi[\ov\Psi -s \, \ov\psi] \right) - 4\,i\,{\cal U} \left(\chi_i \ov\chi^i - \ov\chi_i \chi^i \right)  \nonumber\\
& & \hskip-0.5cm + 2\, i\,{\cal U} (\chi_0 \ov\chi^0 - \ov\chi_0 \chi^0)   - 2\, i\, u^i (\chi_i \ov\chi_0 - \ov\chi_i \chi_0) + 2\, i\,{\cal U} \sigma_i (\chi^i \ov\chi^0 - \ov\chi^i \chi^0) + 2\,i \, u^i\, \sigma_j (\chi_i \ov\chi^j - \ov\chi_i \chi^j) \biggr], \nonumber
\eea
where the $\Phi[y]$ functional is simplified to the form $\Phi[y] = y_0 - i\,  u^i \, y_i\, - y^i \,\sigma_i + i\,{\cal U}\, y^0$. For simple models where $\gamma = 0$, say for example the current toroidal setups with untwisted sector, this large complex structure limit is directly applicable, and we find that master formula (\ref{eq:V_gen5}) precisely reproduces the scalar potential with 76276 terms in usual fluxes, and 2816 terms while using {\it new} axionic fluxes as mentioned in Table \ref{tab_term-counting1}.


\subsection{STU-like model in the isotropic limit}


Despite having a drastic reduction in the number of scalar potential terms from 76276 to 2816, one can argue that even 2816 terms obtained after using the {\it new} axionic fluxes are still hard to handle for making any phenomenological use of the potential. In this regard, we simplify the setup by imposing the isotropic limit defined by the following identifications among the various respective moduli and the fluxes,
\bea
\label{eq:IIBiso}
& & {U}^1 = {U}^2 = {U}^3 = {U}, \quad {T}_1 = {T}_2= {T}_3 = {T}, \quad {\tau}_1 = {\tau}_1 = {\tau}_1 \equiv {\tau},\\
& & \rho_1 = \rho_2 = \rho_3 \equiv \rho\,, \quad {u}^1 = {u}^2 = {u}^3 = u, \quad {v}^1 = {v}^2 = {v}^3 \equiv v,  \nonumber\\
& & \nonumber\\
& F/H: & F_1 = F_2 = F_3, \quad F^1 = F^2 = F^3, \quad {H}_1 = {H}_2 = {H}_3, \quad {H}^1 = {H}^2 = {H}^3, \nonumber\\
& Q: & {Q}^1{}_0 ={Q}^2{}_0 = {Q}^3{}_0, \quad {Q}^{1}{}_{1} = {Q}^{2}{}_{2} = {Q}^{3}{}_{3}, \quad {Q}^{1}{}_{2} = {Q}^{1}{}_{3} = {Q}^{2}{}_{1} = {Q}^{2}{}_{3} = {Q}^{3}{}_{1} = {Q}^{3}{}_{2}, \nonumber\\
& & {Q}^{10} ={Q}^{20} = {Q}^{30}, \quad {Q}^{11} = {Q}^{22} = {Q}^{33}, \quad {Q}^{12} = {Q}^{13} = {Q}^{21} = {Q}^{23} = {Q}^{31} = {Q}^{32}\,, \nonumber\\
& P: & {P}^1{}_0 ={P}^2{}_0 = {P}^3{}_0, \quad {P}^{1}{}_{1} = {P}^{2}{}_{2} = {P}^{3}{}_{3}, \quad {P}^{1}{}_{2} = {P}^{1}{}_{3} = {P}^{2}{}_{1} = {P}^{2}{}_{3} = {P}^{3}{}_{1} = {P}^{3}{}_{2}, \nonumber\\
& & {P}^{10} ={P}^{20} = {P}^{30}, \quad {P}^{11} = {P}^{22} = {P}^{33}, \quad {P}^{12} = {P}^{13} = {P}^{21} = {P}^{23} = {P}^{31} = {P}^{32}\,, \nonumber\\
& P': & {P'}^{12}{}_0 ={P'}^{23}{}_0 = {P'}^{31}{}_0, \quad {P'}^{12}{}_{3} = {P'}^{23}{}_{1} = {P'}^{31}{}_{2}, \quad {P'}^{120} ={P'}^{230} = {P'}^{310}, \nonumber\\
& & {P'}^{123} = {P'}^{231} = {P'}^{312}, \quad {P'}^{12}{}_{1} = {P'}^{12}{}_{2} = {P'}^{23}{}_{2} = {P'}^{23}{}_{3} = {P'}^{31}{}_{1} = {P'}^{31}{}_{3}, \nonumber\\
& & {P'}^{121} = {P'}^{122} = {P'}^{232} = {P'}^{233} = {P'}^{311} = {P'}^{313}\,, \nonumber\\
& Q': & {Q'}^{12}{}_0 ={Q'}^{23}{}_0 = {Q'}^{31}{}_0, \quad {Q'}^{12}{}_{3} = {Q'}^{23}{}_{1} = {Q'}^{31}{}_{2}, \quad {Q'}^{120} ={Q'}^{230} = {Q'}^{310},\nonumber\\
& & {Q'}^{123} = {Q'}^{231} = {Q'}^{312}, \quad {Q'}^{12}{}_{1} = {Q'}^{12}{}_{2} = {Q'}^{23}{}_{2} = {Q'}^{23}{}_{3} = {Q'}^{31}{}_{1} = {Q'}^{31}{}_{3}, \nonumber\\
& & {Q'}^{121} = {Q'}^{122} = {Q'}^{232} = {Q'}^{233} = {Q'}^{311} = {Q'}^{313}\,, \nonumber\\
& H': & {H'}^{123}_1 = {H'}^{123}_2 = {H'}^{123}_3, \quad {H'}^{1231} = {H'}^{1232} = {H'}^{1233}, \nonumber\\
& F': & {F'}^{123}_1 = {F'}^{123}_2 = {F'}^{123}_3, \quad {F'}^{1231} = {F'}^{1232} = {F'}^{1233}. \nonumber
\eea
Subsequently, taking the isotropic limit implies that one has a total of six moduli/axions, namely $\{\tau,  \rho,  u,  v,  s,  c_0\}$ and 40 flux components given as
\bea
& & \hskip-0.5cm F_0, \quad F_1, \quad F^1, \quad F_0, \quad {H}_0, \quad {H}_1, \quad {H}^1, \quad {H}^0, \\
& & \hskip-0.5cm {Q}^1{}_0, \quad {Q}^{1}{}_{1}, \quad {Q}^{1}{}_{2}, \quad {Q}^1{}^0, \quad {Q}^{11}, \quad {Q}^{12}, \quad {P}^1{}_0, \quad {P}^{1}{}_{1}, \quad {P}^{1}{}_{2}, \quad {P}^1{}^0, \quad {P}^{11}, \quad {P}^{12}\,, \nonumber\\
& & \hskip-0.5cm {P'}^{12}{}_0, \, {P'}^{12}{}_{3}, \, {P'}^{12}{}_{1}, \, {P'}^{120}, \, {P'}^{123}, \, {P'}^{121}, \quad {Q'}^{12}{}_0, \, {Q'}^{12}{}_{3}, \, {Q'}^{12}{}_{1}, \, {Q'}^{120}, \, {Q'}^{123}, \, {Q'}^{121}, \nonumber\\
& & \hskip-0.5cm {H'}^{123}{}_0, \quad {H'}^{123}{}_1, \quad {H'}^{1231}, \quad {H'}^{1230}, \quad F'^{123}{}_0, \quad F'^{123}{}_1, \quad F'^{1231}, \quad F'^{1230}. \nonumber
\eea
Now, the flux superpotential (\ref{eq:W-toroidal}) after imposing the isotropic conditions simplifies as
\bea
\label{eq:WgenIIB-iso}
& & \hskip-1cm W_{\rm iso} = \Bigl({F}_0 + 3\, U \, {F}_1 + 3\, U^2\, F^1 - U^3\, F^0 \Bigr) - \, S \Bigl({H}_0 + 3 \, U\, {H}_1 + 3\, U^2 \, H^1 - U^3 \, H^0 \Bigr) \\
& &  + 3 \, T \Bigl({{Q}}^1{}_0 + \, U \, ({{Q}}^1{}_1 + 2\, {{Q}}^1{}_2) + U^2\, ({Q}^{11} + 2\, {Q}^{12}) - U^3 \, {Q}^{10} \Bigr)\,\nonumber\\
& &  - 3\,S \, T \Bigl({{P}}^1{}_0 + \, U \, ({{P}}^1{}_1 + 2\, {{P}}^1{}_2) + U^2\, ({P}^{11} + 2\, {P}^{12}) - U^3 \, {P}^{10} \Bigr)\,\nonumber\\
& &  + 3 \, T^2\, \Bigl(P'^{12}{}_0 + \, U \, (P'^{12}{}_3+ 2\, P'^{12}{}_1) + \, U^2 \, (P'^{123}+ 2 \, P'^{121}) - U^3\,P'^{120} \Bigr)\,\nonumber\\
& &  - 3 \, S\, T^2\, \Bigl(Q'^{12}{}_0 + \, U \, (Q'^{12}{}_3+ 2\, Q'^{12}{}_1) + \, U^2 \, (Q'^{123}+ 2 \, Q'^{121}) - U^3\,Q'^{120} \Bigr)\,\nonumber\\
& &  + \, T^3\, \Bigl({H'}^{123}_0 + 3\, U \, {H'}^{123}_1 + 3\, U^2\, {H'}^{123}{}^1 - U^3\, {H'}^{123}{}^0 \Bigr)\, \nonumber\\
& &  - S\, T^3\, \Bigl({F'}^{123}_0 + 3\, U \, {F'}^{123}_1 + 3\, U^2\, {F'}^{123}{}^1 - U^3\, {F'}^{123}{}^0 \Bigr)\,. \nonumber
\eea
From the above form of the superpotential (\ref{eq:WgenIIB-iso}), one can observe that it is possible to combine eight pairs of fluxes, e.g. ${{Q}}^1{}_1 + 2\, {{Q}}^1{}_2, \, {Q}^{11} + 2\, {Q}^{12}$ etc. such that only combinations of the two appear in the superpotential, and subsequently reducing the number of `effective' superpotential fluxes from 40 to 32 as we see that
\bea
\label{eq:40to32}
& & {Q}^1{}_1 \to {{Q}}^1{}_1 - 2\, {{Q}}^1{}_2, \qquad {Q}^{11}  \to {Q}^{11} - 2\, {Q}^{12},\\
& & {P}^1{}_1 \to {P}^1{}_1 - 2\, {{P}}^1{}_2, \qquad {P}^{11} \to {P}^{11} - 2\, {P}^{12}, \nonumber\\
& & P'^{12}{}_3 \to P'^{12}{}_3 - 2\, P'^{12}{}_1, \qquad P'^{123} \to P'^{123} - 2 \, P'^{121}, \nonumber\\
& & Q'^{12}{}_3 \to Q'^{12}{}_3 - 2\, Q'^{12}{}_1, \qquad Q'^{123} \to Q'^{123} - 2 \, Q'^{121}. \nonumber
\eea
Subsequently the simplified isotropic superpotential $W_{\rm iso}$ after taking care of the shifts in the respective fluxes as menioned in (\ref{eq:40to32}) reduces to the following form,
\bea
\label{eq:WgenIIB-iso-simp}
& & \hskip-0.5cm W_{\rm iso} = \biggl[\left({F}_0 + 3 U {F}_1 + 3 U^2 F^1 - U^3 F^0 \right) - S \left({H}_0 + 3  U {H}_1 + 3 U^2 H^1 - U^3 H^0 \right)\biggr] \\
& & \hskip0cm + 3 T \biggl[\left({{Q}}^1{}_0 + U {{Q}}^1{}_1 + U^2 {Q}^{11} - U^3 {Q}^{10} \right)- S \left({{P}}^1{}_0 + U {{P}}^1{}_1 + U^2 {P}^{11} - U^3  {P}^{10} \right)\biggr]\nonumber\\
& &  + 3 T^2 \biggl[\left(P'^{12}{}_0 + U P'^{12}{}_3 + U^2 \, P'^{123} - U^3 P'^{120} \right) - S \left(Q'^{12}{}_0 + U Q'^{12}{}_3 + U^2 Q'^{123} - U^3 Q'^{120} \right) \biggr]\nonumber\\
& &  + T^3 \biggl[\left({H'}^{123}_0 + 3 U {H'}^{123}_1 + 3 U^2 {H'}^{123}{}^1 - U^3 {H'}^{123}{}^0 \right) \nonumber\\
& & - S \left({F'}^{123}_0 + 3 U {F'}^{123}_1 + 3 U^2 {F'}^{123}{}^1 - U^3 {F'}^{123}{}^0 \right)\biggr]. \nonumber
\eea
The corresponding simplifications in the taxonomy of the scalar potential pieces are presented in Table \ref{tab_term-counting2}, and in Table \ref{tab_term-counting3} which incorporates the shifts as in (\ref{eq:40to32}). Comparing the respective numbers listed in these two tables we find that the total number of terms in the scalar potential significantly reduces by the shifts, and therefore can be helpful in analyzing the potential for the search of physical flux vacua. However, one would need to readjust the redefinition (\ref{eq:40to32}) in the set of constraints arising from the Bianchi identities. Moreover, the superpotential (\ref{eq:WgenIIB-iso-simp}) corresponds to a typical STU-model with slight rescaling of the fluxes. For a particular case of $\gamma = 0$, such superpotentials may also arise in the type IIB setups using CY orientifold compactifications with $h^{1,1} = 1$ as we will discuss in the next example.

\begin{table}[h!]
\begin{center}
\begin{tabular}{|c||c|c||c|c||c|c|}
\hline
& &&&&&\\
& Fluxes & $\#$(V) & Axionic-fluxes & $\#$(V) & Axionic-fluxes & $\#$(V) \\
& & & (of step-1) & & (of step-2) & \\
\hline
1 & $F$  & 20 & ${\mathbb F}$ & 20  & $f$ & 4  \\
\hline
2 & $F, H$ & 94 & ${\mathbb F}, {\mathbb H}$ & 44  & $f, \, h$ & 12  \\
\hline
3 & $F, H, Q$ & 309 & ${\mathbb F}, {\mathbb H}, {\mathbb Q}$ & 112  & $f, \, h, \, q$ & 40  \\
\hline
4 & $F, H, Q, P$ & 858 & ${\mathbb F}, {\mathbb H}, {\mathbb Q}, {\mathbb P}$ & 228  & $f, \, h, \, q, \, p$ & 86  \\
\hline
5 & $F, H, Q, P, P'$  & 1767 & ${\mathbb F}, {\mathbb H}, {\mathbb Q}, {\mathbb P}, {\mathbb P}'$ & 452 & $f, \, h, \, q, \, p, \, p'$ &  156 \\
\hline
6 & $F, H, Q, P,$  & 3396 & ${\mathbb F}, {\mathbb H}, {\mathbb Q}, {\mathbb P},$ & 724  & $f, \, h, \, q, \, p,$ &  244 \\
& $P', Q'$ & &  ${\mathbb P}', {\mathbb Q}'$ & &  $p',\, q'$ & \\
\hline
7 & $F, H, Q, P,$  & 4456 & ${\mathbb F}, {\mathbb H}, {\mathbb Q}, {\mathbb P},$ & 883  & $f, \, h, \, q, \, p,$ &  302 \\
& $P', Q', H'$ & &  ${\mathbb P}', {\mathbb Q}', {\mathbb H}'$ & &  $p', \, q', \, h'$ &\\
\hline
8 & $F, H, Q, P, $  & 6087 & ${\mathbb F}, {\mathbb H}, {\mathbb Q}, {\mathbb P},$ & 1046  & $f, \, h, \, q, \, p,$ & 364  \\
& $P', Q', H', F'$ & & ${\mathbb P}', {\mathbb Q}', {\mathbb H}', {\mathbb F}'$ & & $p', \, q', \, h', \, f'$ &\\
\hline
\end{tabular}
\end{center}
\caption{Counting of scalar potential terms for isotropic cases listed in the Table \ref{tab_term-counting1}.}
\label{tab_term-counting2}
\end{table}

\begin{table}[h!]
\begin{center}
\begin{tabular}{|c||c|c||c|c||c|c|}
\hline
& &&&&&\\
& Fluxes & $\#$(V) & Axionic-fluxes & $\#$(V) & Axionic-fluxes & $\#$(V) \\
& & & (of step-1) & & (of step-2) & \\
\hline
1 & $F$  & 20 & ${\mathbb F}$ & 20  & $f$ & 4  \\
\hline
2 & $F, H$ & 94 & ${\mathbb F}, {\mathbb H}$ & 44  & $f, \, h$ & 12  \\
\hline
3 & $F, H, Q$ & 220 & ${\mathbb F}, {\mathbb H}, {\mathbb Q}$ & 82  & $f, \, h, \, q$ & 28 \\
\hline
4 & $F, H, Q, P$ & 510 & ${\mathbb F}, {\mathbb H}, {\mathbb Q}, {\mathbb P}$ & 142  & $f, \, h, \, q, \, p$ & 52  \\
\hline
5 & $F, H, Q, P, P'$  & 960 & ${\mathbb F}, {\mathbb H}, {\mathbb Q}, {\mathbb P}, {\mathbb P}'$ & 256 & $f, \, h, \, q, \, p, \, p'$ &  88 \\
\hline
6 & $F, H, Q, P,$  & 1754 & ${\mathbb F}, {\mathbb H}, {\mathbb Q}, {\mathbb P},$ & 392  & $f, \, h, \, q, \, p,$ &  132 \\
& $P', Q'$ & &  ${\mathbb P}', {\mathbb Q}'$ & &  $p',\, q'$ & \\
\hline
7 & $F, H, Q, P,$  & 2546 & ${\mathbb F}, {\mathbb H}, {\mathbb Q}, {\mathbb P},$ & 522  & $f, \, h, \, q, \, p,$ &  176 \\
& $P', Q', H'$ & &  ${\mathbb P}', {\mathbb Q}', {\mathbb H}'$ & &  $p', \, q', \, h'$ &\\
\hline
8 & $F, H, Q, P, $  & 3810 & ${\mathbb F}, {\mathbb H}, {\mathbb Q}, {\mathbb P},$ & 656  & $f, \, h, \, q, \, p,$ & 224  \\
& $P', Q', H', F'$ & & ${\mathbb P}', {\mathbb Q}', {\mathbb H}', {\mathbb F}'$ & & $p', \, q', \, h', \, f'$ &\\
\hline
\end{tabular}
\end{center}
\caption{Counting of scalar potential terms for isotropic cases listed in the Table \ref{tab_term-counting2} after imposing Eq.~(\ref{eq:40to32}), which reduces the `effective' flux parameters in the superpotential from 40 to 32 such that all the eight types of fluxes have 4 parameters each.}
\label{tab_term-counting3}
\end{table}


\subsection{Type IIB models using CY orientifold with $h^{1,1}=1$}
\label{sec_modelE}


In this subsection we will demonstrate the use of our master formula for a generic CY threefold with $h^{1,1}= 1$. The Kreutzer-Skarke CY database \cite{Kreuzer:2000xy} consists of four such CY threefolds with trivial fundamental group, and the corresponding topological data relevant for our purpose are collected in Table \ref{tab_Quintic-data} (e.g. see \cite{Conlon:2016aea}).

\begin{table}[H]
\centering
\begin{tabular}{|c||c|c|c|c|c|}
\hline
\# & Hyp.~ in WCP & $\ell_{111}$ & $a_{ij}$ & ${b_i}$ & $\gamma$ \\
\hline
1 & ${\rm WCP}^4[1,1,1,1,1]$  & 5 & $-\frac{11}{2}$ & $\frac{25}{12}$ & $-\frac{25 \zeta[3]}{\pi^3}$ \\
2 & ${\rm WCP}^4[1,1,1,1,2]$  & 3 & $-\frac{9}{2}$ & $\frac{7}{4}$ & $-\frac{51 \zeta[3]}{2\pi^3}$ \\
3 & ${\rm WCP}^4[1,1,1,1,4]$  & 2 & $-3$ & $\frac{11}{6}$ & $-\frac{37 \zeta[3]}{\pi^3}$ \\
4 & ${\rm WCP}^4[1,1,1,2,5]$  & 1 & $-\frac{1}{2}$ & $\frac{17}{12}$ & $-\frac{36 \zeta[3]}{\pi^3}$ \\
\hline
\end{tabular}
\caption{Topological data of the single parameter CY threefolds leading to STU like models}
\label{tab_Quintic-data}
\end{table}


\noindent
The overall volume of such CY threefolds is generically given as,
\bea
& & \hskip-1cm {\cal V} = \frac{1}{6} \kappa_{111} (t^1)^3 = \frac{1}{3}\sqrt{\frac{2}{\kappa_{111}}} \tau_1^{3/2}, \qquad \tau_1 = \frac{1}{2} \kappa_{111} (t^1)^2.
\eea
For the type IIB model based on orientifolds of CY threefolds with $h^{1,1}_+ = 1 = h^{2,1}_-$, the generalized flux superpotential after the U-dual completion takes the following explicit form,
\bea
\label{eq:W-quintic}
& & \hskip-0.5cm W = \biggl[F_0 + F_1 U^1 +\frac{1}{2} \ell_{111} \left(U^1\right)^2 F^1 -F^0 \left(\frac{1}{6} \ell_{111} \left(U^1\right)^3+i \gamma \right) \biggr]\\
& & -\, S\, \biggl[F_0 + F_1 U^1 +\frac{1}{2} \ell_{111} \left(U^1\right)^2 F^1 -F^0 \left(\frac{1}{6} \ell_{111} \left(U^1\right)^3+i \gamma \right)\biggr] \nonumber\\
& & + \, T_1\, \biggl[Q^1_0 + Q^1_1 U^1 +\frac{1}{2} \ell_{111} \left(U^1\right)^2 Q^{11} -Q^{10} \left(\frac{1}{6} \ell_{111} \left(U^1\right)^3+i \gamma \right) \biggr]\nonumber\\
& & - \, S \, T_1\, \biggl[P^1_0 + P^1_1 U^1 +\frac{1}{2} \ell_{111} \left(U^1\right)^2 P^{11} -P^{10} \left(\frac{1}{6} \ell_{111} \left(U^1\right)^3+i \gamma \right) \biggr]\nonumber\\
& & + \frac{1}{2}\, T_1^2\, \biggl[P'^{11}_0 + P'^{11}_1 U^1 +\frac{1}{2} \ell_{111} \left(U^1\right)^2 P'^{111} -P'^{110} \left(\frac{1}{6} \ell_{111} \left(U^1\right)^3+i \gamma \right) \biggr]\nonumber\\
& & - \frac{1}{2}\, S\, T_1^2\, \biggl[Q'^{11}_0 + Q'^{11}_1 U^1 +\frac{1}{2} \ell_{111} \left(U^1\right)^2 Q'^{111} -Q'^{110} \left(\frac{1}{6} \ell_{111} \left(U^1\right)^3+i \gamma \right) \biggr]\nonumber\\
& & + \frac{1}{6}\, T_1^3\, \biggl[H'^{111}_0 + H'^{111}_1 U^1 +\frac{1}{2} \ell_{111} \left(U^1\right)^2 H'^{1111} -H'^{1110} \left(\frac{1}{6} \ell_{111} \left(U^1\right)^3+i \gamma \right) \biggr]\nonumber\\
& & - \frac{1}{6}\,S\, T_1^3\, \biggl[F'^{111}_0 + F'^{111}_1 U^1 +\frac{1}{2} \ell_{111} \left(U^1\right)^2 F'^{1111} -F'^{1110} \left(\frac{1}{6} \ell_{111} \left(U^1\right)^3+i \gamma \right) \biggr].\nonumber
\eea
In addition, the K\"ahler potential takes the following form,
\bea
\label{eq:K-toroidal}
& & \hskip-1.5cm K = -\ln(-i(S-\ov{S})) - \ln\left(-\frac{i\, \ell_{111}}{6}(U^1-\ov{U}^1)^3 + 2\,\gamma\right)\\
& & - 2\ln\left(\frac{1}{6}\sqrt{\frac{1}{\kappa_{111}}}(i\,(T_1-\ov{T}_1))^{3/2}\right). \nonumber
\eea
Using (\ref{eq:Vtot}), this flux superpotential (\ref{eq:W-quintic}) generically results in a huge four-dimensional ${\cal N} = 1$ effective scalar potential, with a total of more than 11000 terms ! However, using the new axionic flux polynomials as defined in Eqns.~(\ref{eq:newOrbits1})-(\ref{eq:newOrbits2}), we find that the same scalar potential can be equivalently expressed by 668 terms ! In order to appreciate the taxonomy of various pieces in the scalar potential we present the numbers of terms for various scenarios in Table \ref{tab_term-counting4}. Furthermore we read-off the full scalar potential with precise 668 terms using our master formula (\ref{eq:master1}) as well as (\ref{eq:V_gen4}) by simply using $K_{\alpha\ov\beta} = \frac{4}{3} \tau_1^2$ for the Quintic-like models with a single K\"ahler modulus.

Moreover, if we set $\gamma = 0$ assuming the large complex-structure limit, we find that this number 668 reduces to 224, which is precisely the same what we have obtained for the isotropic case of the previous toroidal model. In fact, we find that after setting $\gamma = 0$, the corresponding numbers presented in Table \ref{tab_term-counting4} reduce to those of the respective ones mentioned for the isotropic toroidal case in Table \ref{tab_term-counting3}. This observation presents a good consistency check as both models are STU-like models having a single $T$ and single $U$ modulus each, in addition to have the axio-dilaton modulus ! For interested readers we present all these 224 pieces in the appendix \ref{sec_Quintic-pieces} which have explicit dependence on the three saxions $\{s, \tau_1, u^1\}$ while all the axions $\{c_0, \rho_1, v^1\}$ being encoded in the axionic fluxes.

\begin{table}[h!]
\begin{center}
\begin{tabular}{|c||c|c||c|c||c|c|}
\hline
& &&&&&\\
& Fluxes & $\#$(V) & Axionic-fluxes & $\#$(V) & Axionic-fluxes & $\#$(V) \\
& & & (of step-1) & & (of step-2) & \\
\hline
1 & $F$  & 55 & ${\mathbb F}$ & 55  & $f$ & 15  \\
\hline
2 & $F, H$ & 260 & ${\mathbb F}, {\mathbb H}$ & 122  & $f, \, h$ & 36  \\
\hline
3 & $F, H, Q$ & 639 & ${\mathbb F}, {\mathbb H}, {\mathbb Q}$ & 256  & $f, \, h, \, q$ & 91 \\
\hline
4 & $F, H, Q, P$ & 1459 & ${\mathbb F}, {\mathbb H}, {\mathbb Q}, {\mathbb P}$ & 444  & $f, \, h, \, q, \, p$ & 162  \\
\hline
5 & $F, H, Q, P, P'$  & 2773 & ${\mathbb F}, {\mathbb H}, {\mathbb Q}, {\mathbb P}, {\mathbb P}'$ & 763 & $f, \, h, \, q, \, p, \, p'$ &  265 \\
\hline
6 & $F, H, Q, P,$  & 5027 & ${\mathbb F}, {\mathbb H}, {\mathbb Q}, {\mathbb P},$ & 1136  & $f, \, h, \, q, \, p,$ &  390 \\
& $P', Q'$ & &  ${\mathbb P}', {\mathbb Q}'$ & &  $p',\, q'$ & \\
\hline
7 & $F, H, Q, P,$  & 7446 & ${\mathbb F}, {\mathbb H}, {\mathbb Q}, {\mathbb P},$ & 1532  & $f, \, h, \, q, \, p,$ &  523 \\
& $P', Q', H'$ & &  ${\mathbb P}', {\mathbb Q}', {\mathbb H}'$ & &  $p', \, q', \, h'$ &\\
\hline
8 & $F, H, Q, P, $  & 11212 & ${\mathbb F}, {\mathbb H}, {\mathbb Q}, {\mathbb P},$ & 1940  & $f, \, h, \, q, \, p,$ & 668  \\
& $P', Q', H', F'$ & & ${\mathbb P}', {\mathbb Q}', {\mathbb H}', {\mathbb F}'$ & & $p', \, q', \, h', \, f'$ &\\
\hline
\end{tabular}
\end{center}
\caption{Counting of scalar potential terms for models based on CY threefolds with $h^{1,1}=1$. Further, we find that after setting $\gamma = 0$, the corresponding numbers in this table precisely reduces to those of the respective ones mentioned for the isotropic case in Table \ref{tab_term-counting3}.}
\label{tab_term-counting4}
\end{table}


\section{Summary and conclusions}
\label{sec_conclusions}


In this article, we have presented a set of compact and concise formulations of the ${\cal N} = 1$ four-dimensional scalar potentials arising from the generalized flux superpotential with U-dual fluxes. The main starting point is the type IIB holomorphic flux superpotential with four pairs of S-dual fluxes, namely $(F, H), (Q, P), (P', Q')$ and $(H', F')$, which are required for the U-dual completion of the flux superpotential. Such a superpotential is cubic polynomial in terms of the complex-structure moduli ($U^i$) as well as the K\"ahler moduli ($T_\alpha$) while having a linear dependence on the axio-dilaton $S$. It has been notoriously known that the typical scalar potentials arising from such a generalized flux superpotential consist of a huge number of terms and it is hard to handle those for any phenomenological studies such as performing moduli stabilizing and search of physical vacua. For exactly the same reasons despite all the interests in finding the flux vacua the studies have been limited only to numerical analysis \cite{Guarino:2008ik,deCarlos:2009qm,Dibitetto:2011qs,Danielsson:2012by,Damian:2013dq,Damian:2013dwa,Blaback:2015zra,Damian:2023ote}, though some generic analytic studies have been carried out for type IIA constructions in \cite{Shukla:2019akv,Marchesano:2020uqz,Shukla:2022srx,Prieto:2024shz}, and for type IIB constructions in \cite{Shukla:2016xdy,Shukla:2019dqd} and F-theory flux vacua in \cite{Marchesano:2021gyv}. Subsequently it is very important to formulate the scalar potential to some compact and concise form which can be used for this purpose.

For the purpose of writing down the generic scalar potentials in a compact manner we use the so-called axionic flux polynomials. Also we note that we take CY orientifolds such that $h^{1,1}_- = 0 = h^{2,1}_+$ which ensures that there are no odd-moduli and no D-term contributions to the scalar potential.  The main steps for {\it reading-off} the F-term scalar potential pieces after knowing the topological data of the CY threefolds and mirrors are as follows:
\begin{itemize}

\item
Step-1: The number of moduli and flux parameters appearing in the generalized superpotential is determined in terms of the Hodge numbers $h^{1,1}_+$ and $h^{2,1}_-$ counting the $T_\alpha$ and $U^i$ moduli. The fluxes appearing in the holomorphic superpotential are given as:
\bea
\label{eq:}
& & (F_\Lambda, F^\Lambda), \quad (H_\Lambda, H^\Lambda), \quad (Q^\alpha_\Lambda, Q^{\alpha\Lambda}), \quad (P^\alpha_\Lambda, P^{\alpha\Lambda}), \\
& & ({P'}^{\alpha\beta}_\Lambda, {P'}^{\alpha\beta\Lambda}), \quad ({Q'}^{\alpha\beta}_\Lambda, {Q'}^{\alpha\beta\Lambda}), \quad ({H'}^{\alpha\beta\gamma}_\Lambda, {H'}^{\alpha\beta\gamma\Lambda}), \quad ({F'}^{\alpha\beta\gamma}_\Lambda, {F'}^{\alpha\beta\gamma\Lambda}), \nonumber
\eea
where fluxes are counted by indices $\alpha \in h^{1,1}_+$ and $\Lambda \in (1+h^{2,1}_-)$, and the orientifold surviving components of the prime fluxes are determined via (\ref{eq:primefluxIIBc}) using $\kappa^{\alpha\beta\gamma}$ as defined in (\ref{eq:Inv-lijk}). 

\item
Step-2: Write down the axionic flux polynomials for each of the eight types of fluxes $\{F, H, Q, P, P', Q', H', F'\}$ using (\ref{eq:AxionicFlux})-(\ref{eq:AxionicFlux1}) and (\ref{eq:newOrbits1})-(\ref{eq:newOrbits2}) which lead to a set of {\it new axionic flux orbits} denoted as $\{{\rm f}, {\rm h}, {\rm q}, {\rm p}, {\rm p'}, {\rm q'}, {\rm h'}, {\rm f'}\}$ with appropriate indices. For example, ${\rm f'}$ has four types of components, namely $({\rm f'}^{\alpha\beta\gamma}{}_0, {\rm f'}^{\alpha\beta\gamma}{}_i, {\rm f'}^{\alpha\beta\gamma i}, {\rm f'}^{\alpha\beta\gamma 0})$ while any omission of $\alpha$ indices means that they are contracted with $\tau_\alpha$'s; e.g.  ${\rm f'}^{\alpha\beta}{}_0 = \tau_\gamma {\rm f'}^{\alpha\beta\gamma}{}_0$, ${\rm f'}^\alpha{}_0 = \frac{1}{2} \tau_\beta\tau_\gamma {\rm f'}^{\alpha\beta\gamma}{}_0$ and ${\rm f'}_0 = \frac{1}{6} \tau_\alpha\tau_\beta\tau_\gamma {\rm f'}^{\alpha\beta\gamma}{}_0$. 

\item
Step-3: Write down the triplet of combinations $\{\chi, \psi, \Psi^\alpha\}$ defined in (\ref{eq:chi-psi-Psi}) which play the central role in formulating the scalar potential, e.g. using the master formula: 
\bea
& & \hskip-1cm V_{\rm I} = e^K \biggl[K^{i\ov{j}}\,\Theta_i[\chi] \cdot  \ov\Theta_j[\ov\chi] + \frac{4\,{\cal U}+\gamma}{2\,{\cal U}-\gamma} (i\, u^i)\, \left(\Theta_i[\chi] \cdot \ov\Phi[\ov\chi] - \Phi[\chi]\cdot \ov\Theta_i[\ov\chi] \right) \\
& & \hskip-0.5cm + \frac{8\,{\cal U}-\gamma}{2\,{\cal U}-\gamma} \Phi[\chi]\cdot \ov\Phi[\ov\chi] + 4\, s^2\, \Phi[\psi] \cdot \ov\Phi[\ov\psi] + K_{\alpha\ov\beta}\, \Phi[\Psi^\alpha] \cdot \ov\Phi[\ov\Psi^\beta]   \nonumber\\
& & \hskip-0.5cm + \, 2\,i\, \left(\Phi[\Psi -s \, \psi] \cdot \ov\Phi[\ov\chi] - \Phi[\chi]\cdot \ov\Phi[\ov\Psi -s \, \ov\psi] \right) \biggr], \nonumber
\eea
where the linear functional $\Phi[y]$ is defined as $\Phi[y] = \left(y_0 - i\,  u^i \, y_i\, - y^i \,\sigma_i + i\, y^0 \, ({\cal U} + \, \gamma)\right)$ for $y \in \{\chi, \psi, \Psi^\alpha\}$, and $\Theta_i[y] = {y}_i\, -\, i \,\ell_{ij} \, {y}^j - {y}^0 \, \sigma_i$ whereas the moduli space metrics $K^{i\ov{j}}$ and $K_{\alpha\ov\beta}$ are determined by the triple intersection numbers of the CY and its mirror, namely $\kappa_{\alpha\beta\gamma}$ and $\ell_{ijk}$ via using the definition (\ref{eq:Kmetric})-(\ref{eq:InvK}). In fact with these ingredients at hand one can use any of the four master-formulae, given in Eqs.~(\ref{eq:master1}), (\ref{eq:master2}), (\ref{eq:master3}) or (\ref{eq:master4}), to directly write down the scalar potential.

\end{itemize}

\noindent
Following the above strategy we presented four equivalent formulations of the scalar potential based on the compactness and the manifestation of their dependence on CS-moduli. Subsequently we demonstrated the use and process of  applying the master formula in five explicit examples. In this regard we presented the master formula for the standard type IIB superpotential with $(F, H)$ fluxes only. Although this model has been studied in numerous occasions in the past, the studies have been mostly focused on the supersymmetric vacua without explicitly mentioning the scalar potential, like the one presented in (\ref{eq:V_gen-FH}). Subsequently we considered a couple of extensions of this minimal $(F, H)$ flux model by including some non-geometric fluxes, namely $(F, H, Q)$ and $(F, H, Q, P)$ fluxes, and provided the master formulae for their respective induced scalar potentials generalizing the previous results in \cite{Shukla:2019wfo}. This iterative approach of including the additional fluxes manifestly demonstrates the breaking and restoration of the S-duality in the scalar potential as one can see from (\ref{eq:GVW-generic-master-formula}), (\ref{eq:FHQ}) and (\ref{eq:FHQP}).

As a fourth setup we presented a concrete model based on the type IIB superstring compactifications on the toroidal orientifold of the $\mathbb T^6/(\mathbb Z_2 \times \mathbb Z_2)$ orbifold with fluxes of all the type considered in our approach. This model supports 128 fluxes with three $T_\alpha$ and three $U^i$ moduli each along with the axio-dilaton and subsequently one has 76276 terms in the scalar potential induced via the flux superpotential. It has been recently shown that this number can be reduced to 10888 via using a set of axionic fluxes constructed by fluxes and RR axions \cite{Leontaris:2023lfc,Leontaris:2023mmm}. Taking one step further, in the current work we have shown that this number can be reduced to 2816 if we use a {\it new} set of axionic fluxes which absorb all the axionic dependence including the complex structure axions. In order to further reduce the scalar potential size we have taken the isotropic limit and found that 2816  reduces to 224. This way we have significantly reduced the size of the scalar potential from 76276 to 224 which can be used for analytic exploration of vast landscape of the non-geometric flux vacua \cite{AbdusSalam:2024arh}. 

For the illustration of the application of our master formulae to beyond toroidal cases we consider another class of models which can arise by type IIB compactifications on the orientifolds of CY threefolds with $h^{1,1} = h^{1,1}_+ = 1$.  Similar to the previous isotropic case it corresponds to an STU-like model of the toroidal case with untwisted sector. However unlike the toroidal case $\gamma$ is generically non-zero. We find that a direct computation of scalar potential from the flux superpotential leads to 11212 terms. To begin with it is not pragmatic at all to think about using this scalar potential for any phenomenological purpose such as moduli stabilization etc. However, the internal structure among the pieces as reflected via the axionic flux combinations helped us not only to reduce the size of the scalar potential to just 668 terms but also provided a master formula which can be used to {\it read-off} the scalar potential pieces beyond the simple toroidal like models. Moreover this number 668 reduces to 224 for $\gamma = 0$ which can be justified in the large complex-structure limit.

Given that the number of terms in the scalar potential has been reduced to 224 after considering the large complex-structure limit to ignore the non-perturbative effects and setting via $\gamma = 0$, it creates a hope that the scalar potential which, to begin with appeared to be too huge to think of extracting any phenomenological application, may be used for model building purpose. We plan to report such an application of our master formulae in a future work \cite{AbdusSalam:2024arh}.


\section*{Acknowledgments}
We would like to thank Shehu AbdusSalam, Xin Gao, David Prieto and Joan Quirant  for useful discussions on related topics. PS would like to thank the {\it Department of Science and Technology (DST), India} for the kind support.


\newpage
\appendix
\setcounter{equation}{0}


\section{Some explicit formulations of the scalar potential}
\label{sec_AlternateFormulations}


In the section we present a couple of more formulations or master formulae for the scalar potential. These are slight complicated than the previous ones due to explicit nature of the pieces, however they can be useful for learning more insights of the potential and their any possible use in phenomenological model building.

\subsection*{Third formulation:}
Using the combinations $\psi, \chi$ and $\Psi$ defined in (\ref{eq:chi-psi-Psi}), the generic scalar potential (\ref{eq:V_gen}) can also be expressed as,
\bea
\label{eq:master3}
& & \hskip-1cm V_{\rm III} = e^K \biggl[V_{\rm III}^{(1)} + V_{\rm III}^{(2)} + V_{\rm III}^{(3)} + V_{\rm III}^{(4)} + V_{\rm III}^{(5)} \biggr],
\eea
where the five pieces take the following explicit expressions,
\bea
& & \hskip-0.5cm V_{\rm III}^{(1)} = \frac{8\,{\cal U} - \gamma}{2\,{\cal U} - \gamma} \chi_0 \ov{\chi}_0 + \left(K^{i\ov{j}} - \frac{3\,\gamma u^i u^j}{2{\cal U} -\gamma} \right)\chi_i \, \ov{\chi}_j + \left((4{\cal U} + \gamma)^2 K_{i\ov{j}}+ \frac{3 \gamma\, \sigma_i \, \sigma_j}{2{\cal U}-\gamma}\right)\chi^i \, \ov{\chi}^j \\
& & + \frac{8\,{\cal U}^2 + \gamma\, {\cal U} + 2\, \gamma^2}{2} \chi^0\, \ov{\chi}^0 + \frac{3 \, \gamma\, \sigma_i}{2{\cal U} - \gamma} \left(\chi_0  \ov\chi^i + \ov\chi_0 \, \chi^i \right) - \frac{3\gamma\, u^i}{2} \left(\chi_i \ov\chi^0 + \chi^0 \ov\chi_i\right) \nonumber\\
& & + 4\, s^2 \, \left((\psi_0 - \sigma_i \psi^i)(\ov\psi_0 - \sigma_j \ov\psi^j)+ (u^i\, \psi_i - ({\cal U}+\gamma)\psi^0)(u^j\,\ov\psi_j\,- ({\cal U}+\gamma)\ov\psi^0) \right)\nonumber\\
& & + K_{\alpha\ov\beta}\left((\Psi^\alpha_0 - \sigma_i \, \Psi^{\alpha i})(\ov\Psi^\beta_0 - \sigma_j \, \ov\Psi^{\beta j}) + (u^i\, \Psi^\alpha_i -({\cal U} + \gamma)\, \Psi^{\alpha0})\,(u^j \, \ov\Psi^\beta_j \,-({\cal U} + \gamma)\, \ov\Psi^{\beta0}) \right), \nonumber\\
& & \nonumber\\
& & \hskip-0.5cm V_{\rm III}^{(2)} =  2\, s \left(u^i(\psi_0 \ov\chi_i + \chi_i \ov\psi_0 - \chi_0 \ov\psi_i    - \psi_i \ov\chi_0) + ({\cal U} + \gamma) \sigma_i (\chi^0 \ov\psi^i + \psi^i \ov\chi^0 - \chi^i \ov\psi^0 - \psi^0 \ov\chi^i)\right)\nonumber\\
& & - 2 \left(u^i(\Psi_0 \ov\chi_i  + \chi_i \ov\Psi_0 - \chi_0 \ov\Psi_i - \Psi_i \ov\chi_0) - ({\cal U} + \gamma) \sigma_i (\chi^0 \ov\Psi^i + \Psi^i \ov\chi^0 - \chi^i \ov\Psi^0 - \Psi^0 \ov\chi^i)\right) \nonumber\\
& & + 2\,s \, u^i \sigma_j (\psi_i \ov\chi^j  + \ov\psi_i \chi^j - \chi_i \ov\psi^j - \ov\chi_i \psi^j) - 2\, u^i \sigma_j (\Psi_i \ov\chi^j  + \ov\Psi_i \chi^j - \chi_i \ov\Psi^j - \ov\chi_i \Psi^j), \nonumber\\
& & \nonumber\\
& & \hskip-0.5cm V_{\rm III}^{(3)} = 2\, ({\cal U} + \gamma) \left(s \left(\chi_0 \ov\psi^0 + \ov\chi_0 \psi^0 - \psi_0 \ov\chi^0 - \ov\psi_0 \chi^0\right) - \left(\chi_0 \ov\Psi^0 + \ov\chi_0 \Psi^0 - \Psi_0 \ov\chi^0 - \ov\Psi_0 \chi^0\right)\right)\nonumber\\
& & + i\, (2\,{\cal U} - \gamma) \left(\chi_0 \ov\chi^0 - \ov\chi_0 \chi^0 \right) - i\, (4\,{\cal U} + \gamma) \left(\chi_i \ov\chi^i - \ov\chi_i \chi^i \right), \nonumber\\
& & \nonumber\\
& & \hskip-0.5cm V_{\rm III}^{(4)} = - 4\, i\, s^2 \, \left((u^i\, \psi_i - ({\cal U}+\gamma)\psi^0)\, (\ov\psi_0 - \sigma_j \ov\psi^j) - (\psi_0 - \sigma_i \psi^i)\,(u^j\,\ov\psi_j\,- ({\cal U}+\gamma)\ov\psi^0) \right)\nonumber\\
& & - \, i\, K_{\alpha\ov\beta}\left((u^i\, \Psi^\alpha_i -({\cal U} + \gamma)\, \Psi^{\alpha0}) \,(\ov\Psi^\beta_0 - \sigma_j \, \ov\Psi^{\beta j}) - \,(\Psi^\alpha_0 - \sigma_i \, \Psi^{\alpha i}) \, (u^j \, \ov\Psi^\beta_j \,-({\cal U} + \gamma)\, \ov\Psi^{\beta0}) \right), \nonumber\\
& & \nonumber\\
& & \hskip-0.5cm V_{\rm III}^{(5)} = -2\, i \left(u^i\, (\chi_0 \ov\chi_i - \chi_0 \ov\chi_i) + ({\cal U} + \gamma) \sigma_i (\chi^0 \ov\chi^i - \ov\chi^0 \chi^i) + \sigma_i u^j (\chi^i \ov\chi_j- \chi_j \ov\chi^i) \right) \nonumber\\
& & +2\, i\, s\, (\psi_0(-\ov\chi_0+\sigma_j \ov\chi^j) + \psi_i u^i (({\cal U} + \gamma) \ov\chi^0 - u^j \ov\chi_j) - (({\cal U} + \gamma) \ov\psi^0 - u^j \ov\psi_j)(\chi_i u^i - ({\cal U} + \gamma) \chi^0) \nonumber\\
& & +(\ov\psi_0 - \sigma_j \ov\psi^j) (\chi_0 - \sigma_i \chi^i) + ({\cal U} + \gamma) ((-{\cal U} + \gamma) \ov\chi^0 + \ov\chi_j u^j) \psi^0 + \sigma_i \psi^i (\ov\chi_0 - \sigma_j \ov\chi^j) \nonumber\\
& & -2\, i\, (\Psi_0(-\ov\chi_0+\sigma_j \ov\chi^j) + \Psi_i u^i (({\cal U} + \gamma) \ov\chi^0 - u^j \ov\chi_j) - (({\cal U} + \gamma) \ov\Psi^0 - u^j \ov\Psi_j)(\chi_i u^i - ({\cal U} + \gamma) \chi^0) \nonumber\\
& & +(\ov\Psi_0 - \sigma_j \ov\Psi^j) (\chi_0 - \sigma_i \chi^i) + ({\cal U} + \gamma) ((-{\cal U} + \gamma) \ov\chi^0 + \ov\chi_j u^j) \Psi^0 + \sigma_i \Psi^i (\ov\chi_0 - \sigma_j \ov\chi^j). \nonumber
\eea

\subsection*{Fourth formulation:}

Using the combinations $\psi, \chi$ and $\Psi$ defined in (\ref{eq:chi-psi-Psi}), the generic scalar potential (\ref{eq:V_gen}) can also be expressed as,
\bea
\label{eq:master4}
& & V_{\rm IV} = e^K \biggl[A_0 + A_i \, u^i + A^i\, \sigma_i + u^i\sigma_j A_i^j + A_{ij} u^i\, u^j + A^{ij} \sigma_i \sigma_j \biggr],
\eea
where $A$'s are real coefficients given by the following expressions,
\bea
\label{eq:A-gen}
& & \hskip-0.5cm A_{0} = \left(\frac{8\,{\cal U} - \gamma}{2\,{\cal U} - \gamma} \chi_0 \ov{\chi}_0 + \chi_i \, K^{i\ov{j}} \, \ov{\chi}_j + (4 {\cal U} + \gamma)^2\, \chi^i \, K_{i\ov{j}} \, \ov{\chi}^j + \frac{8\,{\cal U}^2 + \gamma\, {\cal U} + 2\, \gamma^2}{2} \chi^0\, \ov{\chi}^0\right) \\
& & + 4\,s^2 (\psi_0 \, \ov\psi_0 + ({\cal U}+ \gamma)^2\, \psi^0 \ov\psi^0) + K_{\alpha\ov\beta} \, \left(\Psi^\alpha_0 \ov\Psi^\beta_0 + ({\cal U} + \gamma)^2\, \Psi^{\alpha0} \ov\Psi^{\beta0}\right) \nonumber\\
& & + \, 2\, ({\cal U} + \gamma) \left(s \left(\chi_0 \ov\psi^0 + \ov\chi_0 \psi^0 - \psi_0 \ov\chi^0 - \ov\psi_0 \chi^0\right) - \left(\chi_0 \ov\Psi^0 + \ov\chi_0 \Psi^0 - \Psi_0 \ov\chi^0 - \ov\Psi_0 \chi^0\right)\right)\nonumber\\
& & +\, i\, (2\,{\cal U} - \gamma) \left(\chi_0 \ov\chi^0 - \ov\chi_0 \chi^0 \right) - i\, (4\,{\cal U} + \gamma) \left(\chi_i \ov\chi^i - \ov\chi_i \chi^i \right) - 4\, i\,s^2 ({\cal U} + \gamma) \left(\psi_0 \ov\psi^0 - \ov\psi_0 \psi^0 \right) \nonumber\\
& & +\, 2 \, i\left(\chi_0\, (s \ov\psi_0 - \ov\Psi_0) - \ov\chi_0 (s \psi_0 - \Psi_0) + ({\cal U}+\gamma)^2 (\chi^0(s\ov\psi^0 - \ov\Psi^0) -\ov\chi^0(s\psi^0 - \Psi^0)) \right) \nonumber\\
& & - i\, ({\cal U}+\gamma)\, K_{\alpha\ov\beta} (\Psi^\alpha_0 \ov\Psi^{\beta0}-\ov\Psi^\alpha_0 \Psi^{\beta0}), \nonumber\\
& & \hskip-0.5cm A_{i} = -\frac{3 \gamma}{2} (\chi_i \ov\chi^0 + \ov\chi_i \chi^0) -4\,s^2 ({\cal U} + \gamma)(\psi_i \ov\psi^0+ \ov\psi_i \psi^0) - ({\cal U} + \gamma)\,K_{\alpha\ov\beta} (\Psi^\alpha_i \ov\Psi^{\beta0} + \ov\Psi^\alpha_i \Psi^{\beta0})\nonumber\\
& & + \, 2\left(\chi_i (s\,\ov\psi_0 - \ov\Psi_0) + \ov\chi_i (s\,\psi_0 - \Psi_0) - \ov\chi_0(s\, \psi_i - \Psi_i) -\chi_0(s\, \ov\psi_i - \ov\Psi_i)\right)\nonumber\\
& & - \, i\left(2\,(\chi_i \ov\chi_0 - \ov\chi_i \chi_0) + 4\, s^2\, (\psi_i \ov\psi_0 - \ov\psi_i \psi_0) + K_{\alpha\ov\beta} (\Psi^\alpha_i \ov\Psi^\beta_0 - \ov\Psi^\alpha_i \Psi^\beta_0) \right) \nonumber\\
& & -\, 2\, i\,({\cal U}+ \gamma) \left(\chi_i(s \ov\psi^0 - \ov\Psi^0) + \ov\chi_i(s \psi^0 - \Psi^0) - \ov\chi^0 (s\psi_i - \Psi_i) - \chi^0 (s\ov\psi_i - \ov\Psi_i)\right), \nonumber\\
& & \hskip-0.5cm A^{i} = \frac{3 \gamma}{(2\, {\cal U} - \gamma)} (\chi^i \ov\chi_0 + \ov\chi^i \chi_0) -4\,s^2 (\psi^i \ov\psi_0+ \ov\psi^i \psi_0) - K_{\alpha\ov\beta} (\Psi^{\alpha i} \ov\Psi^\beta_0 + \ov\Psi^{\alpha i} \Psi^\beta_0)\nonumber\\
& & - \, 2\, ({\cal U} + \gamma)\left(\chi^i (s\,\ov\psi^0 - \ov\Psi^0) + \ov\chi^i (s\,\psi^0 - \Psi^0) - \ov\chi^0(s\, \psi^i - \Psi^i) -\chi^0(s\, \ov\psi^i - \ov\Psi^i)\right)\nonumber\\
& & + \, i\, ({\cal U} + \gamma) \left(2\,(\chi^i \ov\chi^0 - \ov\chi^i \chi^0) + 4\, s^2\, (\psi^i \ov\psi^0 - \ov\psi^i \psi^0) + K_{\alpha\ov\beta} (\Psi^{\alpha i} \ov\Psi^{\beta0} - \ov\Psi^{\alpha i} \Psi^{\beta0})\right) \nonumber\\
& & -\, 2\, i\, \left(\chi^i(s \ov\psi_0 - \ov\Psi_0) + \ov\chi^i(s \psi_0 - \Psi_0) - \ov\chi_0 (s\psi^i - \Psi^i) - \chi_0 (s\ov\psi^i - \ov\Psi^i)\right), \nonumber\\
& & \hskip-0.5cm A_i^j = 2\,(s\, \psi_i - \Psi_i)\ov\chi^j - 2\,\chi_i(s \, \ov\psi^j - \ov\Psi^j) + 2\, (s\, \ov\psi_i - \ov\Psi_i)\chi^j - 2\, \ov\chi_i(s \, \psi^j - \Psi^j)  \nonumber\\
& & + 2\, i (\chi_i \ov\chi^j - \ov\chi_i \chi^j) + 4\, i\, s^2 (\psi_i \ov\psi^j - \ov\psi_i \psi^j) + i K_{\alpha\ov\beta} (\Psi^\alpha_i \ov\Psi^{\beta j} - \ov\Psi^\alpha_i \Psi^{\beta j}), \nonumber\\
& & \hskip-0.5cm A_{ij} = -\frac{3\gamma \chi_i\, \ov\chi_j}{2{\cal U} - \gamma} + 4\,s^2 \, \psi_i \ov\psi_j + K_{\alpha\ov\beta} \Psi^\alpha_i\, \ov\Psi^\beta_j + 2\, i\, \left(\chi_i (s\,\ov\psi_j- \ov\Psi_j) - (s\,\psi_i - \Psi_i)\ov\chi^j \right), \nonumber\\
& & \hskip-0.5cm A^{ij} = \frac{3\gamma \chi^i\, \ov\chi^j}{2{\cal U} - \gamma} + 4\,s^2 \, \psi^i \ov\psi^j + K_{\alpha\ov\beta} \Psi^{\alpha i}\, \ov\Psi^{\beta j} + 2\, i\, \left(\chi^i (s\,\ov\psi^j- \ov\Psi^j) - (s\,\psi^i - \Psi^i)\ov\chi^j \right). \nonumber
\eea
Here we recall that
\bea
& & e^K = \frac{1}{4\,s\,{\cal V}^2\, (4\, {\cal U}+\gamma)}.
\eea


\section{Scalar potential pieces for Quintic-type models}
\label{sec_Quintic-pieces}


For the type IIB models based on orientifold compactifications using a CY with $h^{1,1} = 1$, the generic scalar potential arising from the U-dual completed flux superpotential can be expressed as below,
\bea
\label{eq:Vquintic-taxonomy}
& & V = e^K \biggl[V_{ff} + V_{hh}+ V_{qq}+ V_{pp}+ V_{p'p'} + V_{q'q'} + V_{h'h'}+ V_{f'f'} \\
& & \hskip0.5cm + V_{fh}+ V_{fq}+ V_{fp}+ V_{fp'}+ V_{fq'}+ V_{fh'}+ V_{ff'}+ V_{hq} + V_{hp}+ V_{hp'} \nonumber\\
& & \hskip0.5cm + V_{hq'} + V_{hh'} + V_{hf'}+ V_{qp}+ V_{qp'}+ V_{qq'} + V_{qh'}+ V_{qf'}+ V_{pp'} + V_{pq'} \nonumber\\
& & \hskip0.5cm + V_{ph'}+ V_{pf'} + V_{p'q'} + V_{p'h'}+ V_{p'f'} + V_{q'h'} + V_{q'f'} + V_{h'f'}\biggr]\,, \nonumber
\eea
where the eight terms in the first line are what we call as `diagonal terms' while the remaining 28 terms are called as `cross-terms'. Also we have the following overall factor in $V$,
\bea
& & e^K = \frac{1}{4\, s\, {\cal V}^2 \, (4\, {\cal U} + \gamma)} \simeq \frac{27\,\kappa_{111}}{16\, s\, \ell_{111} \, (u^1)^3 \tau_1^3}.
\eea
The generic expressions in $V$ are very involved and not illuminating to present them here, however considering $\gamma = 0$ one can capture all the leading order pieces in the large complex-structure limit resulting in a total of only 224 terms arising from the 36 pieces of (\ref{eq:Vquintic-taxonomy}). The details about each of these 36 terms are given as below:
\bea
\label{eq:diag-1a}
& (1): & \quad V_{ff} = 4 f_0^2+\frac{4}{3} f_1^2 \left(u^1\right)^2+\frac{1}{3} \ell_{111}^2 \left(f^1\right)^2 \left(u^1\right)^4+\frac{1}{9} \ell_{111}^2 \left(f^0\right)^2 \left(u^1\right)^6,\\
& (2): & \quad V_{hh} = 4 s^2 h_0^2+\frac{4}{3} s^2 h_1^2 \left(u^1\right)^2+\frac{1}{3} s^2 \ell_{111}^2 \left(h^1\right)^2 \left(u^1\right)^4+\frac{1}{9} s^2 \ell_{111}^2 \left(h^0\right)^2 \left(u^1\right)^6,\nonumber\\
& (3): & \quad V_{qq} = \frac{4}{3} \tau_1^2 \left(q_0^1\right)^2 +\frac{1}{27} \ell_{111}^2 \tau_1^2 \left(q^{10}\right)^2 \left(u^1\right)^6+\frac{8}{3} \ell_{111} \tau_1^2 q^{11} \left(u^1\right)^2 q_0^1\nonumber\\
& & \hskip1.2cm -\frac{1}{3} \ell_{111}^2 \tau_1^2 \left(q^{11}\right)^2 \left(u^1\right)^4 +\frac{8}{9} \ell_{111} \tau_1^2 q^{10}\left(u^1\right)^4 q_1^1-\frac{4}{3} \tau_1^2 \left(u^1\right)^2 \left(q_1^1\right)^2,  \nonumber\\
& (4): & \quad V_{pp} = \frac{4}{3} s^2 \tau_1^2 \left(p_0^1\right)^2 +\frac{1}{27} s^2 \ell_{111}^2 \tau_1^2 \left(p^{10}\right)^2 \left(u^1\right)^6+\frac{8}{3} s^2 \ell_{111} \tau_1^2 p^{11} \left(u^1\right)^2 p_0^1\nonumber\\
& & \hskip1cm -\frac{1}{3} s^2 \ell_{111}^2 \tau_1^2 \left(p^{11}\right)^2 \left(u^1\right)^4 +\frac{8}{9} s^2 \ell_{111} \tau_1^2 p^{10} \left(u^1\right)^4 p_1^1-\frac{4}{3} s^2 \tau_1^2 \left(u^1\right)^2 \left(p_1^1\right)^2, \nonumber\\
& (5): & \quad V_{p'p'} = \frac{1}{3} \tau_1^4 \left({p'}_0^{11}\right)^2 -\frac{1}{12} \ell_{111}^2 \tau_1^4 \left({p'}^{111}\right)^2 \left(u^1\right)^4+\frac{2}{3} \ell_{111} \tau_1^4 {p'}^{111} \left(u^1\right)^2 {p'}_0^{11}\nonumber\\
& & \hskip0.5cm +\frac{2}{9} \ell_{111} \tau_1^4 {p'}^{110} \left(u^1\right)^4 {p'}_1^{11}-\frac{1}{3} \tau_1^4 \left(u^1\right)^2 \left({p'}_1^{11}\right)^2 +\frac{1}{108} \ell_{111}^2 \tau_1^4 \left({p'}^{110}\right)^2 \left(u^1\right)^6, \nonumber\\
& (6): & \quad V_{q'q'} = \frac{1}{3} s^2 \tau_1^4 \left({q'}_0^{11}\right)^2-\frac{1}{12} s^2 \ell_{111}^2 \tau_1^4 \left({q'}^{111}\right)^2 \left(u^1\right)^4+\frac{2}{3} s^2 \ell_{111} \tau_1^4 {q'}^{111} \left(u^1\right)^2 {q'}_0^{11}\nonumber\\
& & \hskip0.5cm +\frac{2}{9} s^2 \ell_{111} \tau_1^4 {q'}^{110} \left(u^1\right)^4 {q'}_1^{11}-\frac{1}{3} s^2 \tau_1^4 \left(u^1\right)^2 \left({q'}_1^{11}\right)^2+\frac{1}{108} s^2 \ell_{111}^2 \tau_1^4 \left({q'}^{110}\right)^2 \left(u^1\right)^6,  \nonumber\\
& (7): & \quad V_{h'h'} = \frac{1}{108} \ell_{111}^2 \tau_1^6 \left({h'}^{1111}\right)^2 \left(u^1\right)^4+\frac{1}{324} \ell_{111}^2 \tau_1^6 \left({h'}^{1110}\right)^2 \left(u^1\right)^6\nonumber\\
& & \hskip1.5cm +\frac{1}{9} \tau_1^6 \left({h'}_0^{111}\right)^2+\frac{1}{27} \tau_1^6 \left(u^1\right)^2 \left({h'}_1^{111}\right)^2, \nonumber\\
& (8): & \quad V_{f'f'} = \frac{1}{108} s^2 \ell_{111}^2 \tau_1^6 \left({f'}^{1111}\right)^2 \left(u^1\right)^4+\frac{1}{324} s^2 \ell_{111}^2 \tau_1^6 \left({f'}^{1110}\right)^2 \left(u^1\right)^6\nonumber\\
& & \hskip1.5cm +\frac{1}{9} s^2 \tau_1^6 \left({f'}_0^{111}\right)^2+\frac{1}{27} s^2 \tau_1^6 \left(u^1\right)^2 \left({f'}_1^{111}\right)^2,  \nonumber\\
& (9): & \quad V_{fh} = \frac{4}{3} \, s \, \ell_{111}\,\left(u^1\right)^3 \left(h_0 f^0-h_1 f^1-f_0 h^0+f_1 h^1\right), \nonumber\\
& (10): & \quad V_{fq} = -\, \frac{4}{3} \, \tau_1\, \ell_{111}  \left(u^1\right)^3 \left(f_0 q^{10}-f_1 q^{11}-f^0 q_0^1+f^1 q_1^1\right), \nonumber
\eea
\bea
\label{eq:cross1a}
& (11): & \quad V_{fp} = -4 s f_0 \ell_{111} \tau_1 p^{11} \left(u^1\right)^2-\frac{4}{3} s f_1 \ell_{111} \tau_1 p^{10} \left(u^1\right)^4+\frac{4}{3} s \ell_{111}^2 \tau_1 f^1 p^{11} \left(u^1\right)^4\nonumber\\
& & \hskip1.5cm -4 s \ell_{111} \tau_1 f^1 \left(u^1\right)^2 p_0^1+\frac{16}{3} s f_1 \tau_1 \left(u^1\right)^2 p_1^1-\frac{4}{3} s
   \ell_{111} \tau_1 f^0 \left(u^1\right)^4 p_1^1, \nonumber\\
& (12): & \quad V_{fp'} = -2 f_0 \ell_{111} \tau_1^2 {p'}^{111} \left(u^1\right)^2-\frac{2}{3} f_1 \ell_{111} \tau_1^2 {p'}^{110} \left(u^1\right)^4+\frac{2}{3} \ell_{111}^2 \tau_1^2 f^1 {p'}^{111} \left(u^1\right)^4 \nonumber\\
& & \hskip1.5cm -2 \ell_{111} \tau_1^2 f^1 \left(u^1\right)^2 {p'}_0^{11}+\frac{8}{3} f_1 \tau_1^2 \left(u^1\right)^2 {p'}_1^{11}-\frac{2}{3}
   \ell_{111} \tau_1^2 f^0 \left(u^1\right)^4 {p'}_1^{11},\nonumber\\
& (13): & \quad V_{fq'} = \frac{4}{3} s f_0 \ell_{111} \tau_1^2 {q'}^{110} \left(u^1\right)^3-\frac{8}{3} s f_1 \ell_{111} \tau_1^2 {q'}^{111} \left(u^1\right)^3-\frac{1}{3} s \ell_{111}^2 \tau_1^2 f^1 {q'}^{110} \left(u^1\right)^5 \nonumber\\
& & \hskip1.5cm +\frac{1}{3} s \ell_{111}^2 \tau_1^2 f^0 {q'}^{111} \left(u^1\right)^5+4 s f_1 \tau_1^2 u^1
   {q'}_0^{11}-\frac{4}{3} s \ell_{111} \tau_1^2 f^0 \left(u^1\right)^3 {q'}_0^{11}\nonumber\\
& & \hskip1.5cm -4 s f_0 \tau_1^2 u^1 {q'}_1^{11}+\frac{8}{3} s \ell_{111} \tau_1^2 f^1 \left(u^1\right)^3 {q'}_1^{11}, \nonumber\\
& (14): & \quad V_{fh'} = \frac{4}{9} f_0 \ell_{111} \tau_1^3 {h'}^{1110} \left(u^1\right)^3-\frac{8}{9} f_1 \ell_{111} \tau_1^3 {h'}^{1111} \left(u^1\right)^3-\frac{1}{9} \ell_{111}^2 \tau_1^3 f^1 {h'}^{1110} \left(u^1\right)^5\nonumber\\
& & \hskip1.5cm +\frac{1}{9} \ell_{111}^2 \tau_1^3 f^0 {h'}^{1111} \left(u^1\right)^5+\frac{4}{3} f_1 \tau_1^3 u^1
   {h'}_0^{111}-\frac{4}{9} \ell_{111} \tau_1^3 f^0 \left(u^1\right)^3 {h'}_0^{111}\nonumber\\
& & \hskip1.5cm -\frac{4}{3} f_0 \tau_1^3 u^1 {h'}_1^{111}+\frac{8}{9} \ell_{111} \tau_1^3 f^1 \left(u^1\right)^3 {h'}_1^{111}, \nonumber\\
& (15): & \quad V_{ff'} = \frac{4}{3} s f_0 \ell_{111} \tau_1^3 {f'}^{1111} \left(u^1\right)^2+\frac{4}{9} s f_1 \ell_{111} \tau_1^3 {f'}^{1110} \left(u^1\right)^4-\frac{5}{9} s \ell_{111}^2 \tau_1^3 f^1 {f'}^{1111} \left(u^1\right)^4 \nonumber\\
& & \hskip1.5cm -\frac{1}{27} s \ell_{111}^2 \tau_1^3 f^0 {f'}^{1110} \left(u^1\right)^6-\frac{4}{3} s f_0 \tau_1^3
   {f'}_0^{111}+\frac{4}{3} s \ell_{111} \tau_1^3 f^1 \left(u^1\right)^2 {f'}_0^{111}\nonumber\\
& & \hskip1.5cm -\frac{20}{9} s f_1 \tau_1^3 \left(u^1\right)^2 {f'}_1^{111}+\frac{4}{9} s \ell_{111} \tau_1^3 f^0 \left(u^1\right)^4 {f'}_1^{111},  \nonumber\\
& (16): & \quad V_{hq} = 4 \, s \, h_0 \, \ell_{111} \tau_1 q^{11} \left(u^1\right)^2+\frac{4}{3} s h_1 \ell_{111} \tau_1 q^{10} \left(u^1\right)^4-\frac{4}{3} s \ell_{111}^2 \tau_1 h^1 q^{11} \left(u^1\right)^4\nonumber\\
& & \hskip1.5cm +4 s \ell_{111} \tau_1 h^1 \left(u^1\right)^2 q_0^1-\frac{16}{3} s h_1 \tau_1 \left(u^1\right)^2 q_1^1+\frac{4}{3} s \ell_{111} \tau_1 h^0 \left(u^1\right)^4 q_1^1,  \nonumber\\
& (17): & \quad V_{hp} = -\frac{4}{3} s^2 \ell_{111} \tau_1  \left(u^1\right)^3 \left(h_0 p^{10}-h_1 p^{11}-h^0 p_0^1+h^1 p_1^1 \right),  \nonumber\\
& (18): & \quad V_{hp'} = -\frac{4}{3} s h_0 \ell_{111} \tau_1^2 {p'}^{110} \left(u^1\right)^3+\frac{8}{3} s h_1 \ell_{111} \tau_1^2 {p'}^{111} \left(u^1\right)^3+\frac{1}{3} s \ell_{111}^2 \tau_1^2 h^1 {p'}^{110} \left(u^1\right)^5\nonumber\\
& & \hskip1.5cm -\frac{1}{3} s \ell_{111}^2 \tau_1^2 h^0 {p'}^{111} \left(u^1\right)^5-4 s h_1 \tau_1^2 u^1
   {p'}_0^{11}+\frac{4}{3} s \ell_{111} \tau_1^2 h^0 \left(u^1\right)^3 {p'}_0^{11}\nonumber\\
& & \hskip1.5cm +4 s h_0 \tau_1^2 u^1 {p'}_1^{11}-\frac{8}{3} s \ell_{111} \tau_1^2 h^1 \left(u^1\right)^3 {p'}_1^{11},  \nonumber\\
& (19): & \quad V_{hq'} = -2 s^2 h_0 \ell_{111} \tau_1^2 {q'}^{111} \left(u^1\right)^2-\frac{2}{3} s^2 h_1 \ell_{111} \tau_1^2 {q'}^{110} \left(u^1\right)^4+\frac{2}{3} s^2 \ell_{111}^2 \tau_1^2 h^1 {q'}^{111} \left(u^1\right)^4\nonumber\\
& & \hskip1.5cm -2 s^2 \ell_{111} \tau_1^2 h^1 \left(u^1\right)^2 {q'}_0^{11}+\frac{8}{3} s^2 h_1 \tau_1^2 \left(u^1\right)^2
   {q'}_1^{11}-\frac{2}{3} s^2 \ell_{111} \tau_1^2 h^0 \left(u^1\right)^4 {q'}_1^{11},  \nonumber\\
& (20): & \quad V_{hh'} = -\frac{4}{3} s h_0 \ell_{111} \tau_1^3 {h'}^{1111} \left(u^1\right)^2-\frac{4}{9} s h_1 \ell_{111} \tau_1^3 {h'}^{1110} \left(u^1\right)^4+\frac{5}{9} s \ell_{111}^2 \tau_1^3 h^1 {h'}^{1111} \left(u^1\right)^4 \nonumber\\
& & \hskip1.5cm +\frac{1}{27} s \ell_{111}^2 \tau_1^3 h^0 {h'}^{1110} \left(u^1\right)^6+\frac{4}{3} s h_0 \tau_1^3
   {h'}_0^{111}-\frac{4}{3} s \ell_{111} \tau_1^3 h^1 \left(u^1\right)^2 {h'}_0^{111} \nonumber\\
& & \hskip1.5cm +\frac{20}{9} s h_1 \tau_1^3 \left(u^1\right)^2 {h'}_1^{111}-\frac{4}{9} s \ell_{111} \tau_1^3 h^0 \left(u^1\right)^4 {h'}_1^{111}, \nonumber
\eea
\bea
\label{eq:cross1b}
& (21): & \quad V_{hf'} = \frac{4}{9} s^2 h_0 \ell_{111} \tau_1^3 {f'}^{1110} \left(u^1\right)^3-\frac{8}{9} s^2 h_1 \ell_{111} \tau_1^3 {f'}^{1111} \left(u^1\right)^3+\frac{1}{9} s^2 \ell_{111}^2 \tau_1^3 {f'}^{1111} h^0 \left(u^1\right)^5 \nonumber\\
& & \hskip1.5cm -\frac{1}{9} s^2 \ell_{111}^2 \tau_1^3 {f'}^{1110} h^1 \left(u^1\right)^5+\frac{4}{3} s^2 h_1 \tau_1^3
   u^1 {f'}_0^{111}-\frac{4}{9} s^2 \ell_{111} \tau_1^3 h^0 \left(u^1\right)^3 {f'}_0^{111}\nonumber\\
& & \hskip1.5cm -\frac{4}{3} s^2 h_0 \tau_1^3 u^1 {f'}_1^{111}+\frac{8}{9} s^2 \ell_{111} \tau_1^3 h^1 \left(u^1\right)^3 {f'}_1^{111}, \nonumber\\
& (22): & \quad V_{qp} = -\frac{4}{9} s \ell_{111}^2 \tau_1^2 p^{11} q^{10} \left(u^1\right)^5+\frac{4}{9} s \ell_{111}^2 \tau_1^2 p^{10} q^{11} \left(u^1\right)^5+\frac{20}{9} s \ell_{111} \tau_1^2 q^{10} \left(u^1\right)^3 p_0^1\nonumber\\
& & \hskip1.5cm -4 s \ell_{111} \tau_1^2 q^{11} \left(u^1\right)^3 p_1^1-\frac{20}{9} s \ell_{111} \tau_1^2 p^{10}
   \left(u^1\right)^3 q_0^1+\frac{16}{3} s \tau_1^2 u^1 p_1^1 q_0^1\nonumber\\
& & \hskip1.5cm +4 s \ell_{111} \tau_1^2 p^{11} \left(u^1\right)^3 q_1^1-\frac{16}{3} s \tau_1^2 u^1 p_0^1 q_1^1, \nonumber\\
& (23): & \quad V_{qp'} = -\frac{1}{9} \ell_{111}^2 \tau_1^3 {p'}^{111} q^{10} \left(u^1\right)^5+\frac{1}{9} \ell_{111}^2 \tau_1^3 {p'}^{110} q^{11} \left(u^1\right)^5+\frac{8}{9} \ell_{111} \tau_1^3 q^{10} \left(u^1\right)^3 {p'}_0^{11}\nonumber\\
& & \hskip1.5cm -\frac{4}{3} \ell_{111} \tau_1^3 q^{11} \left(u^1\right)^3 {p'}_1^{11}-\frac{8}{9} \ell_{111} \tau_1^3
   {p'}^{110} \left(u^1\right)^3 q_0^1+\frac{4}{3} \tau_1^3 u^1 {p'}_1^{11} q_0^1\nonumber\\
& & \hskip1.5cm +\frac{4}{3} \ell_{111} \tau_1^3 {p'}^{111} \left(u^1\right)^3 q_1^1-\frac{4}{3} \tau_1^3 u^1 {p'}_0^{11} q_1^1, \nonumber\\
& (24): & \quad V_{qq'} = s \ell_{111}^2 \tau_1^3 q^{11} {q'}^{111} \left(u^1\right)^4+\frac{1}{27} s \ell_{111}^2 \tau_1^3 q^{10} {q'}^{110} \left(u^1\right)^6-\frac{8}{3} s \ell_{111} \tau_1^3 {q'}^{111} \left(u^1\right)^2 q_0^1\nonumber\\
& & \hskip1.5cm -\frac{8}{9} s \ell_{111} \tau_1^3 {q'}^{110} \left(u^1\right)^4 q_1^1-\frac{8}{3} s \ell_{111} \tau_1^3 q^{11}
   \left(u^1\right)^2 {q'}_0^{11}+\frac{4}{3} s \tau_1^3 q_0^1 {q'}_0^{11}\nonumber\\
& & \hskip1.5cm -\frac{8}{9} s \ell_{111} \tau_1^3 q^{10} \left(u^1\right)^4 {q'}_1^{11}+4 s \tau_1^3 \left(u^1\right)^2 q_1^1 {q'}_1^{11}, \nonumber\\
& (25): & \quad V_{qh'} = \frac{2}{9} \ell_{111}^2 \tau_1^4 {h'}^{1111} q^{11} \left(u^1\right)^4-\frac{2}{3} \ell_{111} \tau_1^4 q^{11} \left(u^1\right)^2 {h'}_0^{111}-\frac{2}{9} \ell_{111} \tau_1^4 q^{10} \left(u^1\right)^4 {h'}_1^{111}\nonumber\\
& & \hskip1.5cm -\frac{2}{3} \ell_{111} \tau_1^4 {h'}^{1111} \left(u^1\right)^2 q_0^1-\frac{2}{9} \ell_{111} \tau_1^4
   {h'}^{1110} \left(u^1\right)^4 q_1^1+\frac{8}{9} \tau_1^4 \left(u^1\right)^2 {h'}_1^{111} q_1^1, \nonumber\\
& (26): & \quad V_{qf'} = \frac{1}{9} s \ell_{111}^2 \tau_1^4 {f'}^{1111} q^{10} \left(u^1\right)^5-\frac{1}{9} s \ell_{111}^2 \tau_1^4 {f'}^{1110} q^{11} \left(u^1\right)^5-\frac{4}{9} s \ell_{111} \tau_1^4 q^{10} \left(u^1\right)^3 {f'}_0^{111}\nonumber\\
& & \hskip1.5cm +\frac{8}{9} s \ell_{111} \tau_1^4 q^{11} \left(u^1\right)^3 {f'}_1^{111}+\frac{4}{9} s
   \ell_{111} \tau_1^4 {f'}^{1110} \left(u^1\right)^3 q_0^1-\frac{4}{3} s \tau_1^4 u^1 {f'}_1^{111} q_0^1\nonumber\\
& & \hskip1.5cm -\frac{8}{9} s \ell_{111} \tau_1^4 {f'}^{1111} \left(u^1\right)^3 q_1^1+\frac{4}{3} s \tau_1^4 u^1 {f'}_0^{111} q_1^1, \nonumber\\
& (27): & \quad V_{pp'} = -s \ell_{111}^2 \tau_1^3 p^{11} {p'}^{111} \left(u^1\right)^4-\frac{1}{27} s \ell_{111}^2 \tau_1^3 p^{10} {p'}^{110} \left(u^1\right)^6+\frac{8}{3} s \ell_{111} \tau_1^3 {p'}^{111} \left(u^1\right)^2 p_0^1 \nonumber\\
& & \hskip1.5cm +\frac{8}{9} s \ell_{111} \tau_1^3 {p'}^{110} \left(u^1\right)^4 p_1^1+\frac{8}{3} s \ell_{111} \tau_1^3 p^{11}
   \left(u^1\right)^2 {p'}_0^{11}-\frac{4}{3} s \tau_1^3 p_0^1 {p'}_0^{11} \nonumber\\
& & \hskip1.5cm +\frac{8}{9} s \ell_{111} \tau_1^3 p^{10} \left(u^1\right)^4 {p'}_1^{11}-4 s \tau_1^3 \left(u^1\right)^2 p_1^1 {p'}_1^{11}, \nonumber\\
& (28): & \quad V_{pq'} = \frac{1}{9} s^2 \ell_{111}^2 \tau_1^3 p^{11} {q'}^{110} \left(u^1\right)^5-\frac{1}{9} s^2 \ell_{111}^2 \tau_1^3 p^{10} {q'}^{111} \left(u^1\right)^5-\frac{8}{9} s^2 \ell_{111} \tau_1^3 {q'}^{110} \left(u^1\right)^3 p_0^1\nonumber\\
& & \hskip1.5cm +\frac{4}{3} s^2 \ell_{111} \tau_1^3 {q'}^{111} \left(u^1\right)^3 p_1^1+\frac{8}{9} s^2
   \ell_{111} \tau_1^3 p^{10} \left(u^1\right)^3 {q'}_0^{11}-\frac{4}{3} s^2 \tau_1^3 u^1 p_1^1 {q'}_0^{11}\nonumber\\
& & \hskip1.5cm -\frac{4}{3} s^2 \ell_{111} \tau_1^3 p^{11} \left(u^1\right)^3 {q'}_1^{11}+\frac{4}{3} s^2 \tau_1^3 u^1 p_0^1 {q'}_1^{11}, \nonumber\\
& (29): & \quad V_{ph'} = -\frac{1}{9} s \ell_{111}^2 \tau_1^4 {h'}^{1111} p^{10} \left(u^1\right)^5+\frac{1}{9} s \ell_{111}^2 \tau_1^4 {h'}^{1110} p^{11} \left(u^1\right)^5+\frac{4}{9} s \ell_{111} \tau_1^4 p^{10} \left(u^1\right)^3 {h'}_0^{111}\nonumber\\
& & \hskip1.5cm -\frac{8}{9} s \ell_{111} \tau_1^4 p^{11} \left(u^1\right)^3 {h'}_1^{111}-\frac{4}{9} s \ell_{111} \tau_1^4 {h'}^{1110} \left(u^1\right)^3 p_0^1+\frac{4}{3} s \tau_1^4 u^1 {h'}_1^{111} p_0^1\nonumber\\
& & \hskip1.5cm +\frac{8}{9} s \ell_{111} \tau_1^4 {h'}^{1111} \left(u^1\right)^3 p_1^1-\frac{4}{3} s \tau_1^4 u^1 {h'}_0^{111} p_1^1, \nonumber
\eea
\bea
\label{eq:cross1c}
& (30): & \quad V_{pf'} = \frac{2}{9} s^2 \ell_{111}^2 \tau_1^4 {f'}^{1111} p^{11} \left(u^1\right)^4-\frac{2}{3} s^2 \ell_{111} \tau_1^4 p^{11} \left(u^1\right)^2 {f'}_0^{111}-\frac{2}{9} s^2 \ell_{111} \tau_1^4 p^{10} \left(u^1\right)^4 {f'}_1^{111}\nonumber\\
& & \hskip1.5cm -\frac{2}{3} s^2 \ell_{111} \tau_1^4 {f'}^{1111} \left(u^1\right)^2 p_0^1-\frac{2}{9}
   s^2 \ell_{111} \tau_1^4 {f'}^{1110} \left(u^1\right)^4 p_1^1+\frac{8}{9} s^2 \tau_1^4 \left(u^1\right)^2 {f'}_1^{111} p_1^1, \nonumber\\
& (31): & \quad V_{p'q'} = \frac{1}{9} s \ell_{111}^2 \tau_1^4 {p'}^{111} {q'}^{110} \left(u^1\right)^5-\frac{1}{9} s \ell_{111}^2 \tau_1^4 {p'}^{110} {q'}^{111} \left(u^1\right)^5 -\frac{5}{9} s \ell_{111} \tau_1^4 {q'}^{110} \left(u^1\right)^3 {p'}_0^{11}\nonumber\\
& & \hskip1.5cm +s \ell_{111} \tau_1^4 {q'}^{111} \left(u^1\right)^3 {p'}_1^{11}+\frac{5}{9} s \ell_{111} \tau_1^4 {p'}^{110} \left(u^1\right)^3 {q'}_0^{11}-\frac{4}{3} s \tau_1^4 u^1 {p'}_1^{11} {q'}_0^{11}\nonumber\\
& & \hskip1.5cm -s \ell_{111} \tau_1^4 {p'}^{111} \left(u^1\right)^3 {q'}_1^{11}+\frac{4}{3} s \tau_1^4 u^1 {p'}_0^{11} {q'}_1^{11}, \nonumber\\
& (32): & \quad V_{p'h'} = \frac{1}{9} \ell_{111} \tau_1^5  \left(u^1\right)^3 \left({p'}^{110} {h'}_0^{111}-{p'}^{111} {h'}_1^{111}-{h'}^{1110} {p'}_0^{11}+{h'}^{1111} {p'}_1^{11}\right), \nonumber\\
& (33): & \quad V_{p'f'} = \frac{1}{9} s \ell_{111}^2 \tau_1^5 {f'}^{1111} {p'}^{111} \left(u^1\right)^4-\frac{1}{3} s \ell_{111} \tau_1^5 {p'}^{111} \left(u^1\right)^2 {f'}_0^{111} \nonumber\\
& & \hskip1.5cm -\frac{1}{9} s \ell_{111} \tau_1^5 {p'}^{110} \left(u^1\right)^4 {f'}_1^{111}-\frac{1}{3} s \ell_{111} \tau_1^5 {f'}^{1111} \left(u^1\right)^2 {p'}_0^{11} \nonumber\\
& & \hskip1.5cm -\frac{1}{9} s \ell_{111} \tau_1^5 {f'}^{1110} \left(u^1\right)^4 {p'}_1^{11}+\frac{4}{9} s \tau_1^5 \left(u^1\right)^2 {f'}_1^{111} {p'}_1^{11}, \nonumber\\
& (34): & \quad V_{q'h'} = -\frac{1}{9} s \ell_{111}^2 \tau_1^5 {h'}^{1111} {q'}^{111} \left(u^1\right)^4+\frac{1}{3} s \ell_{111} \tau_1^5 {q'}^{111} \left(u^1\right)^2 {h'}_0^{111}\nonumber\\
& & \hskip1.5cm +\frac{1}{9} s \ell_{111} \tau_1^5 {q'}^{110} \left(u^1\right)^4 {h'}_1^{111}+\frac{1}{3} s \ell_{111} \tau_1^5 {h'}^{1111} \left(u^1\right)^2 {q'}_0^{11}\nonumber\\
& & \hskip1.5cm +\frac{1}{9} s \ell_{111} \tau_1^5 {h'}^{1110} \left(u^1\right)^4 {q'}_1^{11}-\frac{4}{9} s \tau_1^5 \left(u^1\right)^2 {h'}_1^{111} {q'}_1^{11}, \nonumber\\
& (35): & \quad V_{q'f'} = \frac{1}{9} \, s^2 \, \ell_{111} \, \tau_1^5  \left(u^1\right)^3 \left({q'}^{110} {f'}_0^{111}-{q'}^{111} {f'}_1^{111}-{f'}^{1110} {q'}_0^{11}+{f'}^{1111} {q'}_1^{11}\right), \nonumber\\
& (36): & \quad V_{h'f'} = \frac{1}{27} \, s \, \ell_{111} \, \tau_1^6 \left(u^1\right)^3 \left({h'}^{1110} {f'}_0^{111}-{h'}^{1111} {f'}_1^{111}-{f'}^{1110} {h'}_0^{111}+{f'}^{1111} {h'}_1^{111}\right). \nonumber
\eea


\bibliographystyle{utphys}
\bibliography{reference}

\end{document}